\documentclass[paper]{JHEP3}
\usepackage{epsfig}
\usepackage{amsmath}
\usepackage{amssymb}
\usepackage{amsbsy}
\usepackage{enumerate}
\usepackage{bbm}
\usepackage{url}

\usepackage[hang,sf]{subfigure}

\newenvironment{enumerateroman}{\begin{enumerate}[i.] }{\end{enumerate}}

\def\beq{\begin{equation}}
\def\beqn{\begin{eqnarray}}
\def\eeq{\end{equation}}
\def\eeqn{\end{eqnarray}}

\newcommand\PYTHIA{{\tt PYTHIA}}

\newcommand\MCFM{{\tt MCFM}}

\newcommand\BlackHat{{\tt BlackHat}}

\newcommand\LambdaQCD{\Lambda_{\rm\scriptscriptstyle QCD}}

\newcommand\MINT{{\tt MINT}}

\def\lq{\left[} 
\def\rq{\right]}

\def\({\left(} 
\def\){\right)}

\newcommand\sss{\mathchoice%
{\displaystyle}%
{\scriptstyle}%
{\scriptscriptstyle}%
{\scriptscriptstyle}%
}

\newcommand\elpelm{\ell^+ \ell^-}

\newdimen\hbigcirc
\newdimen\wbigcirc

\newdimen\figwidth

\figwidth=\textwidth

\newcommand\as{\alpha_{\sss\rm S}}

\newcommand\pt{p_{\sss\rm T}}

\newcommand\kt{k_{\sss\rm T}}

\newcommand\matB{{\cal B}}

\newcommand\MCatNLO{{\tt MC@NLO}}

\newcommand     \MSB            {\ifmmode {\overline{\rm MS}} \else
                                 $\overline{\rm MS}$  \fi}

\newcommand\CA{C_{\sss\rm A}}

\newcommand\CF{C_{\sss\rm F}}

\newcommand\POWHEG{{\tt POWHEG}}
\newcommand\POWHEGBOX{{\tt POWHEG BOX}}

\newcount\minutes 
\newcount\scratch 
 
\def\timestamp{%
\scratch=\time 
\divide\scratch by 60 
\edef\hours{\the\scratch} 
\multiply\scratch by 60 
\minutes=\time 
\advance\minutes by -\scratch 
---$\,$\hours:\null 
\ifnum\minutes< 10 0\fi 
\the\minutes}

\newcommand\ptsupp{{\tt bornsuppfact}}

\newcommand\wneg{{\tt withnegweights}}
\newcommand\kgen{k_{\rm gen}}
\newcommand\kan{k_{\rm an}}

\newcommand\ptZ{p_{\sss T}^{\sss Z}}
\preprint{
DESY 10-158\\
SFB/CPP-10-85\\
IPPP/10/78\\
DCPT/10/156}

\title{Vector boson plus one jet production in
  \POWHEG}  
\author{Simone Alioli\\
  Deutsches Elektronen-Synchrotron DESY\\
  Platanenallee 6, D-15738 Zeuthen, Germany\\
  and INFN, Sezione di Milano-Bicocca \\
  E-mail: \email{simone.alioli@desy.de}}
\author{Paolo Nason\\
  INFN, Sezione di Milano-Bicocca,
  Piazza della Scienza 3, 20126 Milan, Italy\\
  E-mail: \email{Paolo.Nason@mib.infn.it}}
\author{Carlo Oleari\\
  Universit\`a di Milano-Bicocca and INFN, Sezione di Milano-Bicocca\\
  Piazza della Scienza 3, 20126 Milan, Italy\\
  E-mail: \email{Carlo.Oleari@mib.infn.it}}
\author{Emanuele Re\\
  Institute for Particle Physics Phenomenology, Department of Physics\\
  University of Durham, Durham, DH1 3LE, UK\\
  and INFN, Sezione di Milano-Bicocca \\
  E-mail: \email{emanuele.re@durham.ac.uk}} \abstract{
  We present an implementation of the next-to-leading order vector boson plus
  one jet production process in hadronic collision in the framework of
  \POWHEG{}, which is a method to implement NLO calculations within a Shower
  Monte Carlo context. All spin correlations in the vector boson decay
  products have been taken into account.  The process has been implemented in
  the framework of the \POWHEGBOX{}, an automated computer code for turning
  a NLO calculation into a shower Monte Carlo program.  We present
  phenomenological results for the case of the $Z/\gamma$ plus one jet
  production process, obtained by matching the \POWHEG{} calculation with the
  shower performed by \PYTHIA{}, for the LHC, and we compare our results with
  available Tevatron data.
}

\begin{document}

\section{Introduction}

The process of vector boson production in association with jets plays an
important role at present hadron colliders. In the early LHC phase, $Z$ plus
one jet production will play a major role in jet calibration. Furthermore,
vector boson production in association with jets is an important background
to new physics signals. In particular, when the $Z$ decays into neutrinos, it
becomes a source of missing-energy signals. The analogous process of $W$
production, in association with jets, gives origin to an important background
to the production of a lepton in association with missing energy and jets.

$Z$ plus one jet production has a very distinctive signature, when the $Z$
decays into muons or electrons, and several studies have been performed at
the Tevatron, both by the CDF~\cite{Aaltonen:2007cp} and by the D0
Collaboration~\cite{Abazov:2006gs,Abazov:2008ez,Abazov:2009av,Abazov:2009pp}.
They are all carried out by correcting the measured quantities to the
particle level, according to the recommendations developed in the 2007 Les
Houches workshop~\cite{Buttar:2008jx}, and then compared to theoretical
calculations, computed mostly at the next-to-leading order (NLO) level, and
corrected for showering, hadronization and underlying-event effects. These
corrections, in turn, are extracted from shower Monte Carlo~(SMC) programs.

In the present work, we present a calculation for the NLO cross section for
vector boson plus one jet that can be interfaced to a shower Monte Carlo
program, within the \POWHEG{} framework~\cite{Nason:2004rx,Frixione:2007vw}.
This is the first time that such calculation has been performed.  More
specifically, we have implemented this process using the \POWHEGBOX{}, a
general computer code framework for embedding NLO calculations in shower
Monte Carlo programs according to the \POWHEG{} method.  In fact, the
\POWHEGBOX{} framework was developed using the $Z/\gamma+1j$ process as its
first testing example.\footnote{In the rest of the paper, for ease of
  notation, we will refer to the $Z/\gamma+1j$ process simply as ``$Z+1j$
  production'', without mentioning the presence of the photon, whose effects,
  together with all the spin correlations of the decay products, have been
  fully taken into account.}

It is clear that, by using the \POWHEG{} implementation of $V+1j$ production,
the comparison of the theoretical prediction with the experimental results is
eased considerably, and is also made more precise. Rather than estimating
shower and underlying-event corrections using a parton shower program, one
interfaces directly the parton shower to the hard process in question,
yielding an output that can be compared to the experimental results at the
particle level. We will carry out this task for the $Z+1j$ case, and compare
our results to the Tevatron findings.  We remark, however, that a further
improvement to this study could be carried out, by using the \POWHEG{}
program to generate the events that are fed through the detector simulation,
and are directly compared to raw data. We are not, of course, in a position
to perform such a task, that should instead be carried out by the
experimental collaborations.

The paper is organized as follows. In section~\ref{sec:calculation} we give a
brief description of our calculation.  Since we used the \POWHEGBOX{} to
implement our process, we refer the reader to the \POWHEGBOX{}
publication~\cite{Alioli:2010xd}, and report here only a few
details that are particularly relevant to the process in question.  In the
following, we always consider the $Z+1j$ case, since the $W+1j$ is fully
analogous. In section~\ref{sec:validation} we describe the generation of the
$Z+1j$ sample, together with some consistency checks of our calculation, and
new features added to the \POWHEGBOX{}.  In section~\ref{sec:phenomenology},
we present results for the Tevatron and for the LHC at 14~TeV. Comparison
with available data from the CDF and D0 Collaborations are carried
out. Finally, in section~\ref{sec:conclusions}, we give our final remarks.

\section{Description of calculation}
\label{sec:calculation}
We have considered the hadroproduction of a single vector boson plus one jet
at NLO order, with all spin correlations from the decay taken into account.
All fermions (quarks and leptons) have been treated in the massless limit (no
top-quark contributions have been taken into account).

In order to implement a new process at NLO into the \POWHEGBOX{}, we have
to provide the following ingredients:
\begin{enumerateroman}
\item The list of all flavour structures of the Born processes.

\item The list of all flavour structures of the real processes.
 
\item  The Born phase space.
  
\item \label{item:real} The real matrix elements squared for all relevant
  partonic processes.

\item \label{item:virtual} The finite part of the virtual corrections
  computed in dimensional regularization or in dimensional
  reduction.

\item \label{item:born} The Born squared amplitudes ${\cal B}$, the colour
  correlated ones ${\cal B}_{ij}$ and spin correlated ones ${\cal B}_{\mu\nu}$.
    
\item The Born colour structures in the limit of a large  number of colours.
\end{enumerateroman}

For the case at hand, the list of processes is generated going through all
possible massless quarks and gluons that are compatible with the production
of the vector boson plus an extra parton.

The Born phase space for this process poses no challenges: we generate the
momentum of the vector boson distributed according to a Breit-Wigner
function, plus one extra light particle. The vector boson momentum is then
further decayed into two momenta, describing the final-state leptons. At this
stage, the momentum fractions $x_1$ and $x_2$ are also generated and the
momenta of the incoming partons are computed.  
\figwidth=0.3\textwidth
\begin{figure}[htb]
\begin{center}
\subfigure[Born]{
\epsfig{file=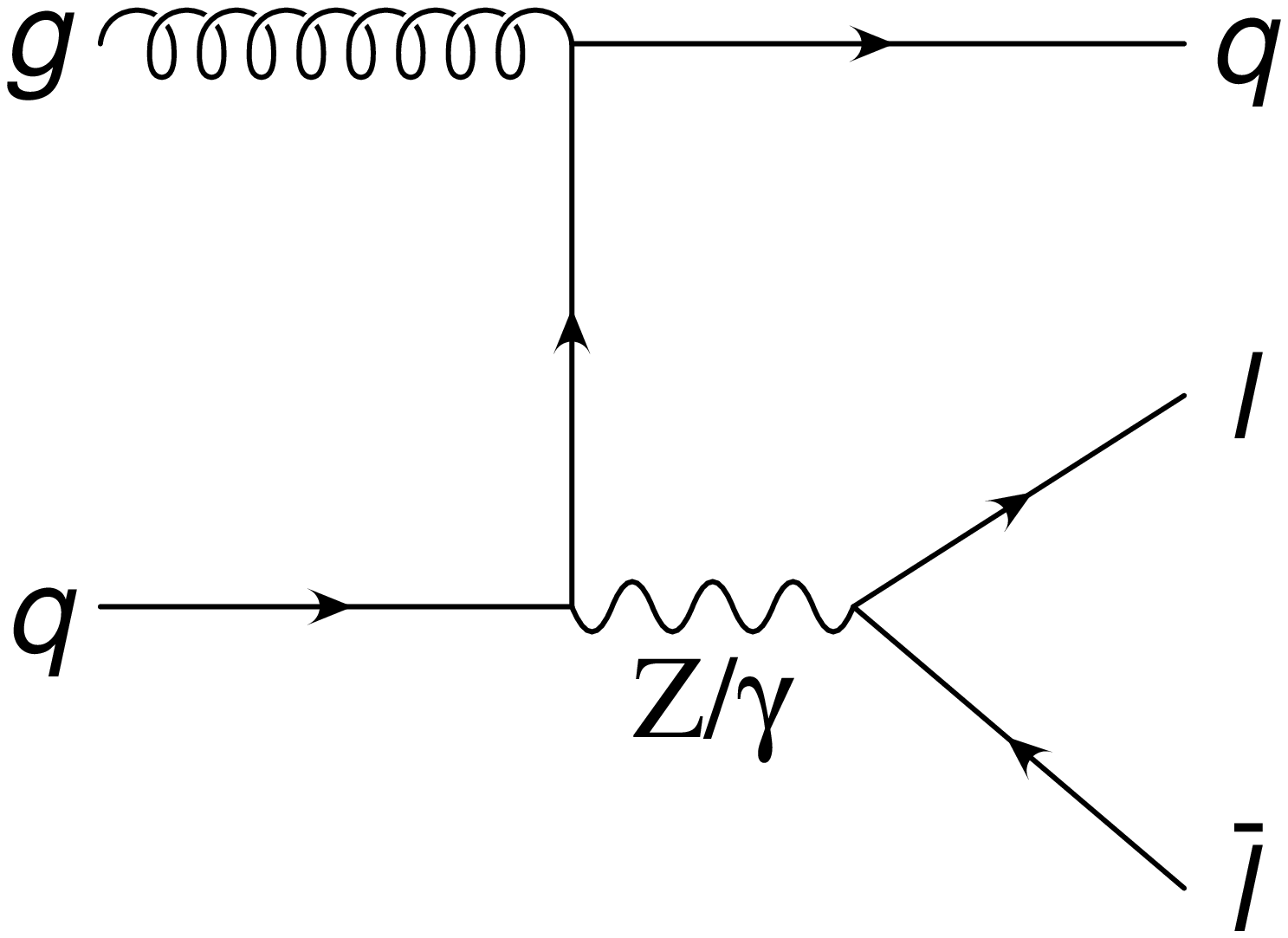,width=\figwidth}}\quad
\subfigure[Born]{
\epsfig{file=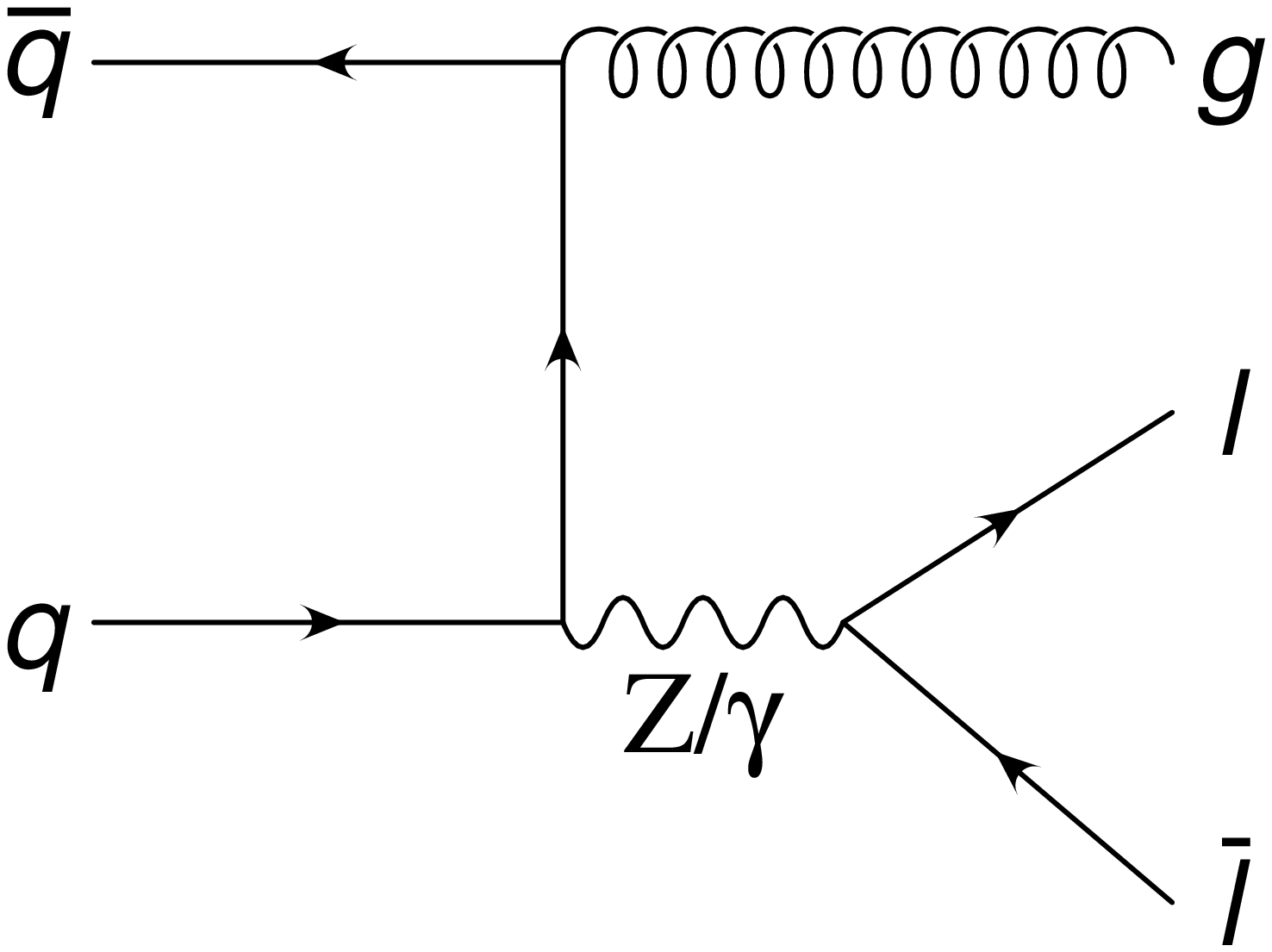,width=\figwidth}}\quad
\subfigure[virtual]{
\epsfig{file=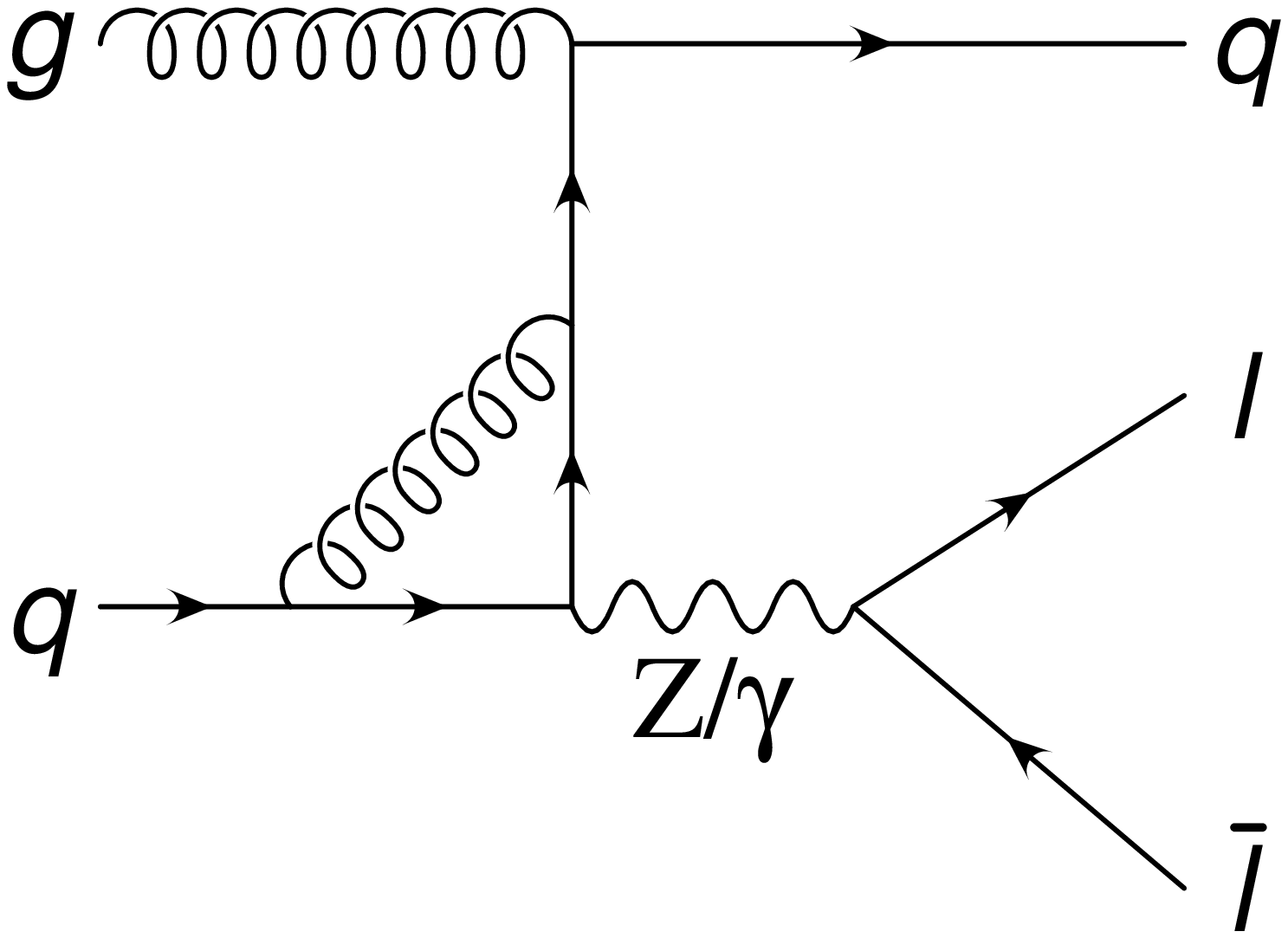,width=\figwidth}}

\vspace{0.3cm}
\subfigure[virtual]{
\epsfig{file=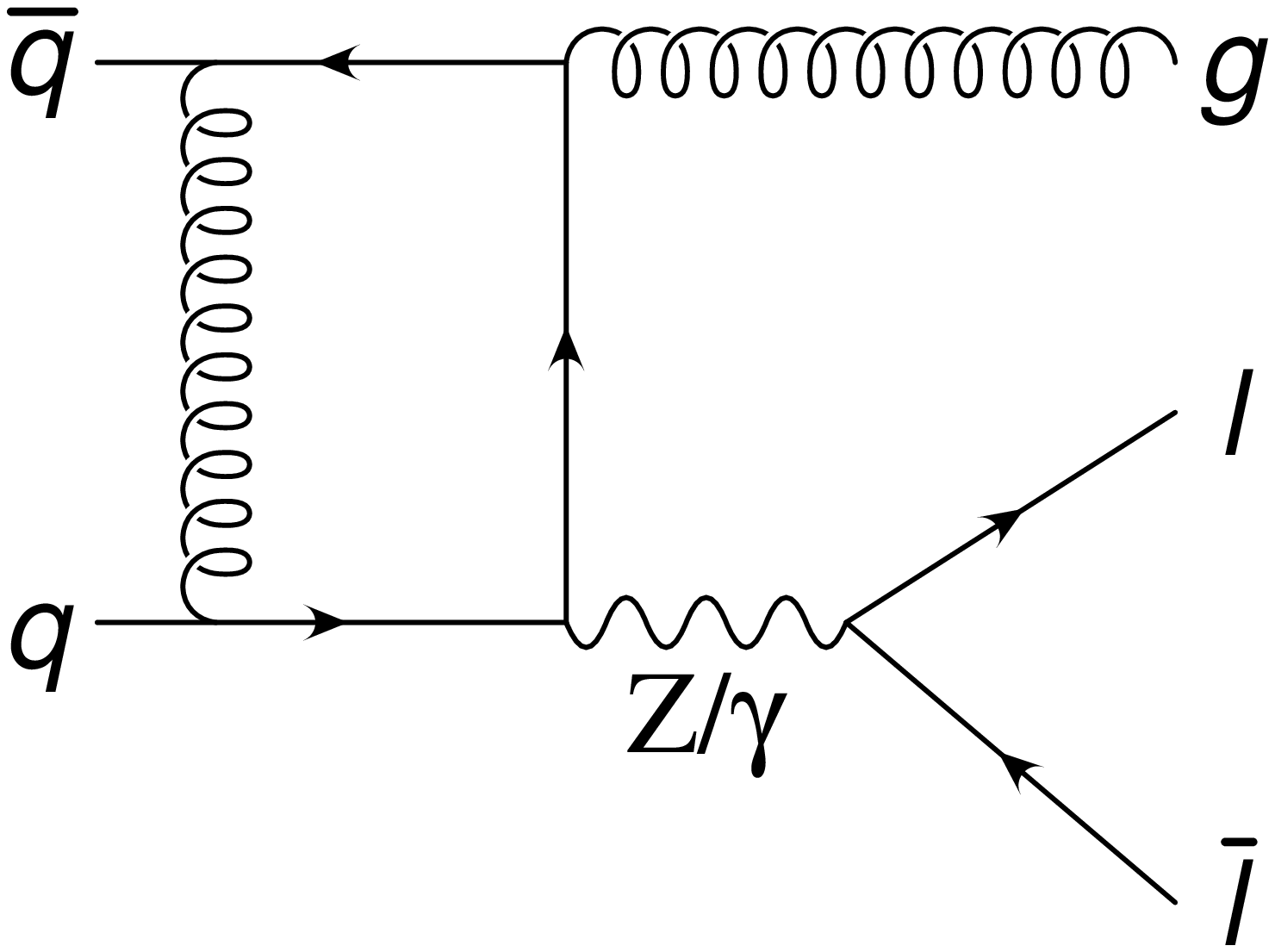,width=\figwidth}}\quad
\subfigure[real]{
\epsfig{file=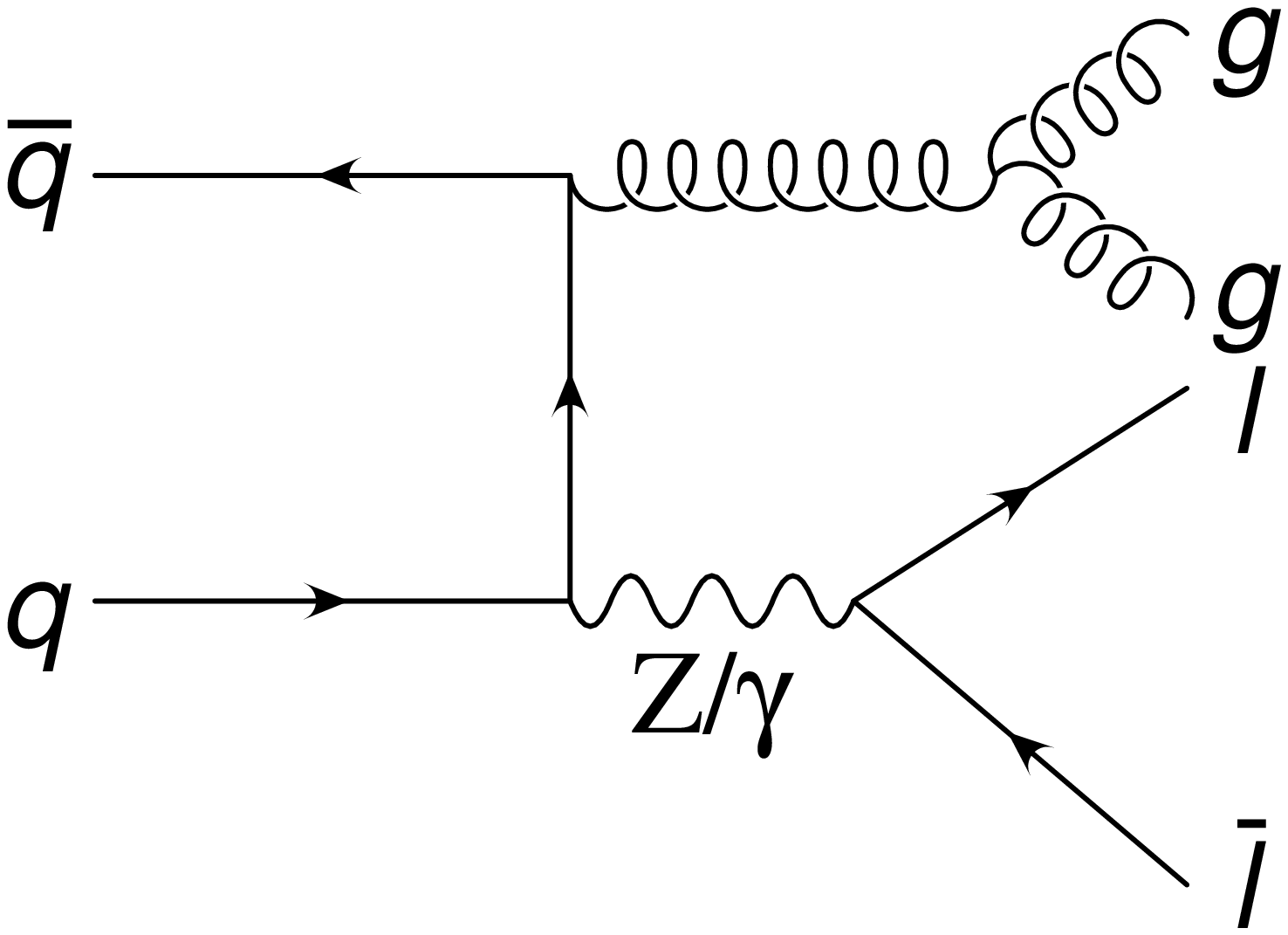,width=\figwidth}}\quad
\subfigure[real]{
\epsfig{file=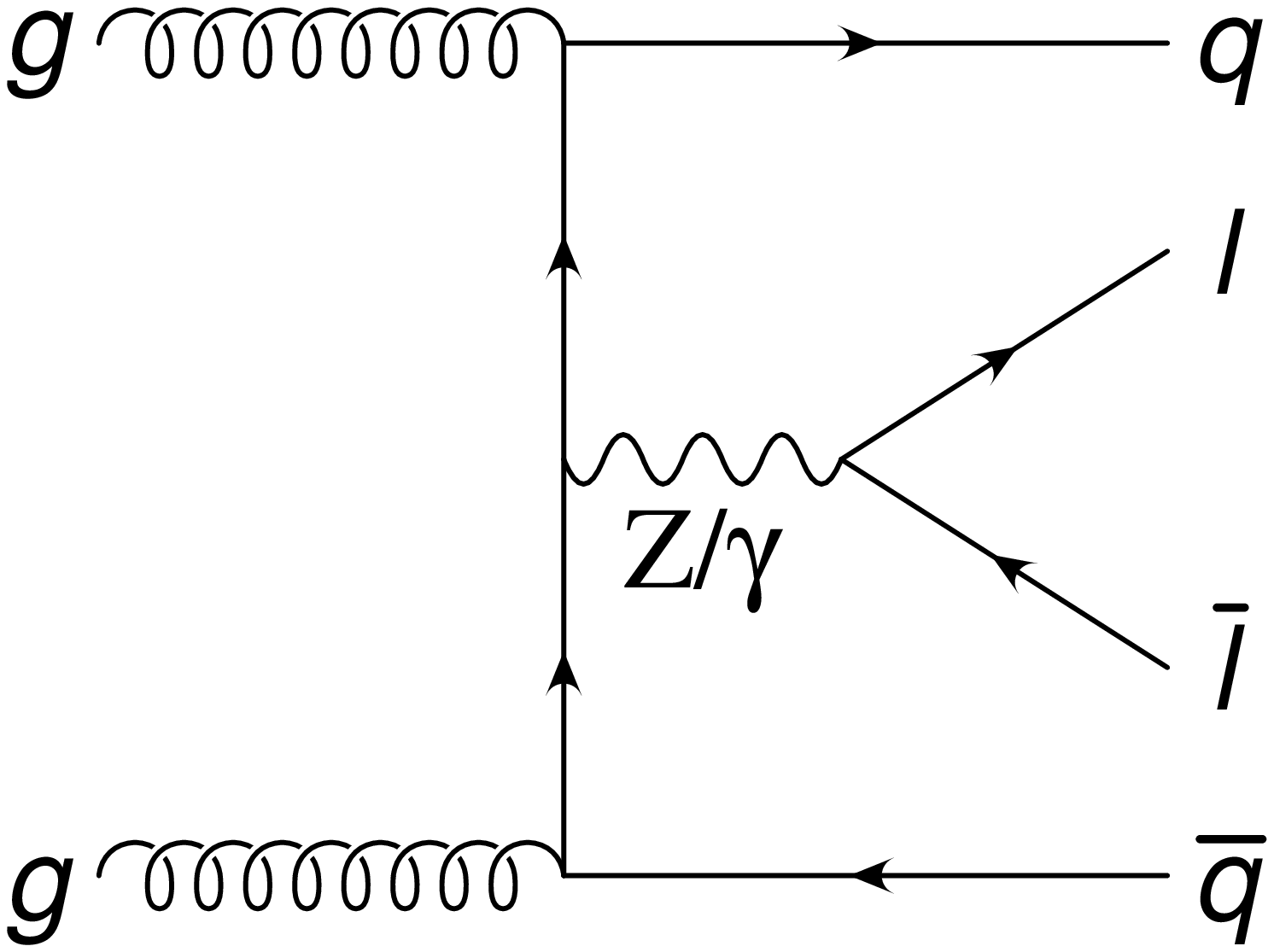,width=\figwidth}}
\end{center}
\caption{\label{fig:graphs} Sample graphs for the Born, virtual and real
  contributions to the $Z/\gamma+1j$ production process.}
\end{figure}

A sample of Feynman diagrams that contribute to the $Z+1j$ process at the
Born level (${\cal B}$) is depicted in panels~(a) and~(b) of
fig.~\ref{fig:graphs}. Together with the Born diagrams, we have to consider
the one-loop corrections to the tree level graphs, and the diagrams with an
extra radiated parton.  A sample of virtual and real contributions is
depicted in panels~(c)--(f) of fig.~\ref{fig:graphs}.

We have computed the Born and real contributions ourselves, using the
helicity-amplitude technique of refs.~\cite{Hagiwara:1985yu,Hagiwara:1988pp}.
The amplitudes are computed numerically in a {\tt fortran} code, as complex
numbers, and squared at the end.  The finite part of the virtual corrections
has been taken from the \MCFM{} program~\cite{Campbell:2002tg}, that uses the
virtual matrix elements given in ref.~\cite{Bern:1997sc} and computed first
in ref.~\cite{Giele:1991vf}.

The use of the helicity-amplitude technique to compute the amplitudes at the
Born level facilitates the calculation of the spin-correlated matrix elements
${\cal B}_{\mu\nu}$. In fact, they are the Born amplitudes just before being
numerically contracted with the polarization vector of the initial- or
final-state gluon.

Since we are dealing with three coloured partons at the Born level, the colour
correlated Born amplitudes ${\cal B}_{ij}$ are all proportional to the Born one.
The two independent colour-correlated matrix elements are given by
\beq
B_{q q'}=\frac{1}{2}\(2 \CF-\CA\) {\cal B}\,,\qquad \qquad B_{q g}=\frac{\CA}{2} {\cal B}\,,
\eeq
and, from  colour conservation,  they satisfy
\begin{equation}\label{eq:colcons}
\sum_{i,i\ne j}\matB_{ij}=C_{f_j}\matB\,,
\end{equation}
where $i$ runs over all coloured particles entering or exiting the process,
and $C_{f_j}$ is the Casimir constant for the colour representation of
particle $j$.

\begin{figure}[htb]
\begin{center}
\subfigure[]{
\epsfig{file=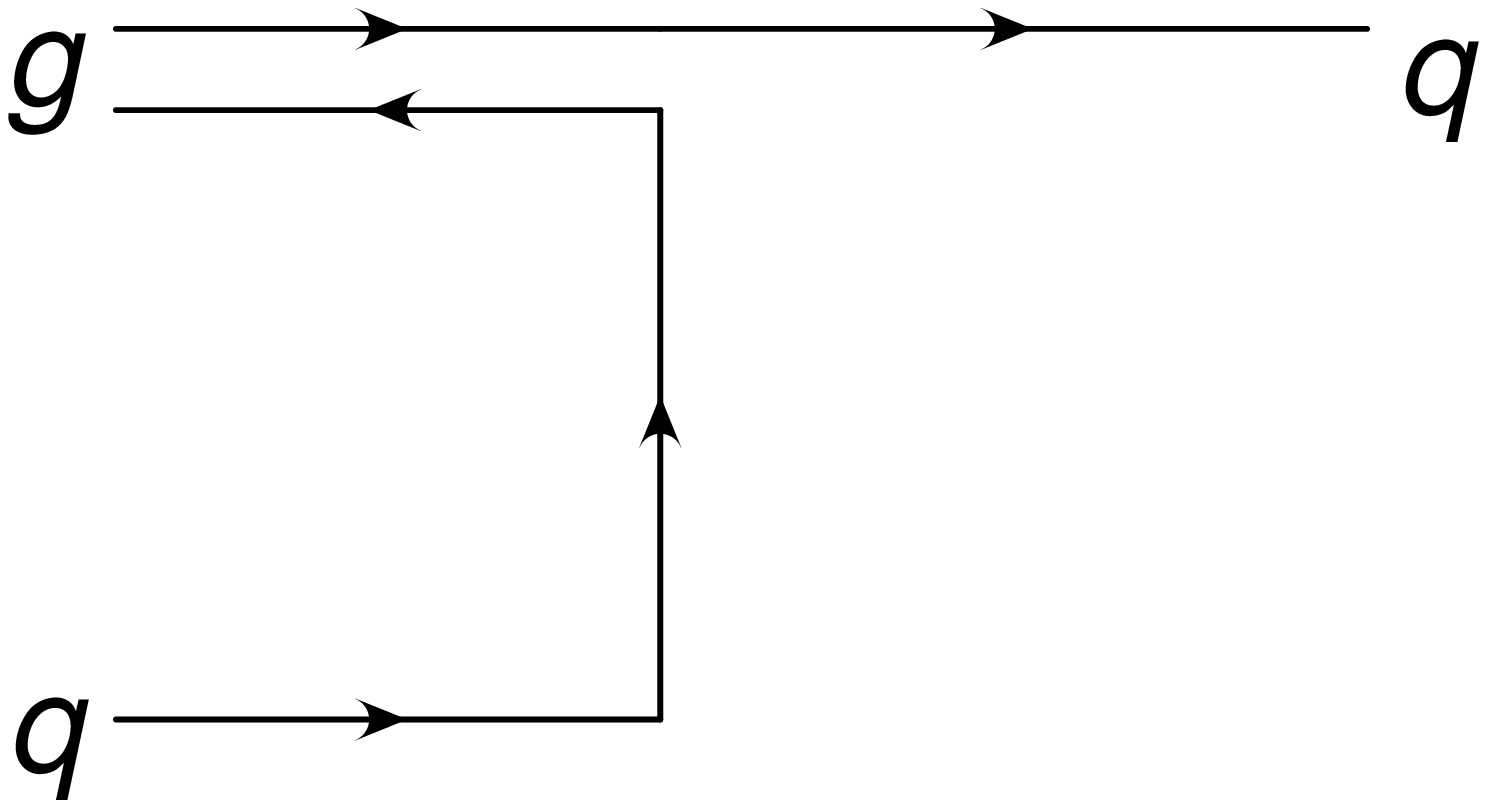,width=\figwidth}}\qquad\qquad
\subfigure[]{
\epsfig{file=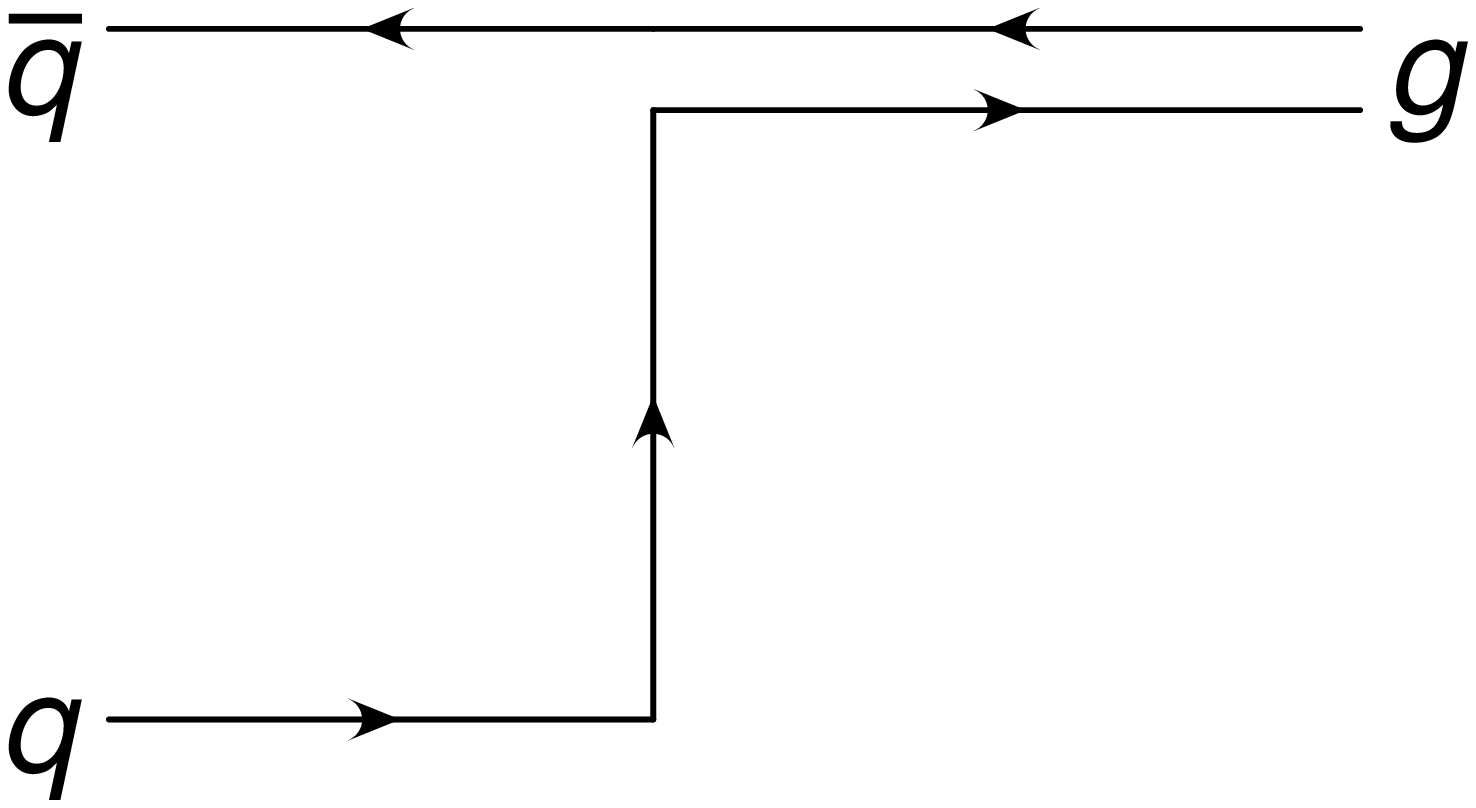,width=\figwidth}}
\end{center}
\caption{\label{fig:colors} Born colour structures of the two
  Feynman diagrams illustrated in panels~(a) and~(b) in fig.~\ref{fig:graphs}.}
\end{figure}

Finally, the Born colour structures, in the limit of a large number of
colours, are straightforward, due to the presence of a single gluon and two
quarks at the Born level. For the two Born diagrams depicted in
fig.~\ref{fig:graphs}, the colour structures are shown in
fig.~\ref{fig:colors}.

\subsection{Generation cut and Born suppression factor}
The $V+1j$ process differs substantially from all processes previously
implemented in \POWHEG{}, in the fact that the Born contribution itself is
collinear and infrared divergent. In all previous implementations, the Born
diagrams were finite, and it was thus possible to generate an unweighted set
of underlying Born configurations covering the whole phase space.  In the
present case, this is not possible, since they would all populate the very
low transverse momentum region. Of course, this problem is also present in
standard shower Monte Carlo programs, where it is dealt with by generating
the Born configuration with a cut $\kgen$ on the transverse momentum of the
$V$ boson. After the shower, one must discard all events that fail some
transverse momentum analysis cut $\kan$ in order to get a realistic
sample. The analysis cut $\kan$ may be applied to the transverse momentum of
the vector boson, or to the hardest jet.  We assume here, for sake of
discussion, that it is applied to the vector boson transverse momentum.

Taking $\kan\gtrsim \kgen$ is not enough to get a realistic sample. In fact,
in an event generated at the Born level with a given $\kt < \kgen$, the
shower may increase the transverse momentum of the jet so that the final
vector-boson transverse momentum $\kt^V$ can be bigger than $\kan$.
Thus, even if the generation cut is below the analysis cut,
it may reduce the number of events that pass the analysis cut.  Of course, as we
lower $\kgen$ keeping $\kan$ fixed, we will reach a point where very few
events below $\kgen$ will pass the analysis cut $\kan$.  In fact, generation
of radiation with transverse momentum larger than $\kgen$ is strongly
suppressed in \POWHEG{}, and, in turn, radiation from subsequent shower is
required to be not harder than the hardest radiation of \POWHEG{}. Thus, if
we want to generate a sample with a given $\kan$ cut, we should choose
$\kgen$ small enough, so that the final sample remains substantially the same
if $\kgen$ is lowered even further.

Together with the option of generating events with a generation cut $\kgen$,
a second option for the implementation of processes with a divergent Born
contribution is also available. It requires that we generate weighted events,
rather than unweighted ones. This is done by using a suppressed cross section
for the generation of the underlying Born configurations
\begin{equation}
\label{eq:supp_fac}
\bar{B}_{\rm supp}=\bar{B}\times F(\kt)\,,
\end{equation}
where $\bar{B}$ is the inclusive NLO cross section at fixed underlying Born
variables, $\kt$ is the transverse momentum of the vector boson in the
underlying Born configuration and $F(\kt)$ is a function that goes to zero in
the $\kt \to 0$ limit fast enough to make $\bar{B}\times F(\kt)$ finite in
this limit.  In this way, $\bar{B}_{\rm supp}$ is integrable, and one can use
it to generate underlying Born configurations according to its value. The
generated events, however, should be given a weight $1/F(\kt)$, rather than
1, in order to compensate for the initial $F(\kt)$ suppression factor. With
this method, events do not concentrate in the low $\kt$ region, although
their weight, in the low $\kt$ region, becomes divergent. After shower, if
one imposes the analysis cut, one gets a finite cross section, since it is
unlikely that events with small transverse momentum at the Born level may
pass the cut after shower.

In recent \POWHEGBOX{} revisions, both methods can be implemented at the
same time. We wanted in fact to be able to implement the following three
options:
\begin{itemize}
\item Generate events using a transverse momentum generation cut $\kgen$.
\item Generate events using a Born suppression factor, and a small transverse
  momentum cut, just enough to avoid unphysical values of the strong coupling
  constant and of the factorization scale that appears in the parton
  distribution functions. In this case, since we are also generating events
  with very small transverse momenta, radiative corrections may become larger
  than the Born term, and negative-weight events could be generated.  We
  should then be prepared to track them. This issue will be further discussed
  in the next section.
\item Apply a Born suppression factor, and set the transverse momentum cut
  $\kgen$ to zero. In this case the program cannot be used to generate
  events.  It can be used, however, to produce NLO fixed order distributions,
  provided the renormalization and factorization scales are set in such a way
  that they remain large enough even at small transverse momentum
  $\kt^V$. This feature is only used for the generation of fixed-order
  distributions.
\end{itemize}
The generation cut is activated by setting the token {\tt bornktmin} to the
desired value in the {\tt powheg.input} file. The Born suppression is
activated by setting the token \ptsupp{} to a positive real value. The
process-specific subroutine {\ttfamily born\_suppression} sets the
suppression factor of eq.~(\ref{eq:supp_fac}) to
\begin{equation}
F(\kt)  = \frac{\kt^2}{\kt^2+\ptsupp^2} \,.  
\end{equation}
If \ptsupp{} is negative, the suppression factor is set to 1.

The need of a transverse momentum cut is not only a technical issue.  The NLO
calculation of $V+1j$ production holds only if the transverse momentum of the
vector boson is not too small. In fact, as the $\kt$ decreases, large Sudakov
logarithms arise in the NLO correction, and the value of the running coupling
increases, up to the point where the cross section, at fixed order, becomes
totally unreliable. These large logarithms should all be resummed in order to
get a sensible answer in this region. In the \POWHEG{} implementation of
single vector-boson production~\cite{Alioli:2008gx}, in fact, these logarithms
are all resummed.  Then it is clear that some sort of merging between the
$V+1j$ and the $V$ production processes should be performed at relatively
small transverse momentum, in order to properly deal with this problem.  Here
we will not attempt to perform such merging, that we leave for future
work. We will simply recall, when looking at our results, that we expect to
get unphysical distributions when the vector boson transverse momentum is too
small.  We will discuss this fact in a more quantitative way in
section~\ref{sec:validation}.

\subsection{Negative-weight events}
In the \POWHEG{} method, negative-weight events can only arise if one is
approaching a region where the NLO computation is no longer feasible.  In our
study for the $V+1j$ process, we approach this region at small transverse
momenta. In order to better understand what happens there, rather than
neglecting negative weights (that is the default behaviour of the
\POWHEGBOX{}), we have introduced a new feature in the program, that allows
one to track also the negative-weight events. This feature is activated by
setting the token \wneg{} to 1~(true). If \wneg{} is set to 1, events with
negative weight can thus appear in the Les Houches event
file~\cite{Boos:2001cv,Alwall:2006yp}.  While we normally set the {\tt
  IDWTUP} flag in the Les Houches interface to 3, in this case we set it to
-4. With this flag, the SMC is supposed to simply process the event, without
taking any other action. Furthermore, the {\tt XWGTUP} (Les Houches) common
block variable is set by the \POWHEGBOX{} to the sign of the event times the
integral of the absolute value of the cross section, in such a way that its
average equals the true total cross section.

We preferred not to use the option {\tt IDWTUP}=-3 for signed events with
constant absolute value. This option is advocated by the Les Houches
interface precisely in such cases, and it requires that the event weight {\tt
  XWGTUP} assumes the values $\pm 1$.  However, the Les Houches interface
does not provide a standard way to store the integral of the absolute value
of the cross section, that would be needed to compute correctly the weight of
the event in this case. In fact, the {\tt XSECUP} variable is reserved for
the true total cross section $ \sigma_{\rm NLO}$.
More specifically, suppose that the total NLO cross section receives
contributions from regions in the phase space where the differential cross
section is positive, and where it is negative, so that we can write
\begin{equation}
 \sigma_{\rm NLO} = \sigma_{(+)}-|\sigma_{(-)}|\,.
\end{equation}
We then generate a sample of $N=N_+ + N_-$, with $N_+$ events with weight
$W_i=+1$ (so that the sum of the $W_i$ on events with positive weights is
$N_+$) and $N_-$ events with weight $W_i=-1$ (so that the sum of the $W_i$ on
events with negative weights is $-N_-$) in such a way that
\begin{equation}
N_+ = \frac{\sigma_{(+)}}{\sigma_{(+)}+|\sigma_{(-)}|} \, N\,, \qquad 
N_- = \frac{\sigma_{(-)}}{\sigma_{(+)}+|\sigma_{(-)}|} \,N\,.
\end{equation}
In order to give the correct NLO cross section, the $N$ events should then be
weighted with the sum of the positive plus the absolute value of the
negative part of the cross section, in such a way that
\begin{equation}
  \frac{1}{N}\sum_{i=1}^N W_i \left(\sigma_{(+)}+|\sigma_{(-)}|\right)=
  \sigma_{(+)}-|\sigma_{(-)}| = \sigma_{\rm NLO}\,.
\end{equation}
Summarizing, since there is no standard way in the Les Houches interface to
pass this absolute value when {\tt IDWTUP}=-3, we set it to -4, and set the
weight {\tt XWGTUP} of the $i$-th event to $W_i \times
\left(\sigma_{(+)}+|\sigma_{(-)}|\right)$, so that, event by event, we have
this information. In this case the average value of the {\tt XWGTUP}
variable is equal to the total cross section, as required by the Les Houches
interface when {\tt IDWTUP}=-4.

Notice that, if \wneg{} is set to true and a Born suppression factor is also
present, the events will have variable {\tt XWGTUP} of either signs. In this
case, {\tt XWGTUP} is set to the sign of the event, times the absolute value
of the cross section, divided by the suppression factor (the output of the
{\ttfamily born\_suppression} routine). Also in this case, the average value
of {\tt XWGTUP} coincides with the true total cross section. 
Weighted events are also useful if one wants to generate a homogeneous sample
from relatively-low up to very-high transverse momenta. In this case, it is
convenient to pick a very large \ptsupp{} value, of the order of the maximum
transverse momentum one is interested in. The large-momentum region will be
more populated in this way. The form of the {\ttfamily born\_suppression}
function can also be changed at will by the user.

\section{Validation of the generated samples}
\label{sec:validation}
\begin{figure}
\epsfig{file=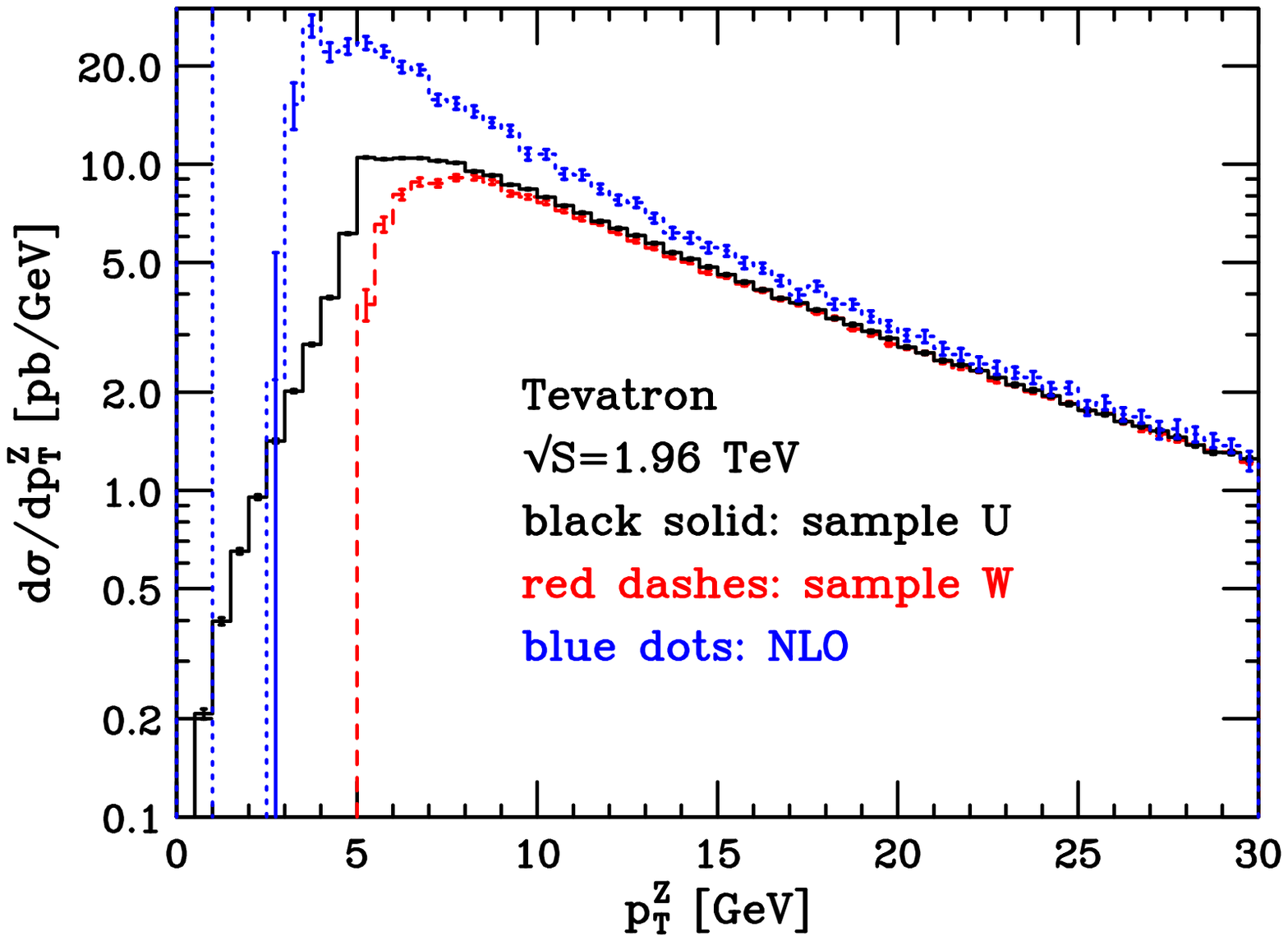,width=0.48\textwidth}
\epsfig{file=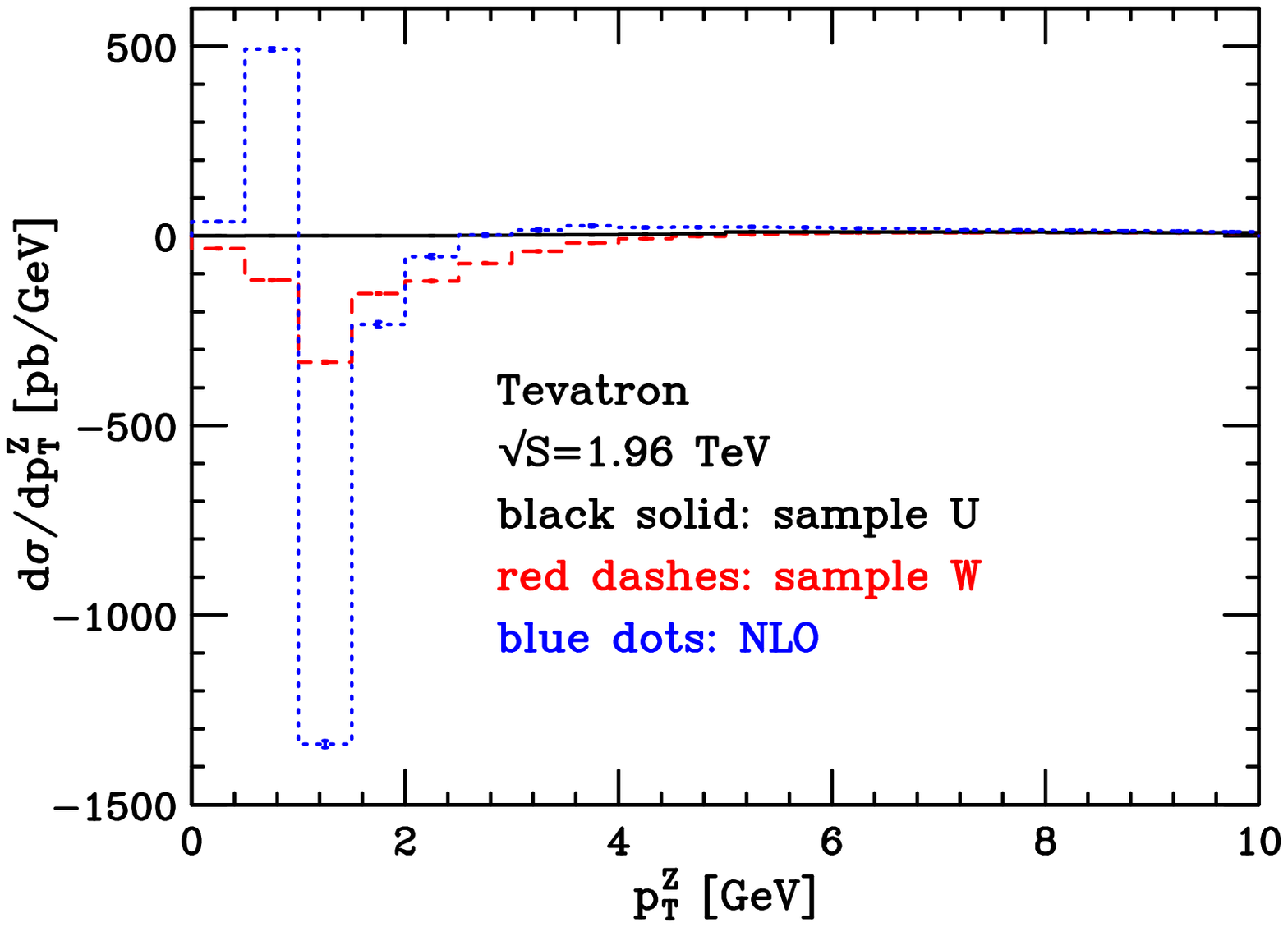,width=0.48\textwidth}
\caption{\label{fig:UW_ptZ}
$Z$ transverse momentum in samples U, W and at fixed NLO order.}
\end{figure}

In this section we discuss the output of the \POWHEGBOX{} for $Z+1j$
production.  We consider here $p\bar{p}$ collisions at $1.96$~TeV. We use the
CTEQ6M pdf set~\cite{Pumplin:2002vw}. The factorization and renormalization
scales (in the computation of the $\bar{B}$ function) are fixed to the
transverse momentum of the vector boson of the underlying Born configuration.
For the generation of radiation, these scales are set by the \POWHEGBOX{},
according to the prescriptions given in~\cite{Alioli:2010xd}.  We have first
produced one sample with a generation cut $\kgen=5$~GeV on the transverse
momentum of the $Z$ boson in the underlying Born configuration. This sample,
that we call sample U (for unweighted), was produced with positive
weights. Then we produced a second sample W, where we used a Born suppression
factor, with $\ptsupp=10$~GeV, and a generation cut of 1~GeV, in order to
avoid unphysical values for the strong coupling and pdf's (we will assume, in
the following, that this tiny generation cut has no effects on distributions
where the $Z$ boson or the hardest jet have transverse momenta of several
GeV).  We have also set \wneg{} to 1, so that we are able to record negative
weighted events that may arise in the region of very small transverse
momenta.  Sample W is thus weighted, with weights of either signs.  We
analyzed the events of the two samples at the \POWHEG{} level, i.e.~without
feeding them to a shower Monte Carlo.  We begin by comparing the transverse
momentum distribution of the $Z$ of the two samples, and of the fixed order
NLO QCD result (obtained as a byproduct of the event generation) in
fig.~\ref{fig:UW_ptZ}.
The figure should be interpreted in the following way. The weighted sample
gives the reference result, since it is unaffected by the generation cut. The
U sample feels clearly the effect of the 5~GeV generation cut. However, for
$\ptZ\gtrsim 10$~GeV, the U and W results coincide, showing that the effect
of the generation cut in this region is fully negligible.  A second important
observation has to do with results in the W sample. It becomes unphysical for $\ptZ
\lesssim 5$~GeV. This indicates that NLO corrections in this region are out
of control, and that our program should not be used for $\ptZ \lesssim
10$~GeV. The fixed order NLO result displays a similarly unphysical behaviour
in the low transverse momentum region. The similarity is more apparent in the
right panel of fig.~\ref{fig:UW_ptZ}, which is given in linear scale. There
is a very large negative value of the NLO cross section in the bin between 1
and 1.5~GeV. The \POWHEG{} result follows the NLO result up to a certain
amount of smearing.

\begin{figure}
\epsfig{file=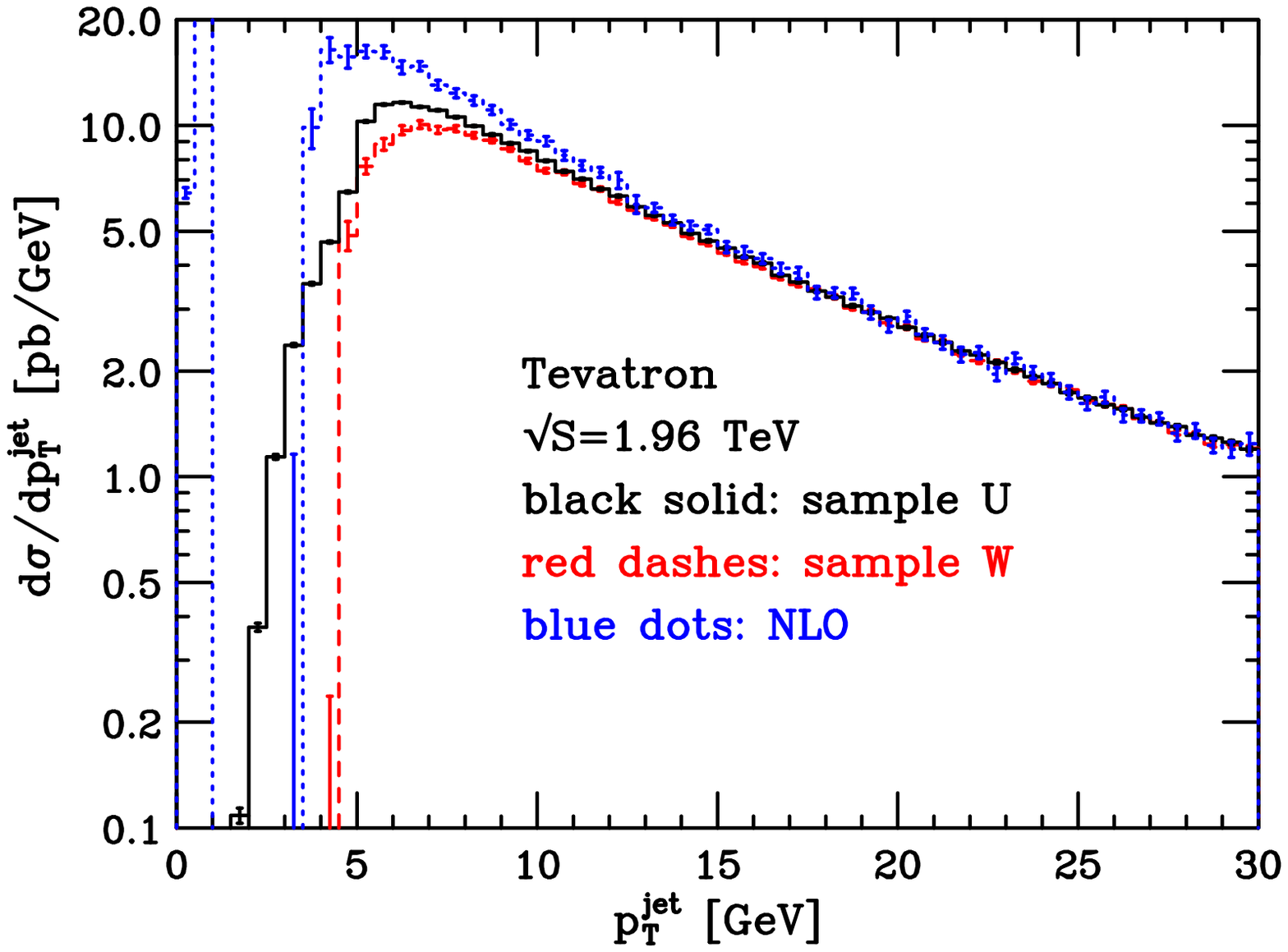,width=0.48\textwidth}
\epsfig{file=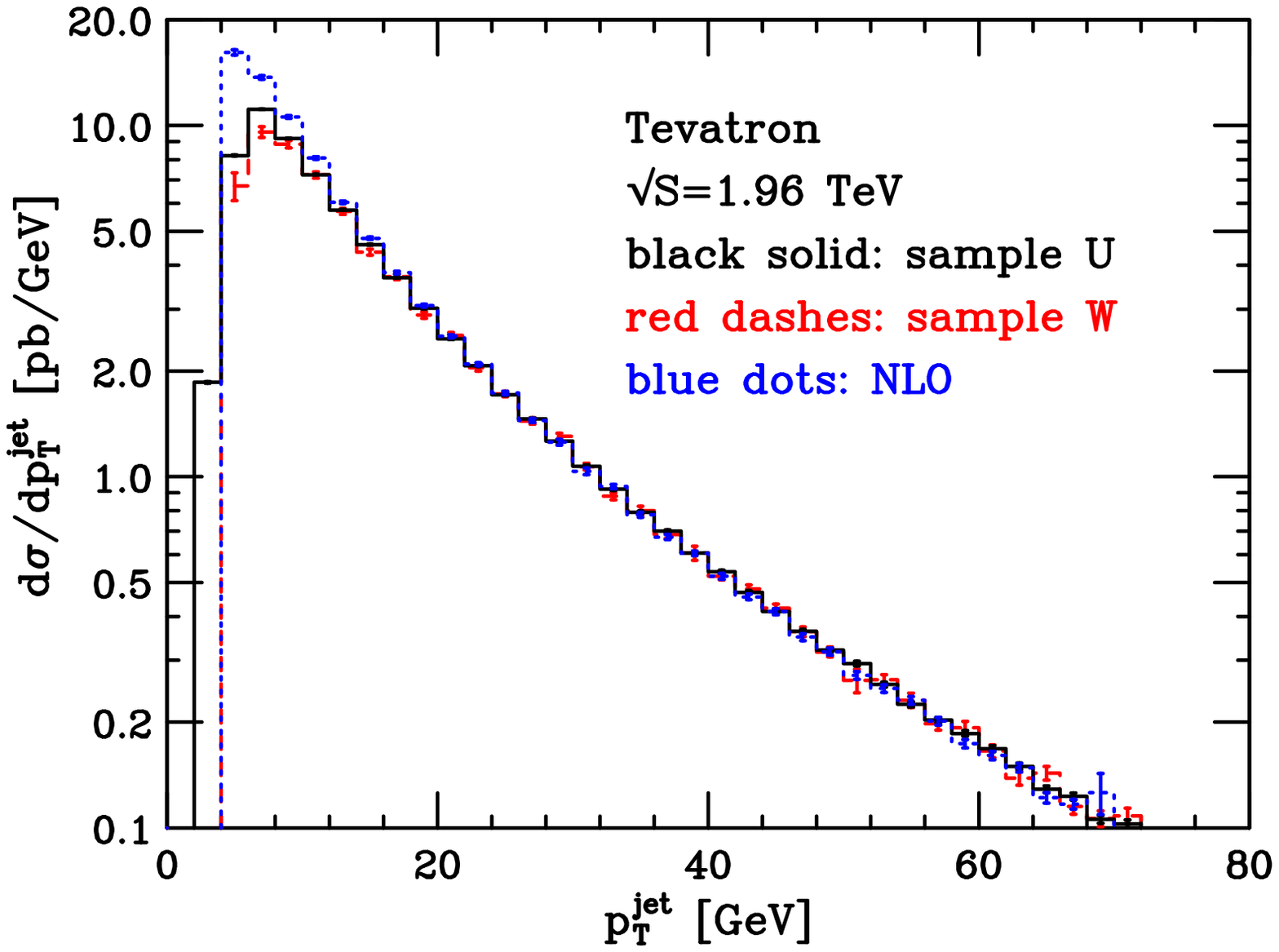,width=0.48\textwidth}
\caption{\label{fig:UW_pt1}
Transverse momentum of the hardest jet in samples U and W.}
\end{figure}

Plots for the transverse momentum of the
hardest jet are shown in fig.~\ref{fig:UW_pt1}.
They are similar to the plots in the $Z$ transverse momentum,
except for a more modest value of the NLO distribution at small
transverse momenta.

\begin{figure}
\epsfig{file=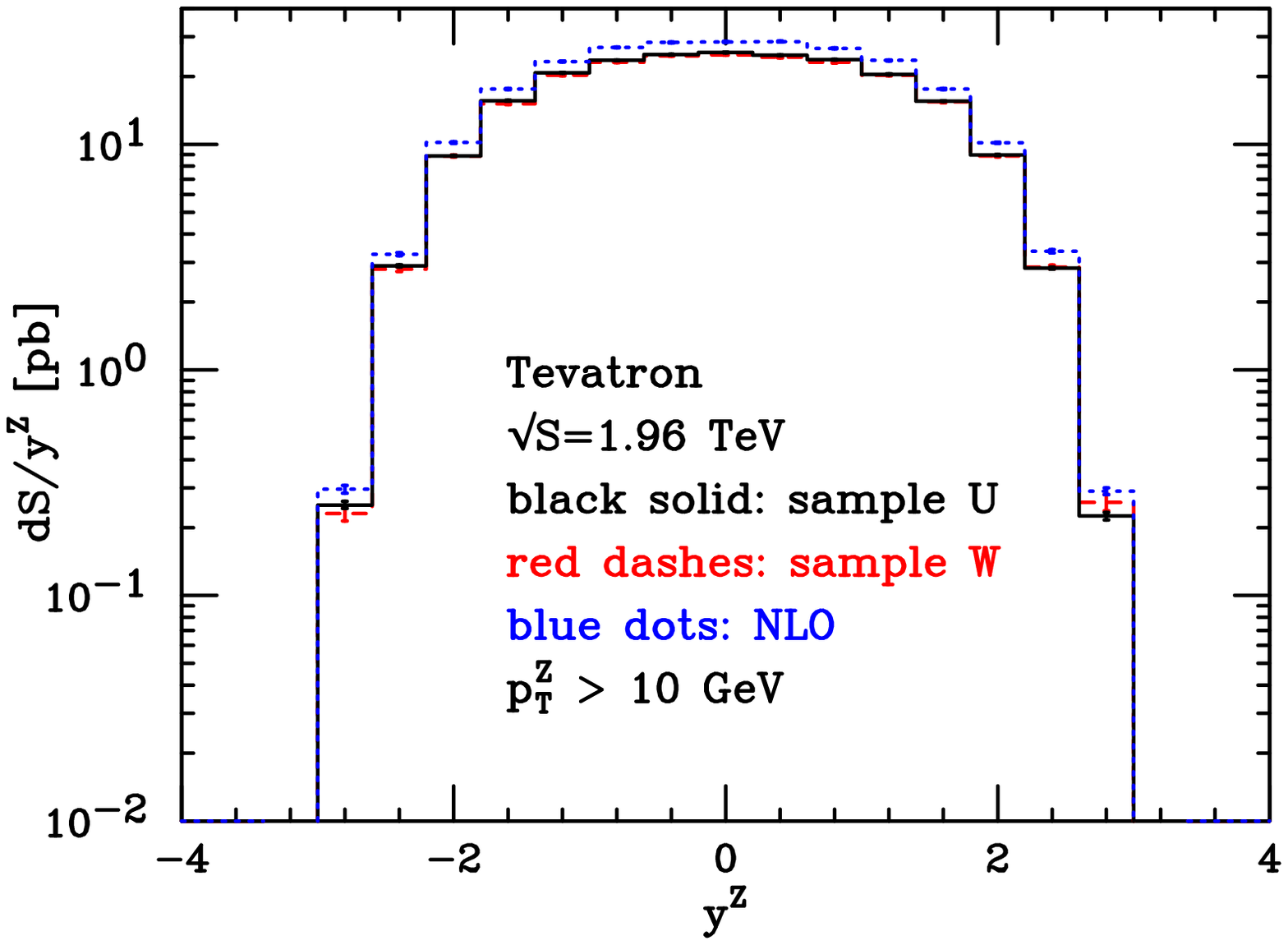,width=0.48\textwidth}
\epsfig{file=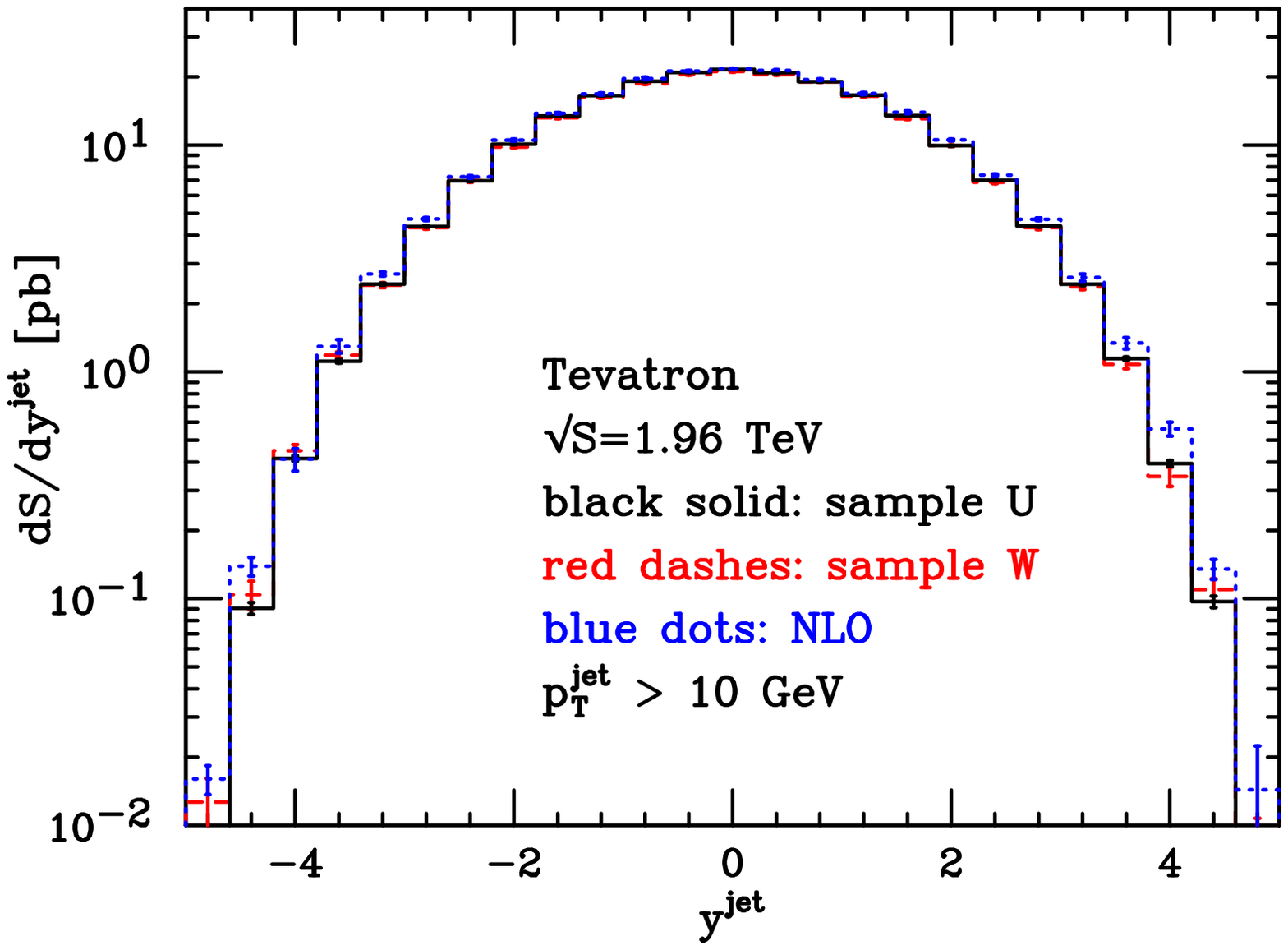,width=0.48\textwidth}
\caption{\label{fig:UW_yZj1_pt10}
Rapidity distribution of the $Z$ and of the hardest jet
in samples U and W, with a $\pt$ cut on the transverse momentum of the
$Z$/jet of 10~GeV}
\end{figure}

Finally, in fig.~\ref{fig:UW_yZj1_pt10}, the rapidity distribution of the $Z$
and of the hardest jet are plotted for the U and W sample and for the NLO
contribution.
A cut of 10~GeV is applied on either the $Z$ or the jet transverse
momentum. We see, again, very good agreement between the U and W samples at
this transverse momentum. We have also analyzed many other distributions for
the U and W sample. They all lead to the conclusion that, when cuts on the
transverse momenta of the order of 10~GeV (twice the generation cut) are
applied, the U sample yields distributions that are equivalent to the W
sample. More specifically, at 10 GeV, in the differential cross section of
the jet or of the $Z$ transverse momentum, the U sample differs from the W
sample by 6\%{}, and the difference decreases with $\pt$.  Thus, we conclude
that for $\pt>10$~GeV, at Tevatron energies, the U sample is substantially
independent of a generation cut less or equal than 5~GeV.

\subsection{Negative weights and folding}
\label{sec:posnegfold}
In order to get a reasonably small fraction of negative-weight events in
the \POWHEGBOX{}, it is at times necessary to increase the folding parameters
{\tt foldcsi}, {\tt foldy} and {\tt foldphi}. These parameters were
introduced in the framework of the \POWHEG{} implementation of heavy-quark
pair production~\cite{Frixione:2007nw}, and are mentioned in the
corresponding manual~\cite{Frixione:2007nu}, and discussed extensively in
ref.~\cite{Alioli:2010xd}. We briefly recall here their genesis and use.  In
\POWHEG{}, the underlying Born configuration for an event is generated
according to the $\bar{B}$ function, that, in the simplest cases, can be
represented by the equation (see ref.~\cite{Frixione:2007vw})
\begin{equation}
\bar{B}(\Phi_B)=B(\Phi_B)+V(\Phi_B)+\int d\Phi_{\rm rad} \lq R\(\Phi_B,\Phi_{\rm
  rad}\)-C\(\Phi_B,\Phi_{\rm rad}\) \rq \,.
\end{equation}
In order to generate $\Phi_B$ distributed according to this probability
density, we actually parametrize the radiation variables in terms of three
variables in the unit cube, $X_i$, and then perform the integral
\begin{equation}
\int d \Phi_B \; d^3 X\; \tilde{B}(\Phi_B,X)\,,
\end{equation}
where
\begin{equation}\label{eq:btilde}
\tilde{B}(\Phi_B,X)=B(\Phi_B)+V(\Phi_B)+
 \left|\frac{\partial \Phi_{\rm rad}}{\partial X}\right| [R(\Phi_B,\Phi_{\rm
   rad})-C(\Phi_B,\Phi_{\rm rad})]\,,
\end{equation}
using the \MINT{} integration program~\cite{Nason:2007vt}.  This program
stores appropriate grids and upper bounds, so that, after the integration, it
is possible to use it to generate points in the integration variables,
i.e.~in $(\Phi_B,X)$, distributed with a probability proportional to
$\tilde{B}$. The $X$ values are ignored, which amounts to integrating over
them, and only $\Phi_B$ is retained. This procedure is sufficient to
generate positive weights if the process in question has a sufficiently-high
scale, like in vector boson, Higgs or $t\bar{t}$ production. In these cases,
$\as$ (present in eq.~(\ref{eq:btilde}) in the virtual $V$, real $R$ and
counterterms $C$ contributions) is small enough so that $B$ is always larger
than the other terms, yielding a positive result for any value of $\Phi_B$
and $X$.  As the scale of the process decreases (and $\as$ increases), it
becomes more likely that $\tilde{B}$ may become negative for some value of
the $X$ parameters, even if $\bar{B}$ is positive for any value of $\Phi_B$.
Of course, this depends upon the way that the $C$ counterterm is defined.  A
brute force remedy to this problem, is to fold the integration of the $(R-C)$
term as many time as necessary to yield a positive $\bar{B}$ (that is to say,
to get a negligible fraction of negative-weight events).  The details of
the folding method are illustrated in the \MINT{} manual. Here we only
illustrate it with an example. If we want to fold the $X_1$ variable twice,
we define
\begin{equation}
\tilde{B}_{\rm folded}(\Phi_B,x_1,X_2,X_3)=
\tilde{B}(\Phi_B,x_1,X_2,X_3)+\tilde{B}(\Phi_B,1/2+x_1,X_2,X_3),
\end{equation}
and integrate in $x_1$ from 0 to $1/2$. This clearly yields the same result
as without the folding. Furthermore, if we increase the number of
foldings for all the three parameters $X_i$, the folded function becomes less
and less dependent upon the $x_i$ values. With this technique, regions of
integration where the function is positive and negative are combined
together. It is clear that, if the $\bar{B}$ function is positive, with large
enough folding numbers, we will achieve the positivity of the folded
$\tilde{B}$ function.

The \MINT{} integrator and the \POWHEGBOX{} can perform the folding
automatically. The user needs only to specify how many times each variables
should be folded. The \MINT{} integrator divides each integration coordinates
in 50 bins, such that each bin contributes equally to the integral of the
absolute value of the $\tilde{B}$ function. Folding is achieved by
overlapping these intervals. Thus, the folding number must be a divisor of
50: 2, 5, 10, 25 or 50. Three tokens in the {\tt powheg.input} file can be
set to the folding number: {\tt ifoldcsi}, {\tt ifoldy} and {\tt ifoldphi},
that refer to the folding of the three radiation variables $\xi$, $y$ and
$\phi$.

\begin{figure}[htb]
\begin{center}
\epsfig{file=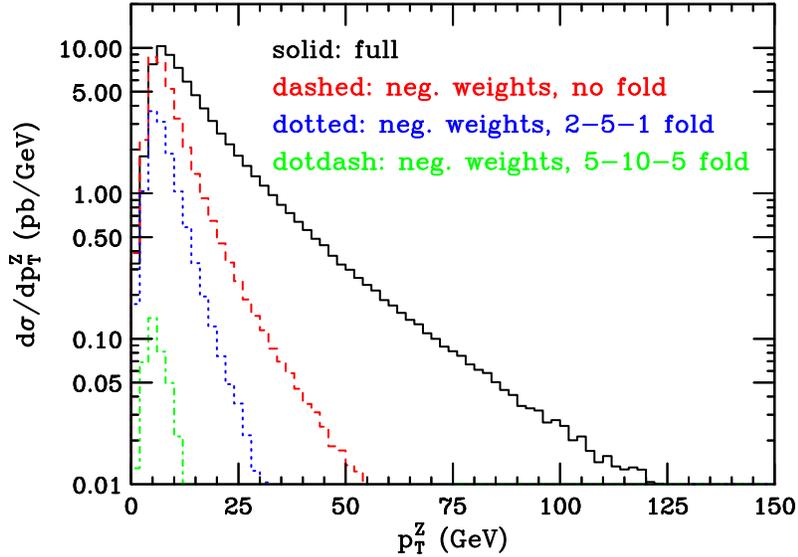,width=0.7\textwidth}
\end{center}
\caption{\label{fig:folding-effects}
The effect of folding on the negative-weight event fraction.}
\end{figure}

In fig.~\ref{fig:folding-effects}, we display the effect of folding on the
amount of negative weights that enter the computation of a physical quantity,
namely the $Z$ transverse momentum (by choosing similar variables, like the
transverse momentum of the hardest jet, we get similar results).
All the histograms in the figure were obtained with the \wneg{} flag set to
true. The solid line was obtained in the standard way, i.e.~by adding
positive and negative contributions. The dashed, dotted and dotdashed lines
were obtained by plotting only the absolute value of the negative
weights. The dashed line was obtained with no folding, the dotted line was
obtained with a 2-5-1 folding (i.e.~{\tt ifoldcsi}$=2$, {\tt ifoldy}$=5$ and
{\tt ifoldphi}$=1$), and the dot-dashes line was obtained with the 5-10-5
folding.  From the figure, we see that the amount of negative weights
increases as the transverse momentum becomes smaller. This can be understood
as being due to the increase of $\as$ at small scales. By increasing the
folding numbers, the number of negative weights is strongly reduced, and,
furthermore, it only affects the region of very small transverse momenta,
where the calculation becomes however unreliable.

Increasing the folding number has a cost on the execution time. As a rule of
thumb, we expect the execution time to increase as the product of the three
folding numbers. Actually, the increase is somewhat less than that, because
the Born and virtual contribution are evaluated only once for each set of
foldings, since they do not depend upon the integration variables. Thus, the
folding should be chosen as a function of the generation cut, and,
ultimately, as a function of the cutoff one is imposing on the hardest-jet
transverse energy or on the $Z$ transverse momentum.

\begin{table}[htb]
\begin{center}
\begin{tabular}{|l|c|c|c|c|c|c|}
\hline
\hfill$\xi$-$y$-$\phi$ folding $\Longrightarrow$&  1-1-1   &2-5-2& 5-5-5 & 5-10-5 & 10-10-10 \\
\hline
LHC, 14~TeV, $\kgen=5$~GeV & 0.25  &0.067 & 0.026 & 0.016  & 0.0074    \\
\hline
LHC, 14~TeV, $\kgen=10$~GeV& 0.17 &0.020 &  0.0029& 0.00036& $3.4\times10^{-5}$ \\
\hline
Tevatron, 1.96~TeV, $\kgen=5$~GeV & 0.23  & 0.029  & 0.0086 & 0.0043 & 0.0028   \\
\hline
\end{tabular}
\end{center}
\caption{\label{table:LHC}
  Negative-weight fractions $\sigma_{(-)}/(\sigma_{(+)}+|\sigma_{(-)}|)$ for given folding numbers.}
\end{table}
As an example, in table~\ref{table:LHC}, we have collected the fraction of
negative weights $\sigma_{(-)}/(\sigma_{(+)}+|\sigma_{(-)}|)$ for the LHC at
14~TeV, with two generation cuts, and for the Tevatron. As can be seen, the
higher is the folding, the smaller is the fraction of negative-weight events.

The typical performance of event generation is, without folding, roughly 1000
events per minute, on typical workstation cpu's.  It is clear that, in the
worse case of very low transverse-momentum jets, assuming that we can
tolerate around $2\%$ negative weights (that would populate, in all cases,
the very small $\pt$ region), by using a 5-5-5 folding we end up generating
of the order of eight events per minute.  It is also clear that we have a
compelling reason to generate independently samples with different generation
cuts. Besides benefiting of a more abundant generation in the high $\pt$
region, we also benefit from the possibility of using smaller folding
numbers.

A clarifying consideration on negative weights in \POWHEG{} can be made in
comparison with \MCatNLO{} \cite{Frixione:2002ik}.  The negative weights that
we get in \POWHEG{} are fully analogous to the negative weights that one gets
in the ${\cal S}$ events in \MCatNLO{}. Also in that case, the number of
negative weights could be reduced or eliminated by using a folding technique,
as in our case.  However, negative weights in the ${\cal H}$ event sample of
\MCatNLO{} cannot be reduced, and so there is no reason to attempt to reduce
the negative weight fraction in the ${\cal S}$ events.

Finally, let us stress again that by performing any analysis on a \POWHEG{}
event sample with signed events, or with weighted events, or with
positive-weight events (using eventually a large enough folding number to get
a negligible fraction of negative weights) yields the same result within
statistics. This property can be easily verified by running the program with
the same {\tt powheg.input} files that differs only for the presence of the
\wneg{} flag set to 1, or for the presence of the \ptsupp{} token, or, again,
differing only in the folding numbers.  In the current version of the
\POWHEGBOX{} we leave this choice to the user.  The performance cost of
getting positive weights may be well balanced if the events have to be run
through costly detector simulators, or if it is undesirable to have negative
weighted events in some complex, multivariate statistical analysis.
The default \POWHEGBOX{} setting is with no negative weights and no
weighted events (i.e.~$F(\kt)=1$).

\subsection{Comparison with \MCFM{}}
As a final check of the code, we have compared the NLO output of the
\POWHEGBOX{} to the output produced by the NLO code from \MCFM~\cite{mcfm},
and total agreement was found. In spite of the fact that we have taken the
virtual correction formulae (that we have checked against the
\BlackHat~\cite{Berger:2008sj} results too) from the \MCFM{} program, this
comparison is a highly non-trivial check, since the subtraction schemes used
by the two programs are completely different.

\section{Phenomenology}
\label{sec:phenomenology}
In this section we directly compare the output of our program to
distributions that have been measured by the CDF and D0 Collaborations at the
Tevatron.  The only aim of this study is to validate, to some extent, our
program.  We believe that a more thorough analysis can only be performed by
the experimental collaborations themselves, using our event generator.  All
the results displayed in this section have been obtained with a sample of
roughly 1.3 million events, at the center-of-mass energy of $1.96$~TeV, in
proton-antiproton collisions. The sample was generated with positive weights,
with the folding parameters 5-10-5 (see section~\ref{sec:posnegfold}). The
generation cut was set to $5$~GeV. The fraction of negative-weight events
generated with this setting of parameters was 0.4\%. As shown in
section~\ref{sec:posnegfold}, this fraction is concentrated in events with
low transverse momenta.  We have used the CTEQ6M pdf set, with the
corresponding value of $\LambdaQCD$. The renormalization and factorization
scales for the calculation of the $\bar{B}$ function are set equal to the $Z$
transverse momentum in the underlying Born configuration.

The events were showered using \PYTHIA~6.4.21~\cite{Sjostrand:2006za}. We have
compared two different choices of tune for \PYTHIA: the tune~A ({\tt
  PYTUNE(100)}), that uses the old shower and underlying event model, and the
Perugia~0 tune ({\tt PYTUNE(320)}), that uses the new transverse-momentum
ordered shower and the new underlying event model. Results for the Tune~A
model will be displayed as blue dashed histograms, while the Perugia~0 will
be shown as red solid lines.  The data will be displayed as simple points
with error bars.

We have switched off photon radiation off leptons ({\tt mstj(41)=3}), so
that, in the case of $Z\to \elpelm$, the lepton energy better represents what
would be measured in an electromagnetic calorimeter.  In addition we have set
the $Z$ mass and width to the values 91.188~GeV and 2.486~GeV respectively,
$\sin^2\theta_W^{\rm eff}=0.2312$ and $\alpha^{-1}_{\rm em}(M_Z)=128.930$.

We applied the jet algorithm to all particles in the event, including all
leptons, except for those coming from $Z$ decay. In other words, when
comparing to experimental results, we assume that the jet energies are fully
corrected to the particle level, including those particles that would not be
visible in the detector.  We have used the jet algorithms as implemented in
the {\tt FASTJET} package~\cite{Cacciari:2005hq}.

\subsection{CDF results}
The CDF Collaboration provided and still provides results for $Z/\gamma\ (\to
e^+ e^-) +1j$ and $Z/\gamma\ (\to \mu^+ \mu^-) +1j$ events.  In order to
perform an analysis as similar as possible to the one done by the CDF
Collaboration, we used the midpoint algorithm~\cite{Abulencia:2005yg} to
combine hadrons (from \POWHEG{} events showered by \PYTHIA{}) into jets, with
cone radius $R=0.7$ and a merging/splitting fraction of 0.75, starting from
seed towers with transverse momenta above 1~GeV.

\subsubsection*{$\boldsymbol{Z/\gamma \ (\to e^+ e^-)}$ + jets}
\begin{figure}[htb]
\begin{center}
\epsfig{file=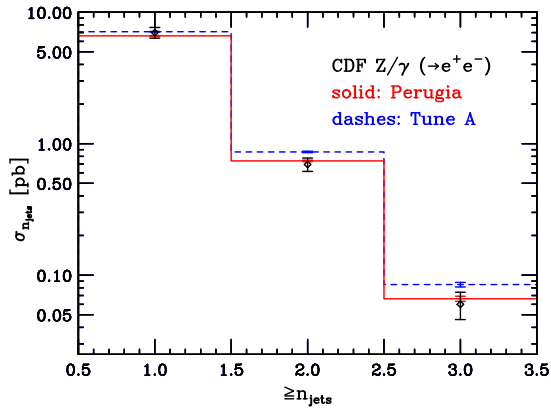,width=0.48\textwidth}
\end{center}
\caption{\label{fig:cdfepem1}
Total cross section for inclusive jet production.}
\end{figure}

\begin{figure}[htb]
\begin{center}
\epsfig{file=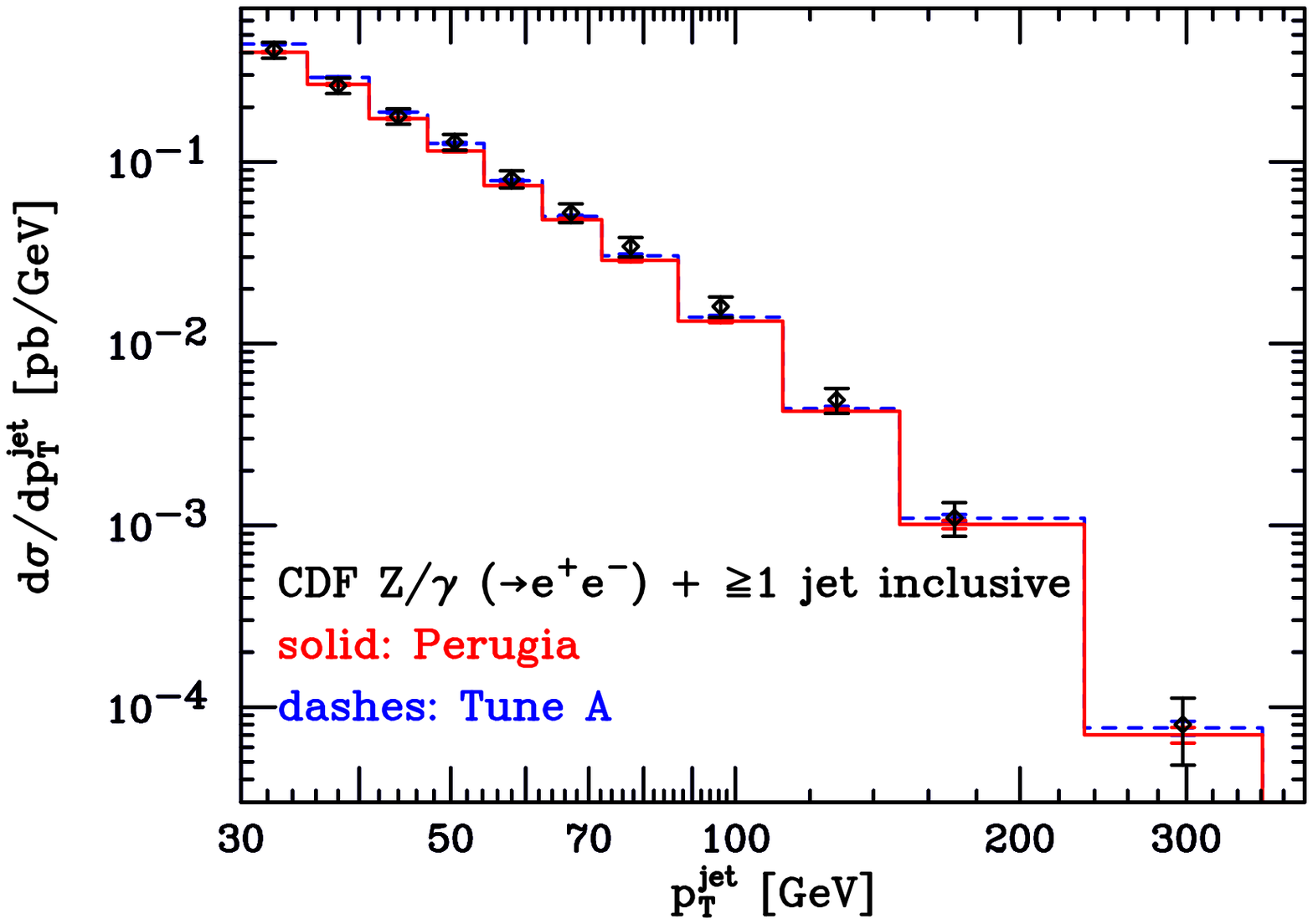,width=0.48\textwidth}
\epsfig{file=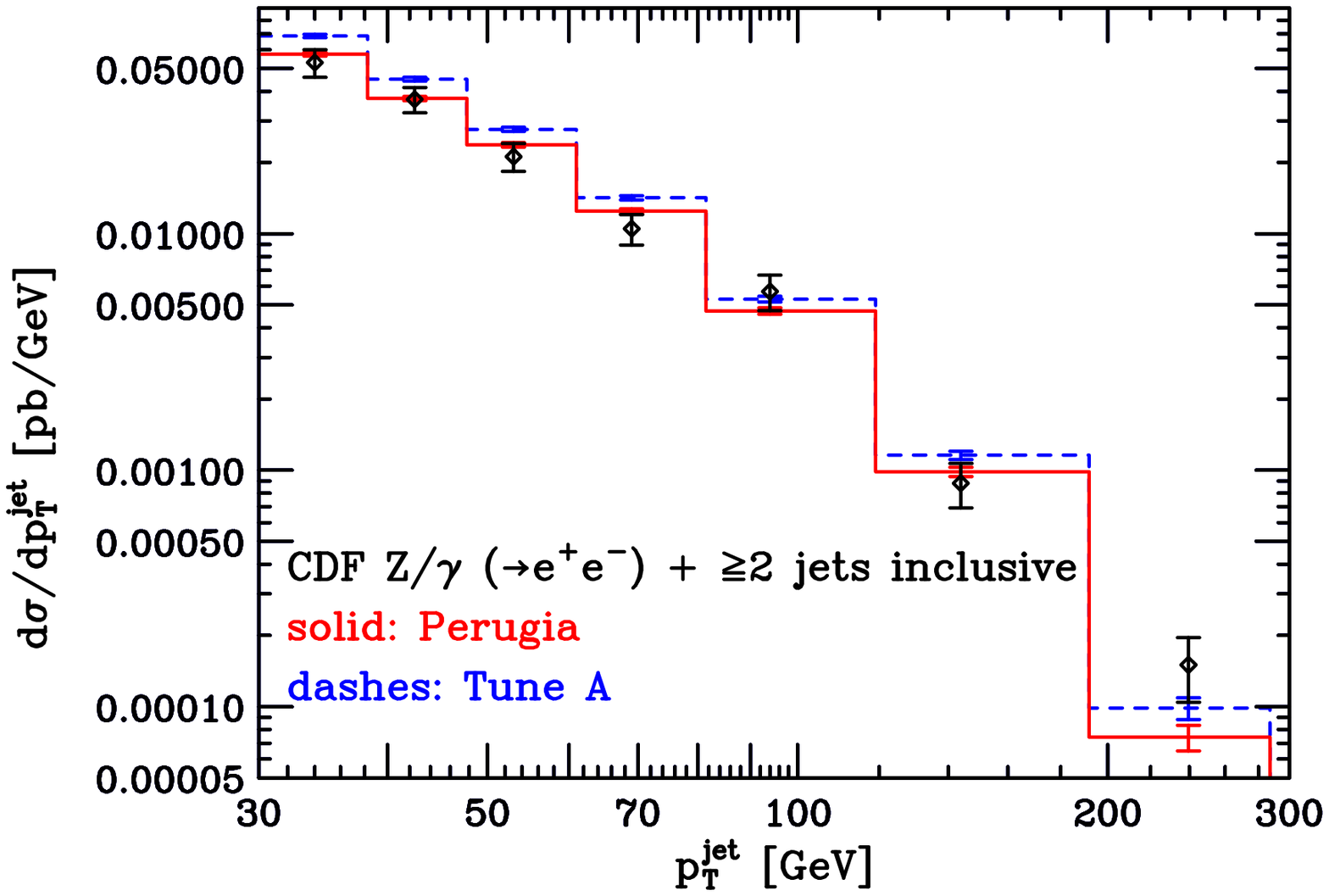,width=0.48\textwidth}
\end{center}
\caption{\label{fig:cdfepem2} The inclusive $\pt$ distributions for events
  with at least one and two jets.}
\end{figure}

\begin{figure}[htb]
\begin{center}
\epsfig{file=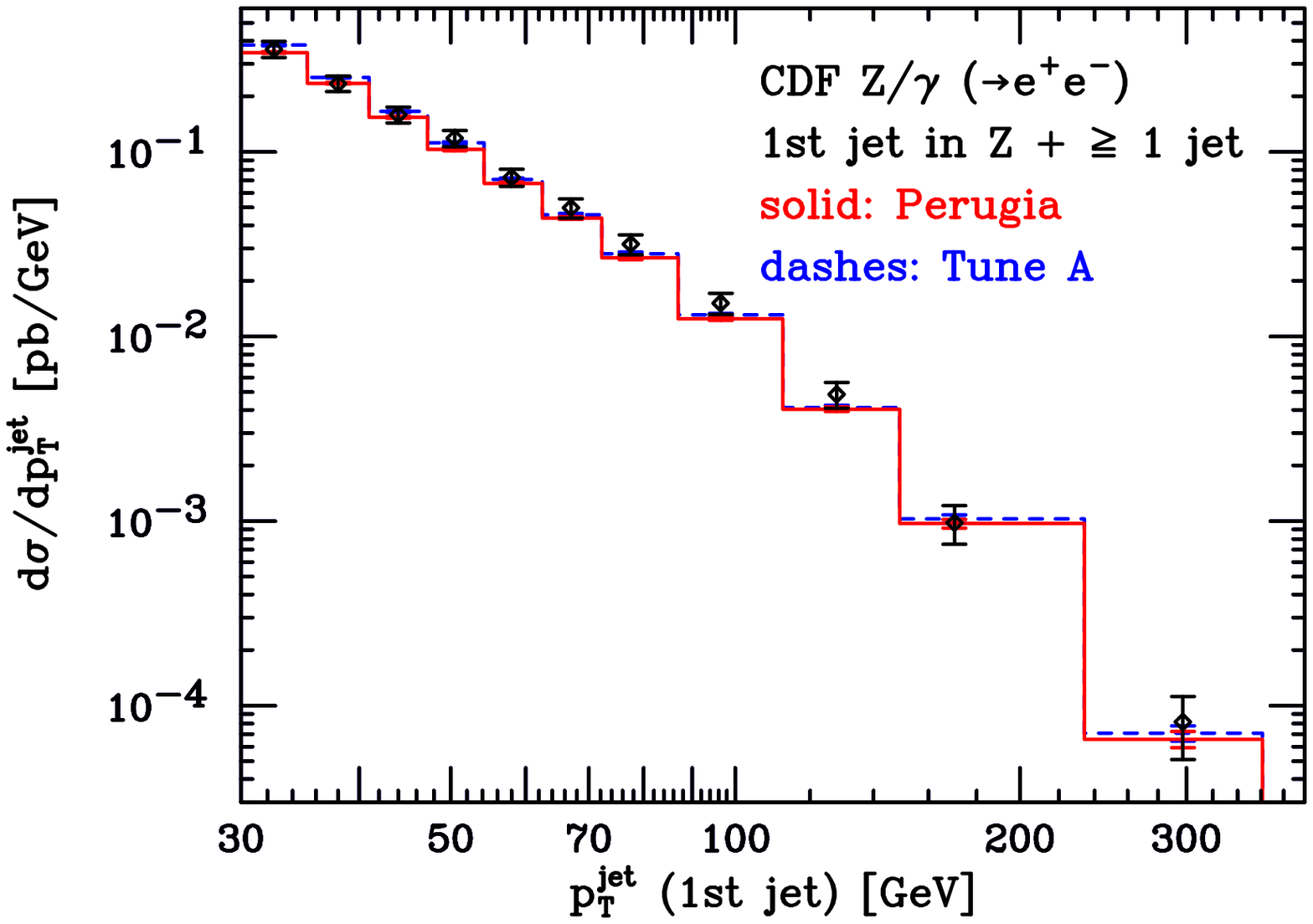,width=0.48\textwidth}
\epsfig{file=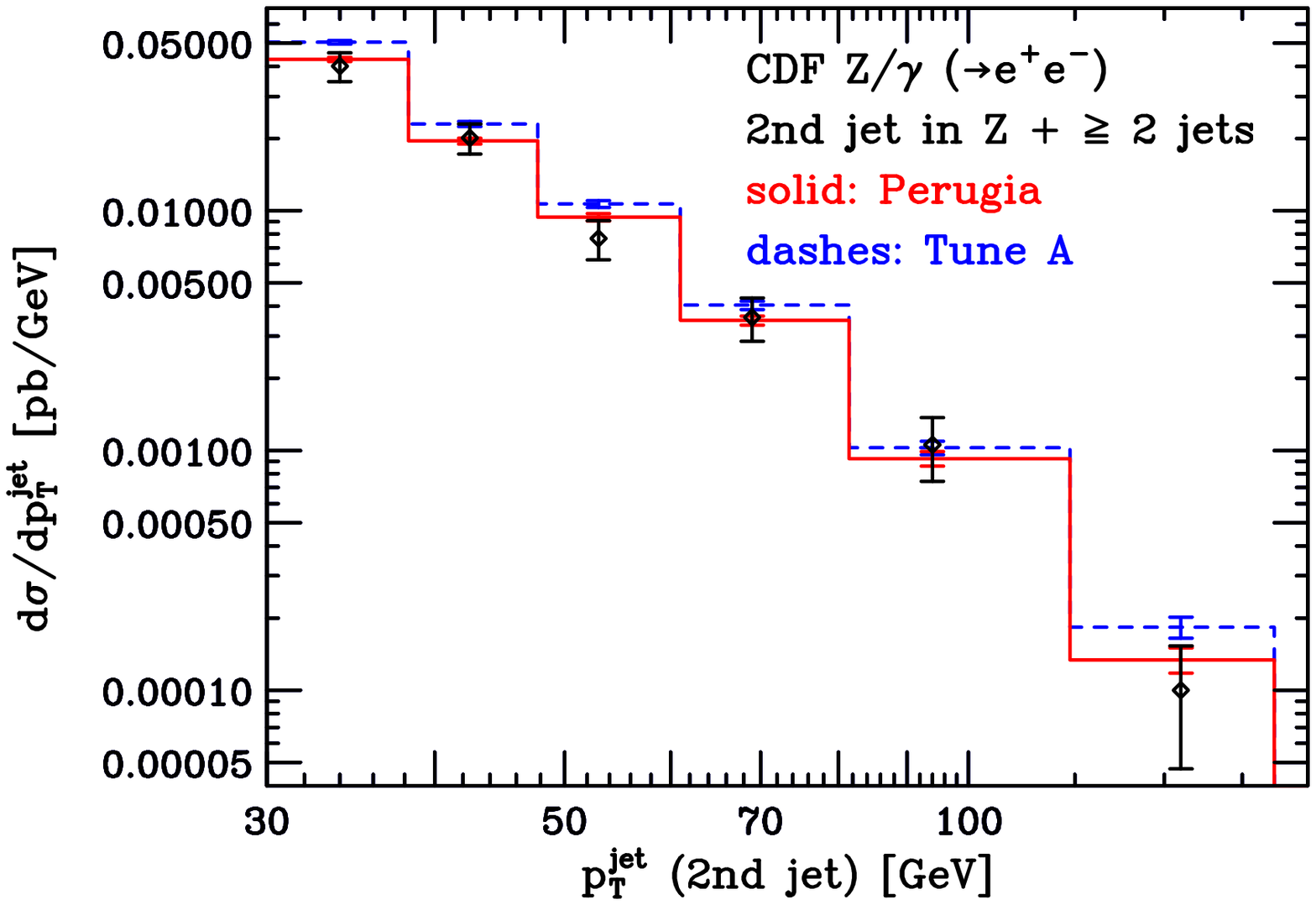,width=0.48\textwidth}\\
\epsfig{file=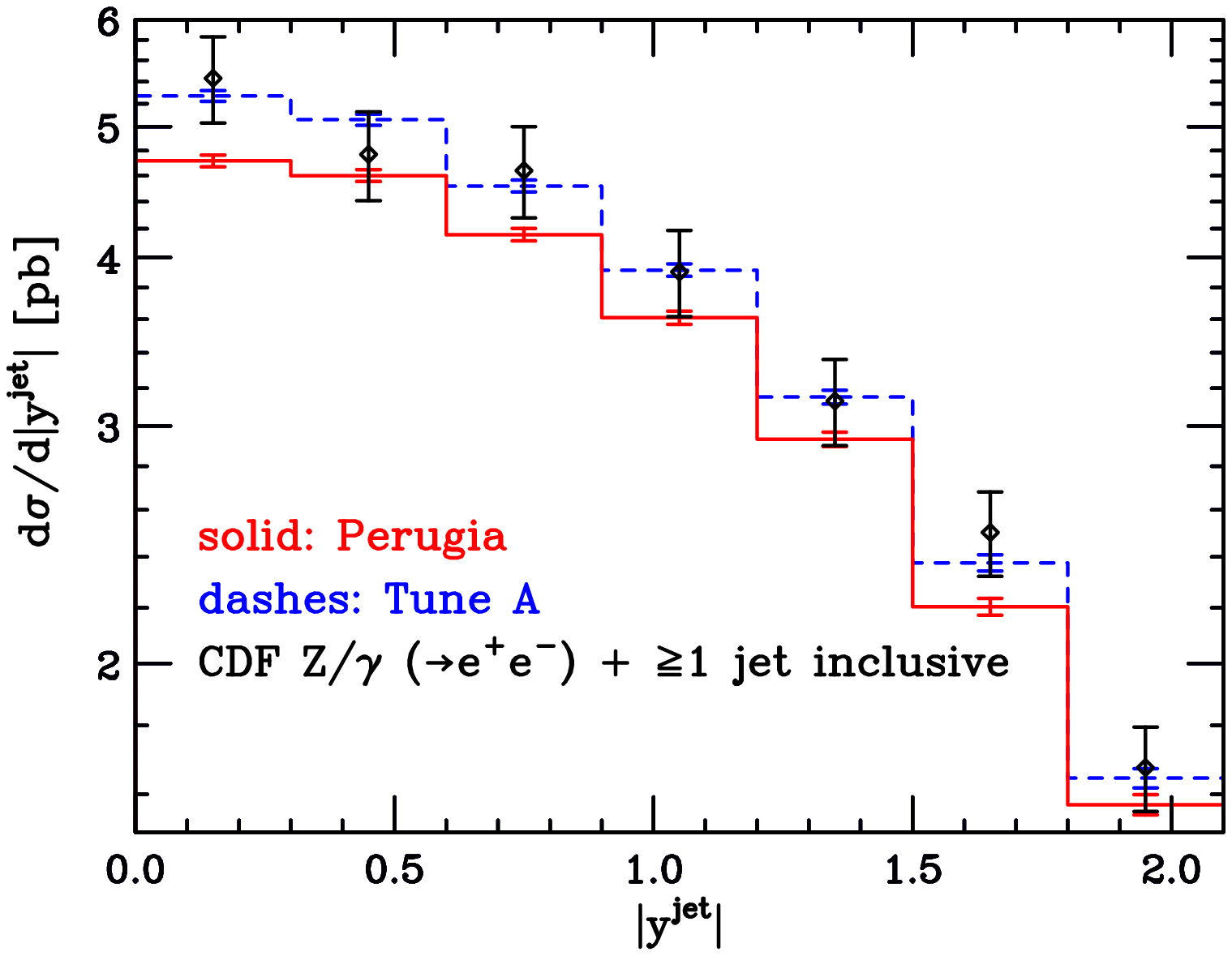,width=0.48\textwidth}
\epsfig{file=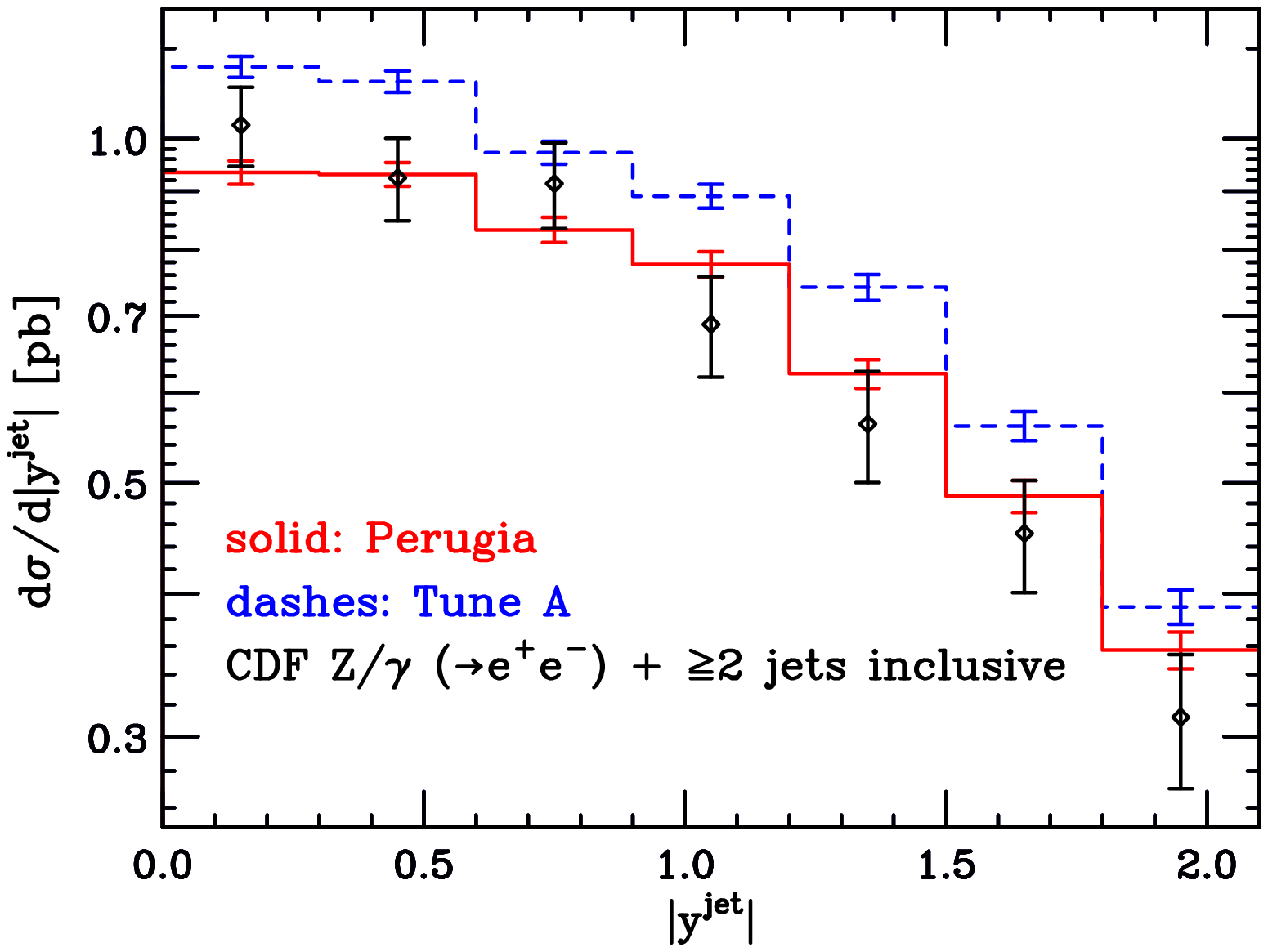,width=0.48\textwidth}
\end{center}
\caption{\label{fig:cdfepem3} $\pt$ distributions of the hardest and
  next-to-hardest jet and the inclusive rapidity distributions for events
  with at least one and two jets.}
\end{figure}

We begin by considering the $Z/\gamma\to e^+ e^-$ results of CDF. In
figures~\ref{fig:cdfepem1}, \ref{fig:cdfepem2} and~\ref{fig:cdfepem3} we
compare the \POWHEG{} results showered by \PYTHIA{} using Tune~A (blue dashed
lines) and Perugia~0 tuning (red solid lines), with the CDF data.

The results in figs.~\ref{fig:cdfepem1} and~\ref{fig:cdfepem2} were published
in~\cite{Aaltonen:2007cp}, while those in fig.~\ref{fig:cdfepem3} were
extracted from the blessed results in~\cite{cdf:webpage}, at an integrated
luminosity of 2.5~fb$^{-1}$.

In figs.~\ref{fig:cdfepem1} and~\ref{fig:cdfepem2} we plot results for the
inclusive total cross section and inclusive $\pt$ distributions for the
production of at least one and two jets.  In fig.~\ref{fig:cdfepem3} we plot
the $\pt$ distribution of the hardest and next-to-hardest jet and the
inclusive rapidity distributions for events with at least one and two jets.

In order to compare with CDF data, we adopted the following cuts 
\beqn
\label{eq:CDF_epem_cuts}
&&66~{\rm GeV} < M_{ee} < 116~{\rm GeV},\quad p_{\sss T}^e > 25~{\rm GeV},
\quad |\eta^{e_1}|< 1.0, \quad 1.2 < |\eta^{e_2}|< 2.8,
\nonumber\\
&&|y^{\rm jet}|<2.1, \quad p_{\sss T}^{\rm jet} > 30~{\rm GeV}, \quad \Delta
R_{e,\, {\rm jet}}>0.7,
\eeqn
where $y$ and $\eta$ represent the rapidity and pseudorapidity of the
specified particles, and where $R$ is the distance in the azimuth-rapidity
plane.

We notice the good agreement between the \POWHEG{} prediction and the data.
It parallels the agreement between data and the NLO \MCFM{} result displayed
in refs.~\cite{Aaltonen:2007cp,cdf:webpage}, despite the fact that, when more
than two jets are considered, \MCFM{} has NLO accuracy, while our generator
is limited to leading order.  However, we emphasize that the \POWHEG{}
results are directly compared to data, while the \MCFM{} ones are first
corrected by parton-to-hadron correction factors, as detailed
in~\cite{Aaltonen:2007cp}. Notice also the dependence of the results from the
chosen tune of \PYTHIA{}. The Perugia~0 tune seems to give a slightly better
agreement with data. We point out that the differences between the \POWHEG{}
results and the data is of the same order of the differences between the two
tunes, thus suggesting that, by directly tuning the \POWHEG{} results to
data, one may get an even better agreement.

\subsubsection*{$\boldsymbol{Z/\gamma\ (\to \mu^+ \mu^-)}$ + jets}
\begin{figure}[htb]
\begin{center}
\epsfig{file=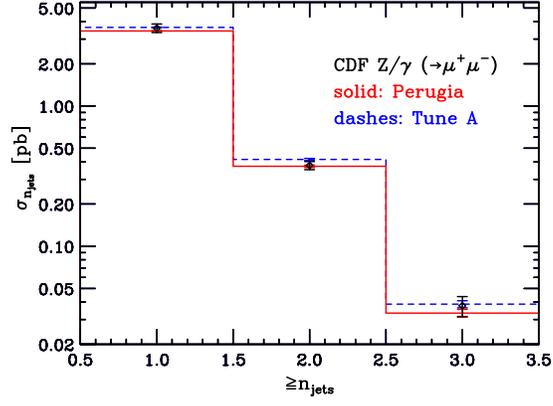,width=0.48\textwidth}
\end{center}
\caption{\label{fig:cdfmupmum1}
Total cross section for inclusive jet production.}
\end{figure}
\begin{figure}[htb]
\begin{center}
\epsfig{file=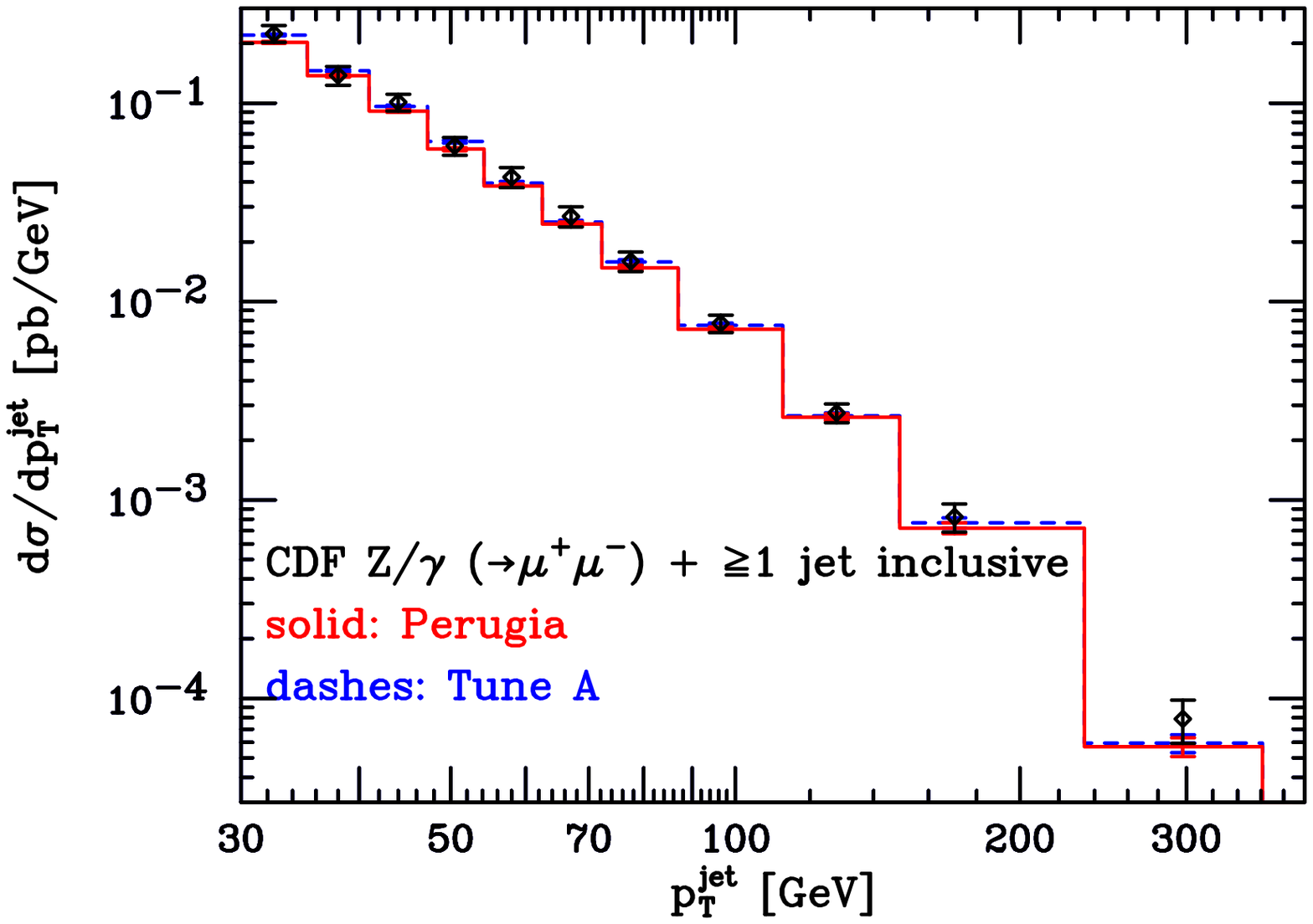,width=0.48\textwidth}
\epsfig{file=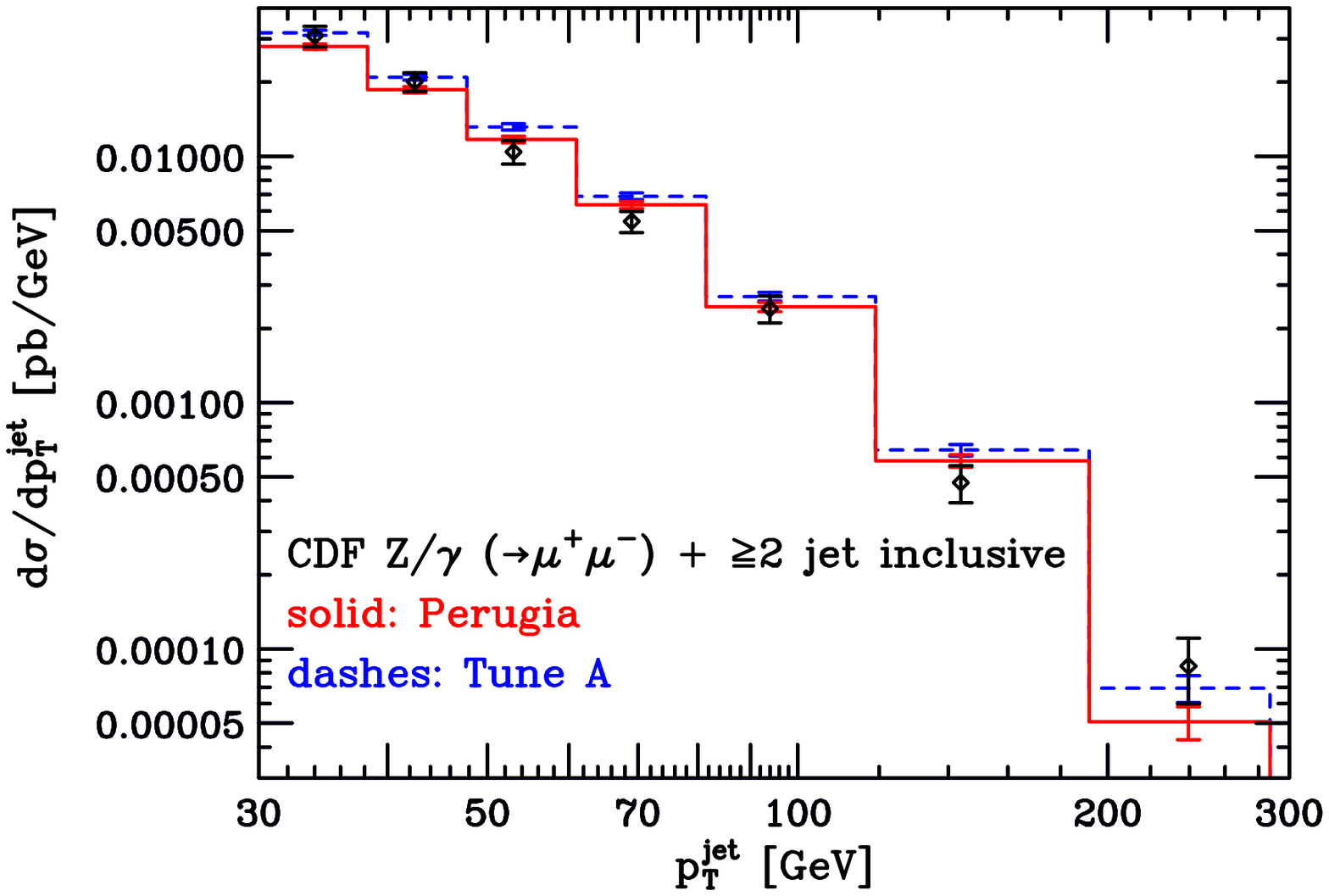,width=0.48\textwidth}\\
\epsfig{file=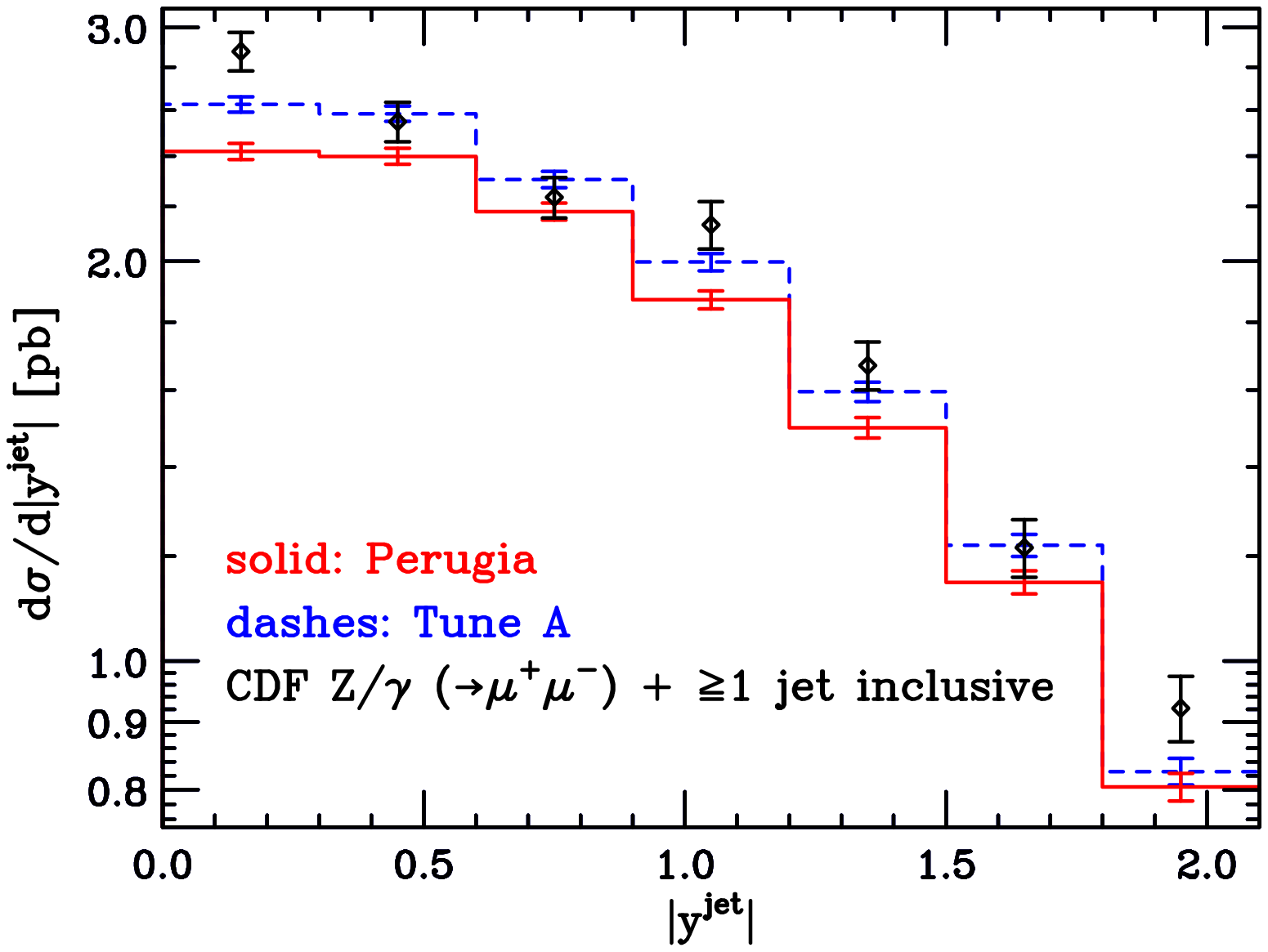,width=0.48\textwidth}
\epsfig{file=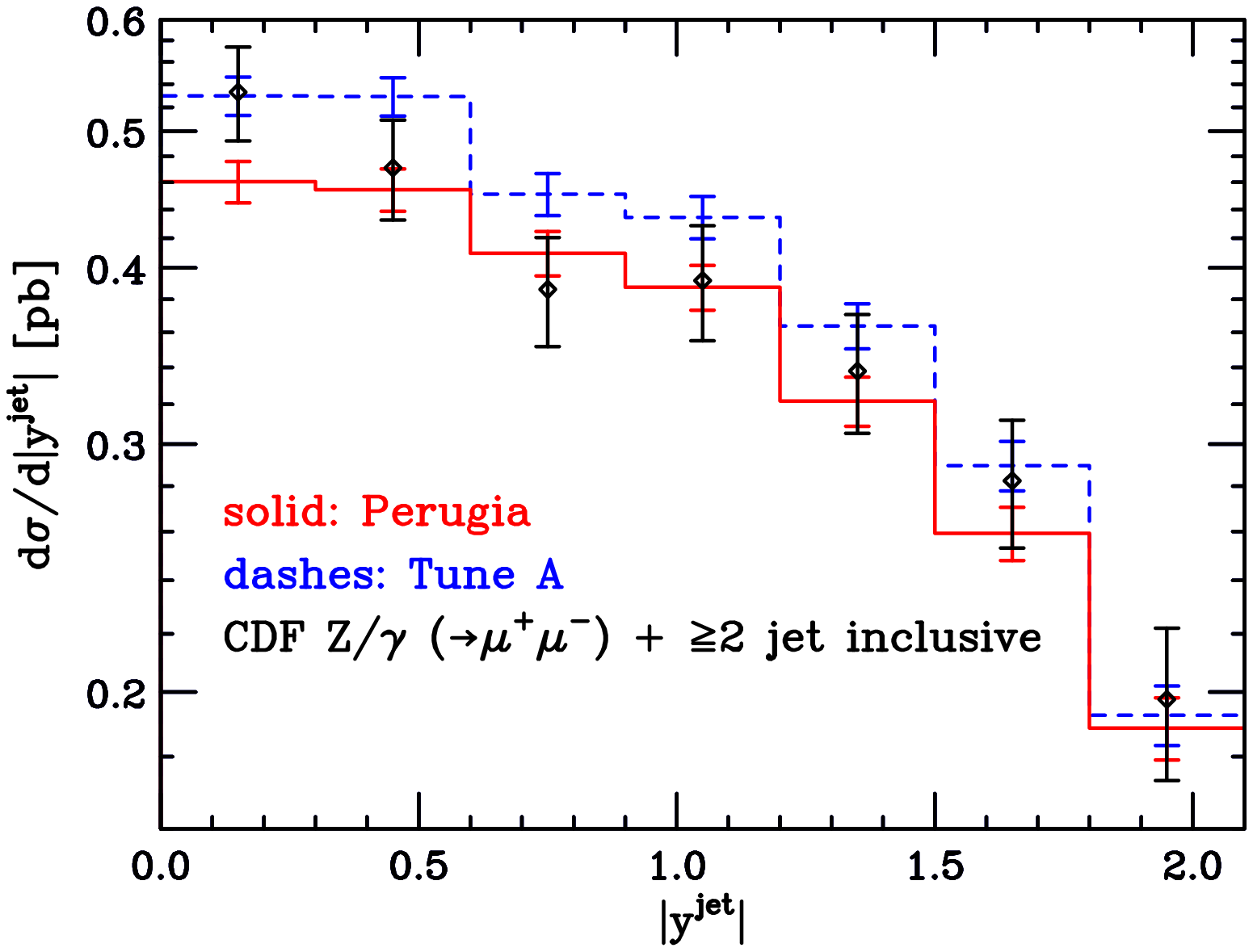,width=0.48\textwidth}
\end{center}
\caption{\label{fig:cdfmupmum2}
The inclusive $\pt$ and rapidity distributions for events with at least one
and two jets.} 
\end{figure}

Similar studies for the $Z/\gamma$ decaying in the $\mu^+ \mu^-$ channel
were also performed by CDF.  In figs.~\ref{fig:cdfmupmum1}
and~\ref{fig:cdfmupmum2} we display the total cross section for inclusive jet
production and the inclusive $\pt$ and rapidity distributions for events with
at least one and two jets.  In order to perform an analysis as close as
possible to the CDF experimental settings, we have applied the following cuts
\beqn
\label{eq:CDF_mupmum_cuts}
&&66~{\rm GeV} < M_{\mu\mu} < 116~{\rm GeV},\quad p_{\sss T}^\mu > 25~{\rm GeV},
\quad |\eta^{\mu}|< 1.0,
\nonumber\\
&&|y^{\rm jet}|<2.1, \quad p_{\sss T}^{\rm jet} > 30~{\rm GeV}, \quad \Delta
R_{\mu,\, {\rm jet}}>0.7\,.
\eeqn
In this case, no clear conclusions can be
drawn on which of the two chosen tunes better reproduces the data.

\subsection{D0 results}
The D0 Collaboration performed analyses similar to those done by the CDF
Collaboration, focusing on more exclusive jet cross sections, and
considering also some angular distributions.

To perform an analysis as similar as possible to the one done by the D0
Collaboration, we used the D0 Run~II iterative seed-based cone jet
algorithm~\cite{D0:runIIjetalgo} in order to recombine hadrons into jets,
with a splitting/merging fraction of 0.5 and a cone radius $R=0.5$.

\subsubsection*{$\boldsymbol{Z/\gamma \ (\to \mu^+ \mu^-)}$ + jets}
In ref.~\cite{Abazov:2008ez,Abazov:2009pp}, several distributions were
studied by the D0 experiment using the set of cuts
\beqn
\label{eq:D0_mupmum_cuts}
&&65~{\rm GeV} < M_{\mu\mu} < 115~{\rm GeV},\quad p_{\sss T}^\mu > 15~{\rm GeV},
\quad |\eta^{\mu}|< 1.7,
\nonumber\\
&&|y^{\rm jet}|<2.8, \quad p_{\sss T}^{\rm jet} > 20~{\rm GeV},
\quad \Delta
R_{\mu,\, {\rm jet}}>0.5\; .
\eeqn
\begin{figure}[htb]
\begin{center}
\epsfig{file=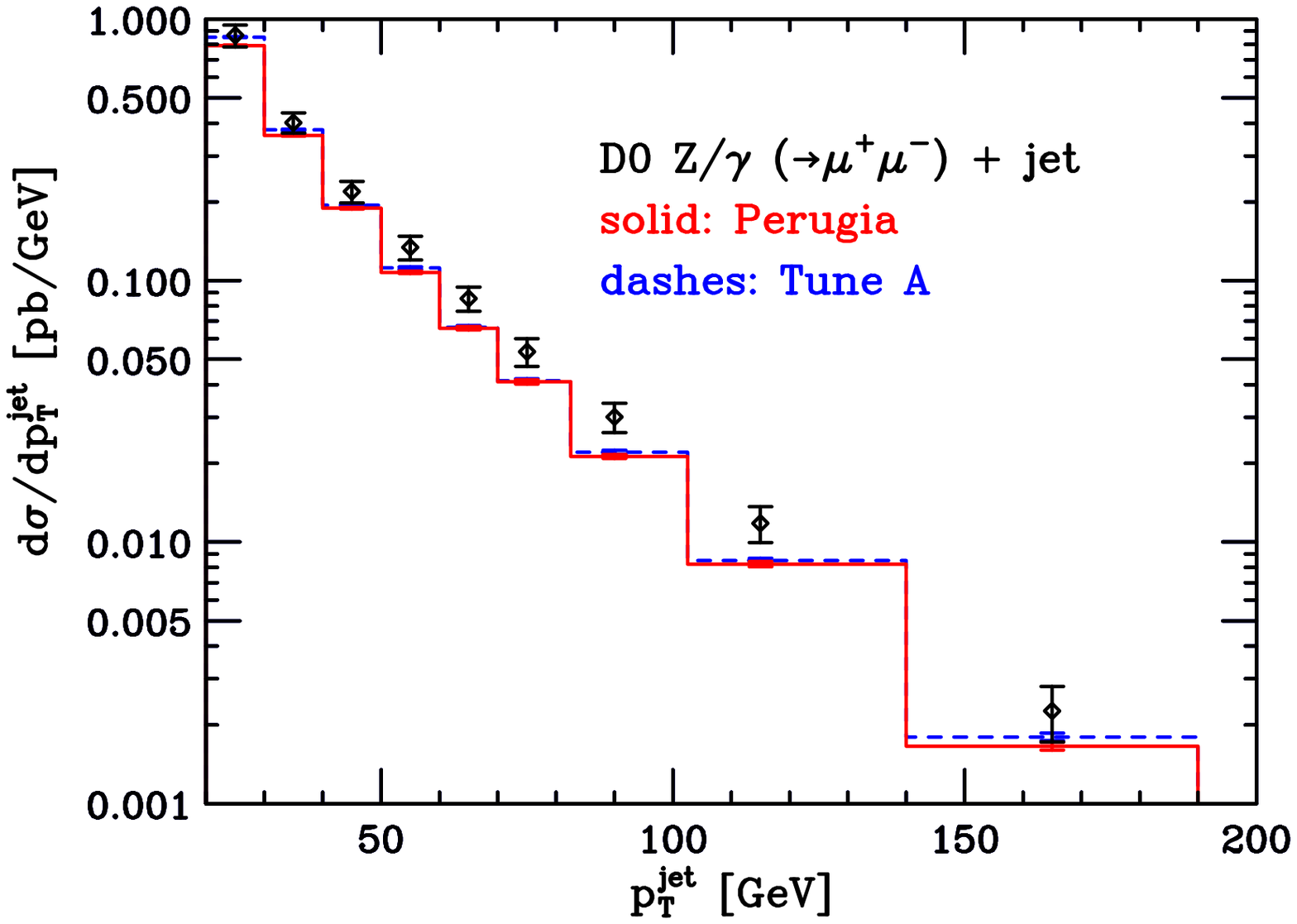,width=0.48\textwidth}
\epsfig{file=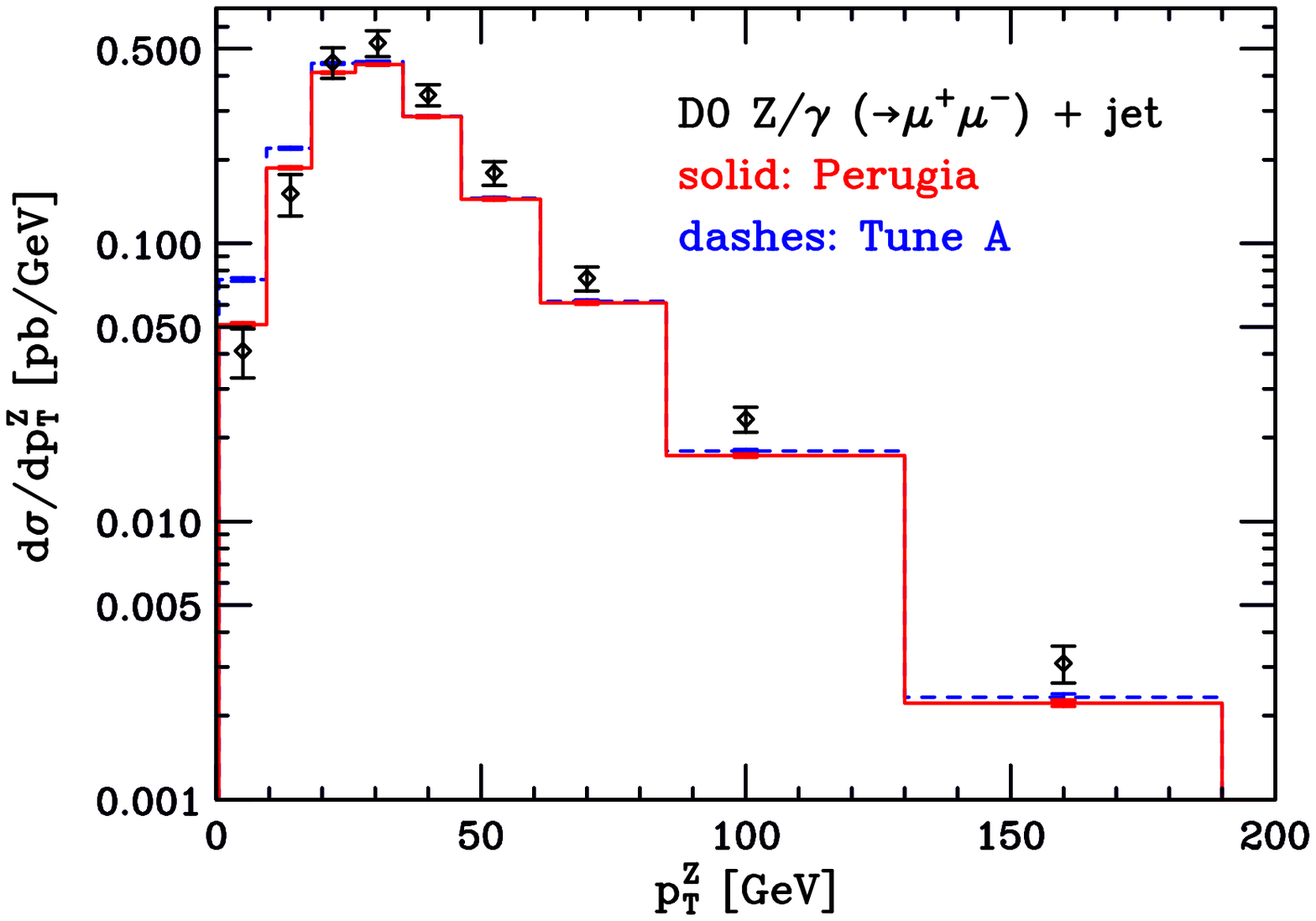,width=0.48\textwidth}\\
\epsfig{file=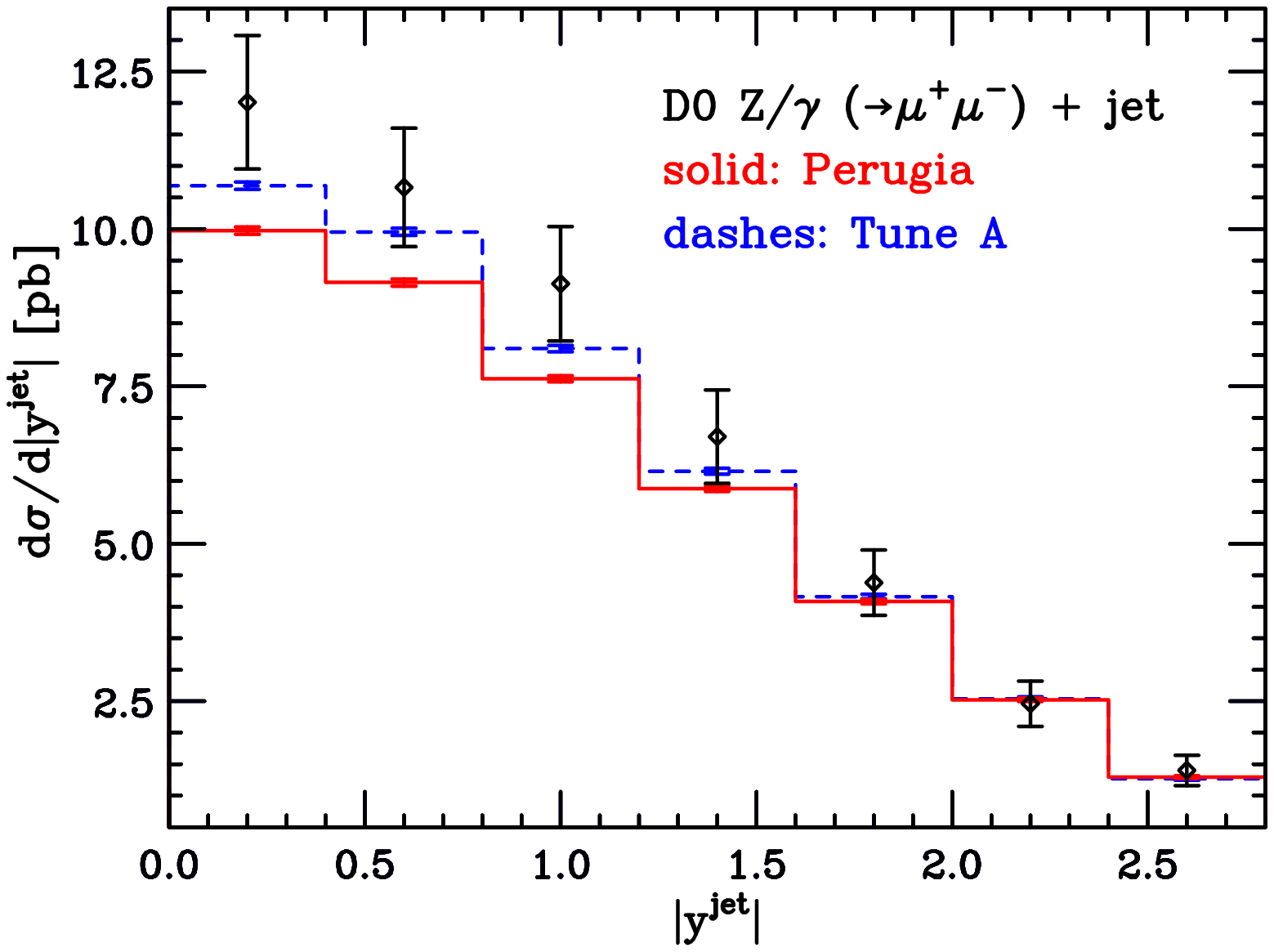,width=0.48\textwidth}
\epsfig{file=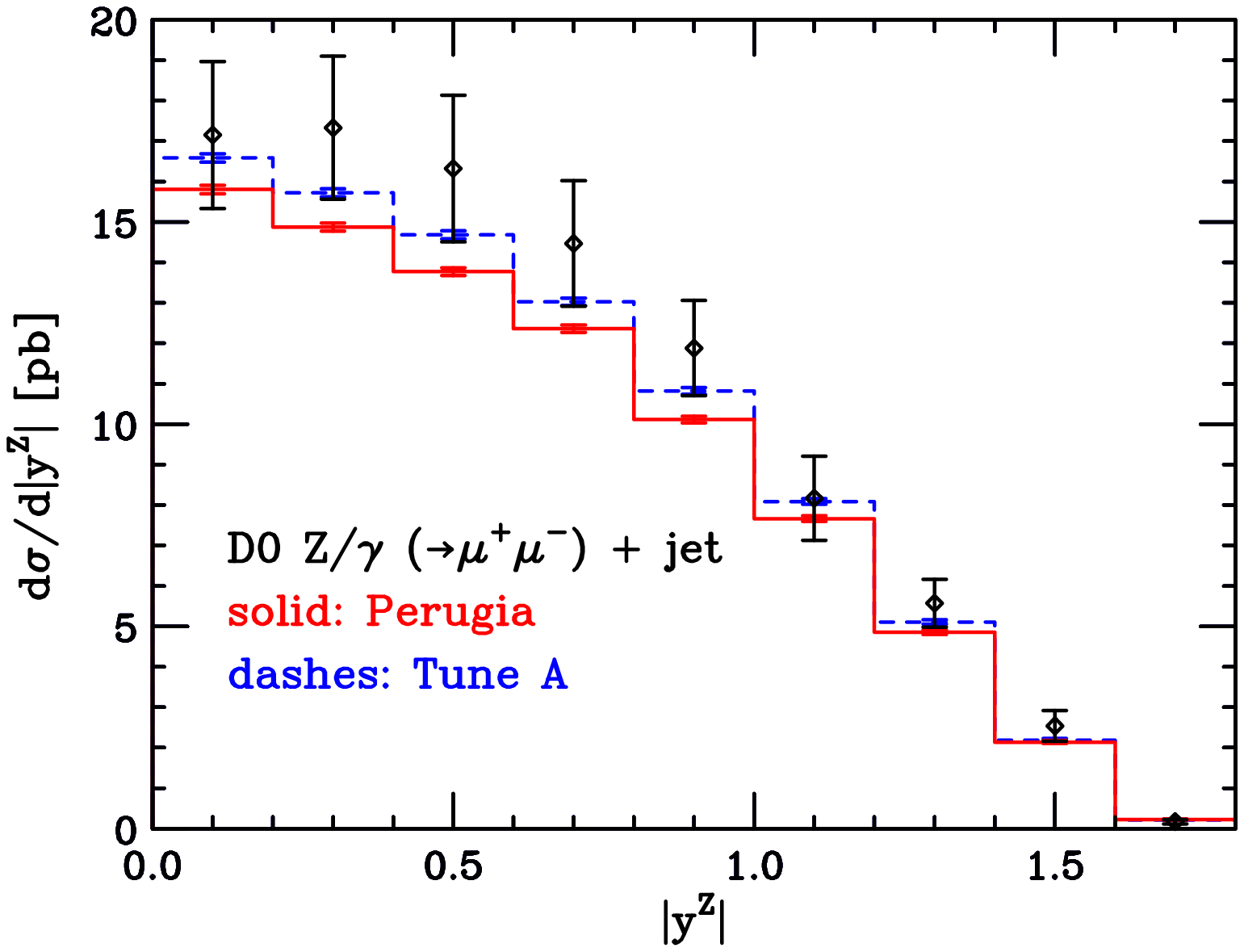,width=0.48\textwidth}
\end{center}
\caption{\label{fig:D0mupmum1} Distributions of the transverse momentum and
  rapidity of the hardest jet and of the $Z$ boson.}
\end{figure}
\begin{figure}[htb]
\begin{center}
\epsfig{file=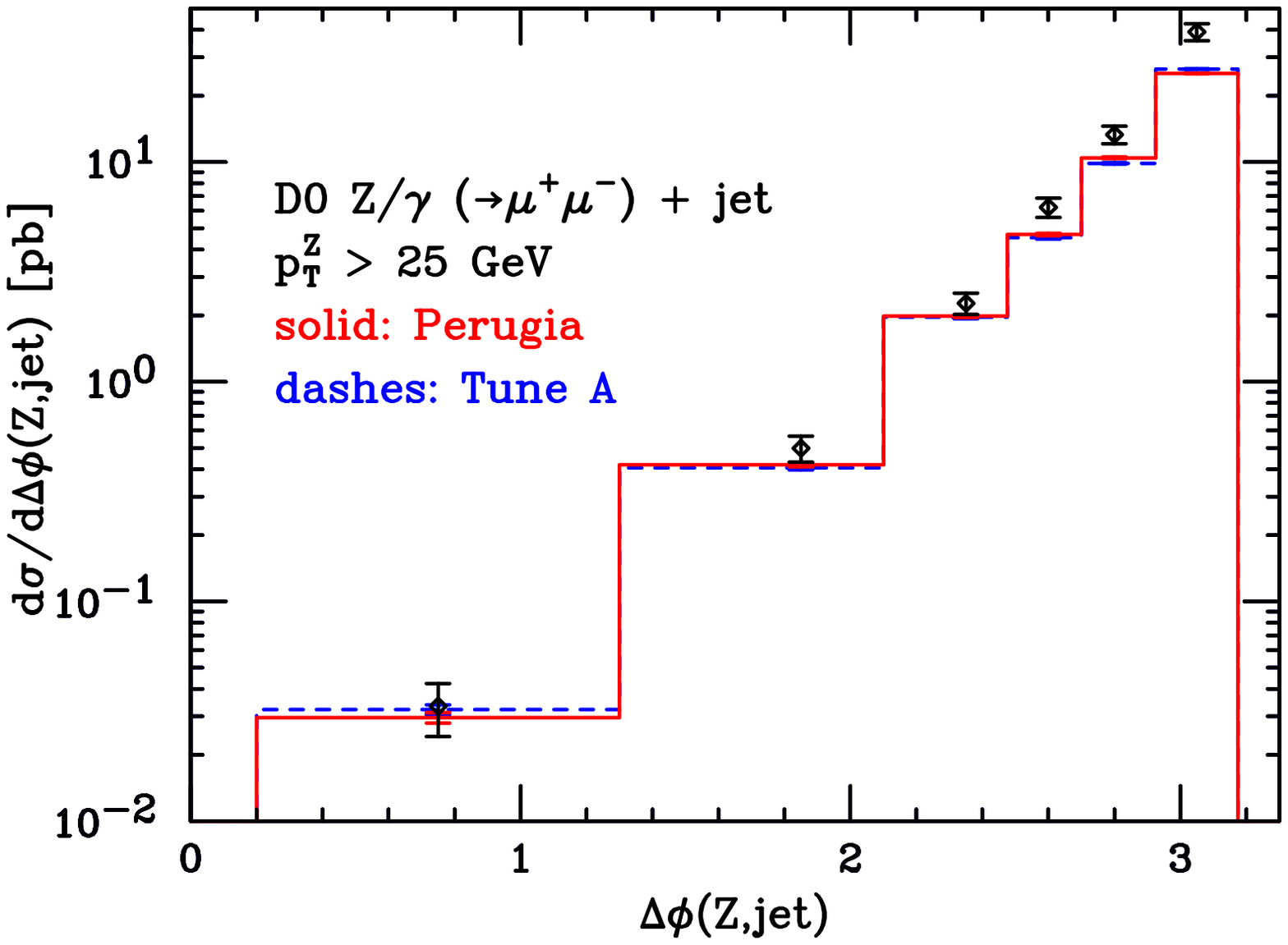,width=0.48\textwidth}
\epsfig{file=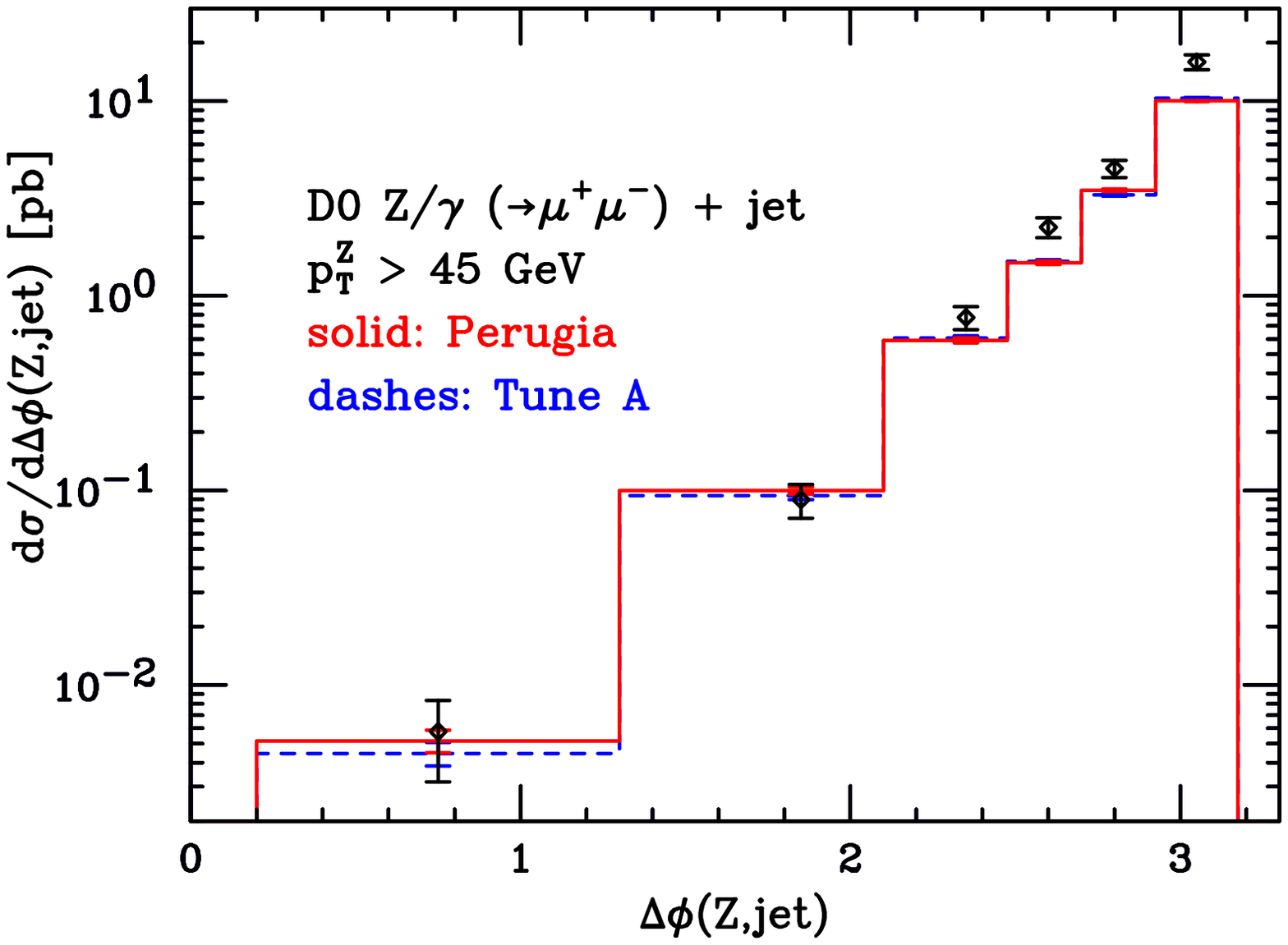,width=0.48\textwidth}\\
\epsfig{file=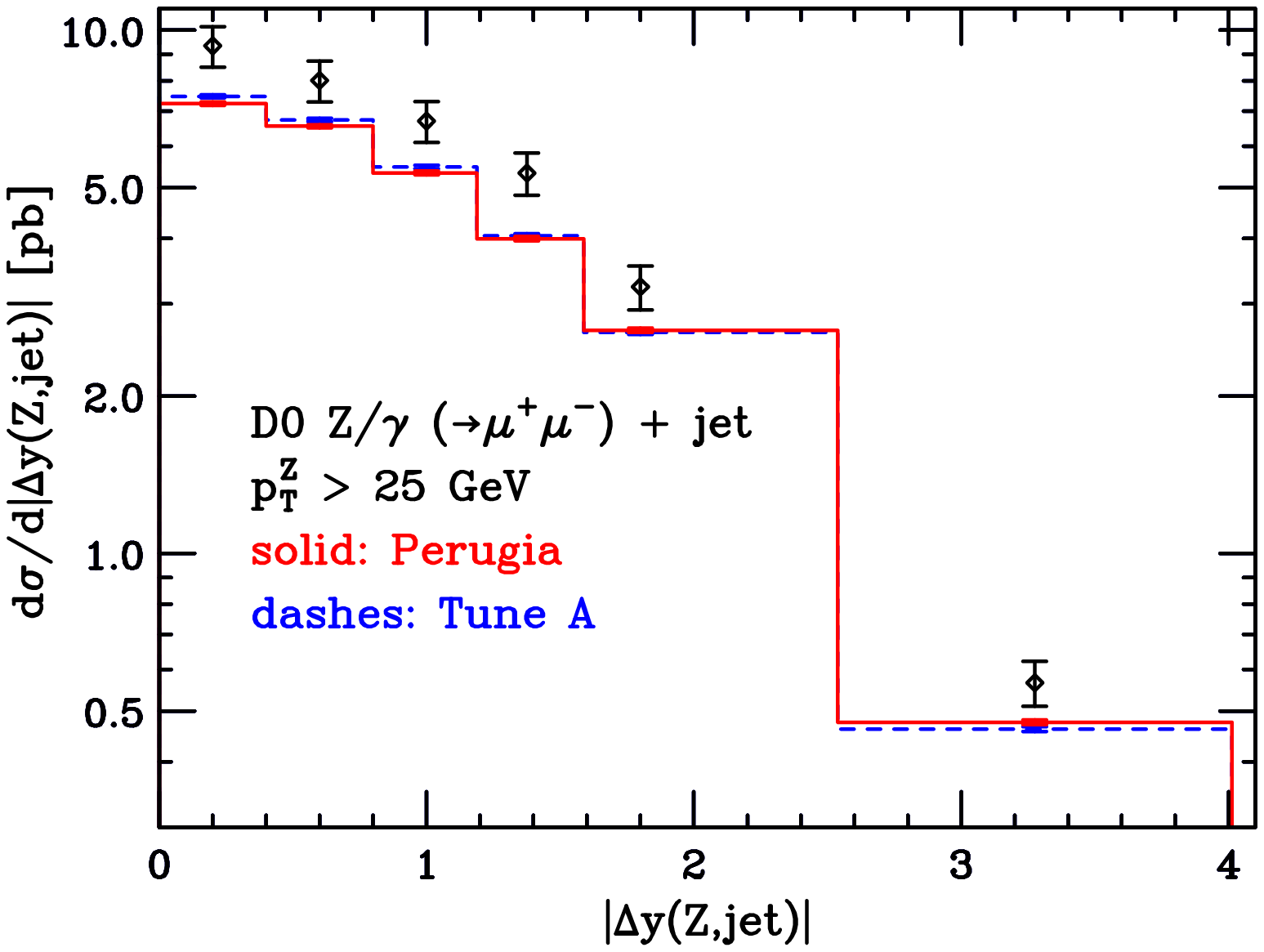,width=0.48\textwidth}
\epsfig{file=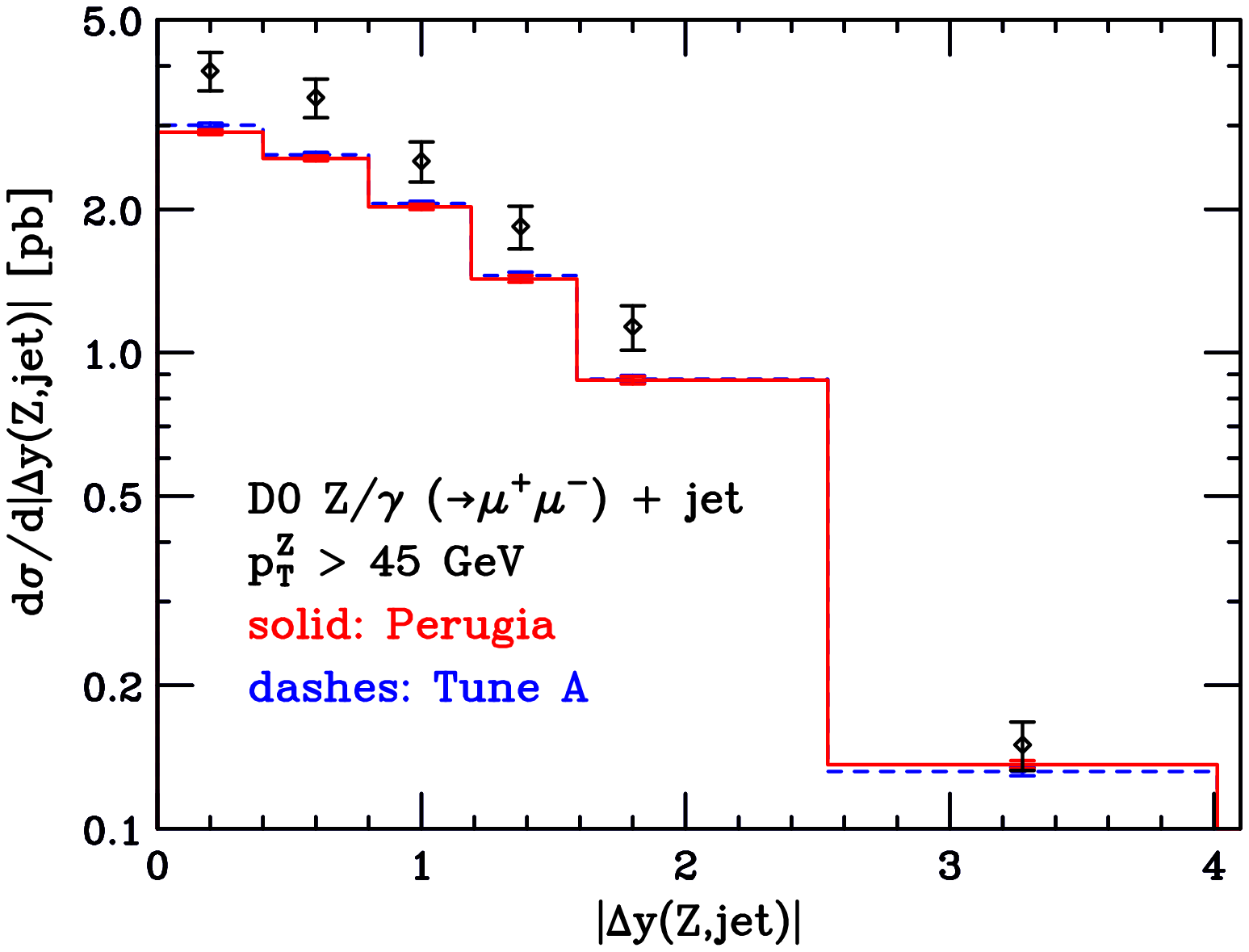,width=0.48\textwidth}
\end{center}
\caption{\label{fig:D0mupmum2}
Azimuthal separation and rapidity separation between the $Z$ boson and the
leading jet, for events with $p_{\sss T}^Z > 25$~GeV and $p_{\sss T}^Z >
45$~GeV.} 
\end{figure}
\begin{figure}[htb]
\begin{center}
\epsfig{file=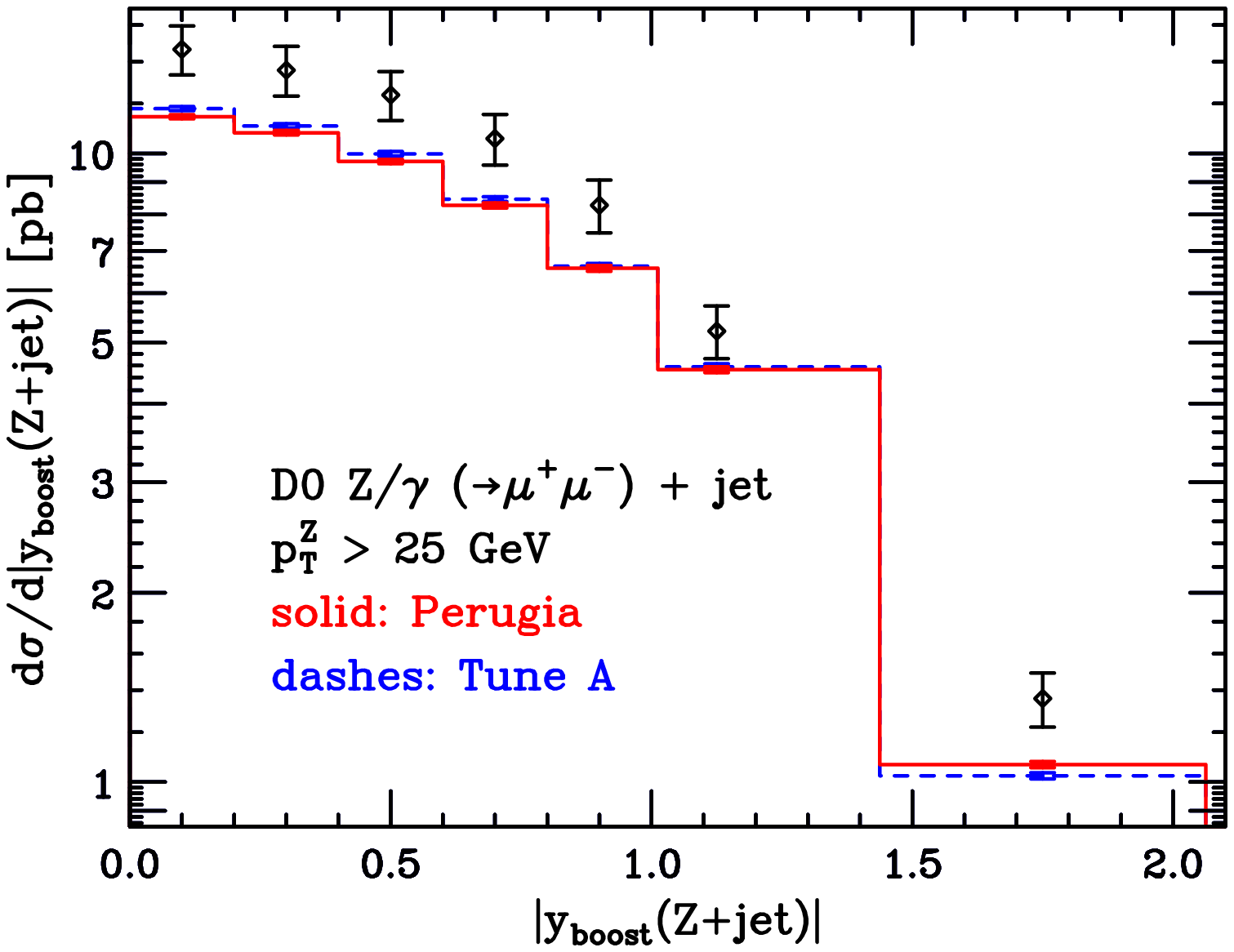,width=0.48\textwidth}
\epsfig{file=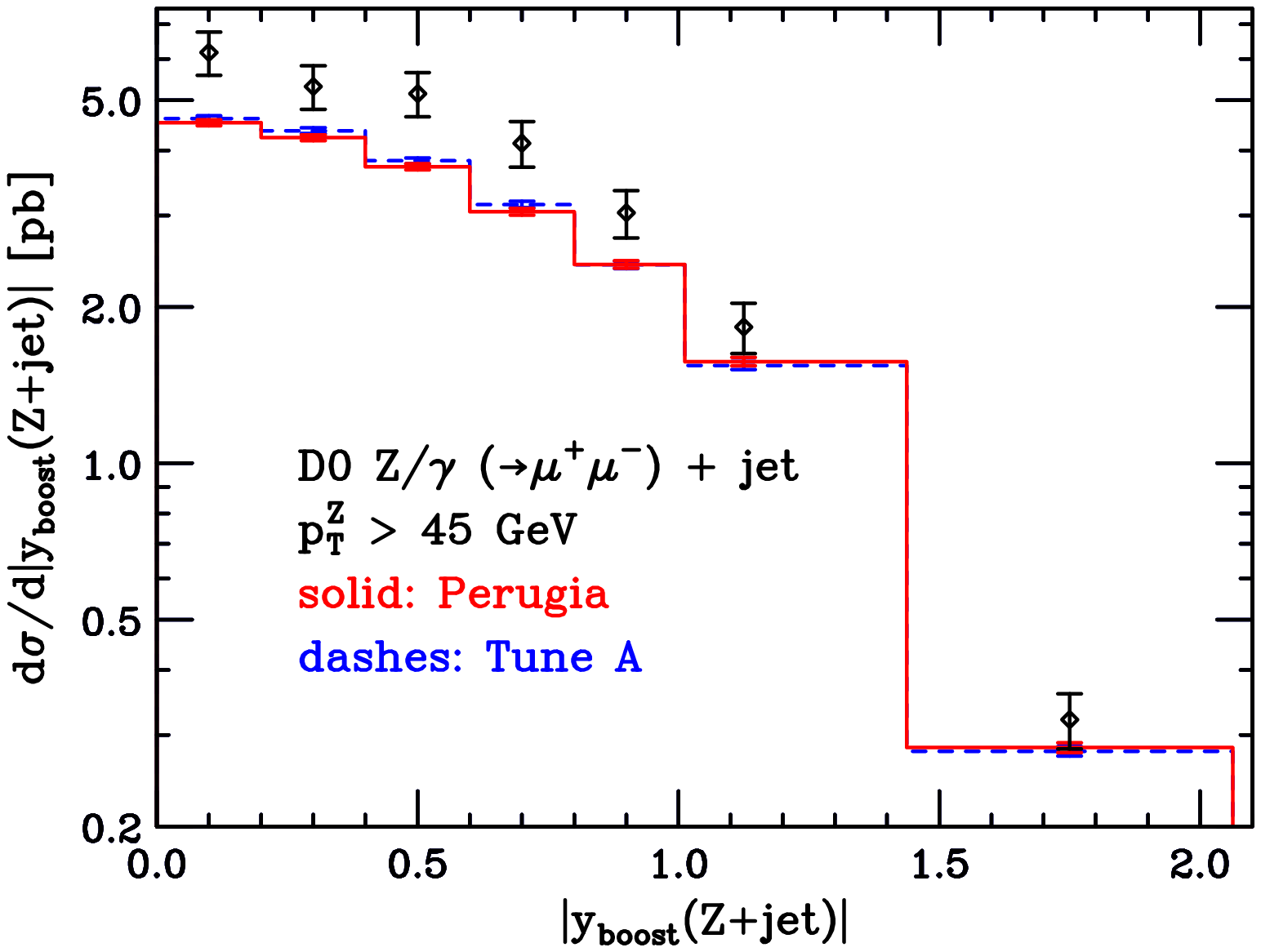,width=0.48\textwidth}
\end{center}
\caption{\label{fig:D0mupmum3} Absolute value of the average rapidity of the
  $Z$ and of the leading jet, $|y_{\rm boost}(Z+{\rm jet})|$, for events with
  $p_{\sss T}^Z > 25$~GeV and $p_{\sss T}^Z > 45$~GeV.}
\end{figure}
In fig.~\ref{fig:D0mupmum1}, we plot distributions of the transverse momentum
and rapidity of the hardest jet and of the $Z$ boson.  In
fig.~\ref{fig:D0mupmum2}, we show the azimuthal separation and rapidity
separation between the $Z$ boson and the leading jet, for events with
$p_{\sss T}^Z > 25$~GeV and $p_{\sss T}^Z > 45$~GeV, while in
fig.~\ref{fig:D0mupmum3} we plot the absolute value of the average rapidity
of the $Z$ and of the leading jet, $|y_{\rm boost}(Z+{\rm jet})|$, for events
with $p_{\sss T}^Z > 25$~GeV and $p_{\sss T}^Z > 45$~GeV, where
\beq
y_{\rm
  boost}(Z+{\rm jet}) = \frac{1}{2}\(y^Z + y^{\rm jet}\)\;,
\eeq
$y^{\rm jet}$ being the rapidity of the leading jet.

In ref.~\cite{Abazov:2009pp}, the D0 Collaboration plotted the normalized
differential cross sections, in order to reduce the associated error bars.
We have, instead, preferred to plot the differential cross sections itself,
in order to appreciate how well \POWHEG{} reproduce not only the shapes but
also the overall normalization. The total cross section they use to normalize
is given in~\cite{Abazov:2008ez} and is equal to 
\beq
\label{eq:sigZ}
\sigma_Z=118\pm 0.5\,({\rm stat.})\pm 4\,({\rm syst.})\pm 4\,({\rm muon})
\pm 7\,({\rm lumi.}) {\rm \ pb}\,.
\eeq
In addition, we have added, in quadrature, a flat systematic luminosity error
of 8\% to every bin.

We see a noticeable discrepancy between data and the \POWHEG{} prediction in
the shape of the transverse-momentum spectra of the jet and of the $Z$.  This
discrepancy also manifest itself in differences in the normalization of the
angular distributions in figs.~\ref{fig:D0mupmum2} and~\ref{fig:D0mupmum3},
since transverse momentum cuts are applied there.  We do not wish to comment
further on this problem, that, on the other hand, is not present in the CDF
analyses, and in the D0 analysis of the $Z/\gamma\to e^+e^-$ channel.  We
remark that the problem may well be due to our failure to understand some
features of the D0 analysis, rather than to problems in the data or in
\POWHEG{}. We also notice that the angular correlation between the $Z$ and
the jet $\pt$ is only qualitatively described by \POWHEG{}. We remind the
reader, however, that \POWHEG{} has only LO accuracy for this quantity, that
is determined by the emission of a second hard parton not included in the
hardest jet, and that near the back-to-back region, Sudakov resummation
effects, as well as non perturbative effects, become determinant.

\subsubsection*{$\boldsymbol{Z/\gamma\ (\to e^+ e^-)}$ + jets}
D0 has published studies for the $Z/\gamma\ (\to e^+ e^-)+1j$ channel
in ref.~\cite{Abazov:2009av}. 
\begin{figure}[htb]
\begin{center}
\epsfig{file=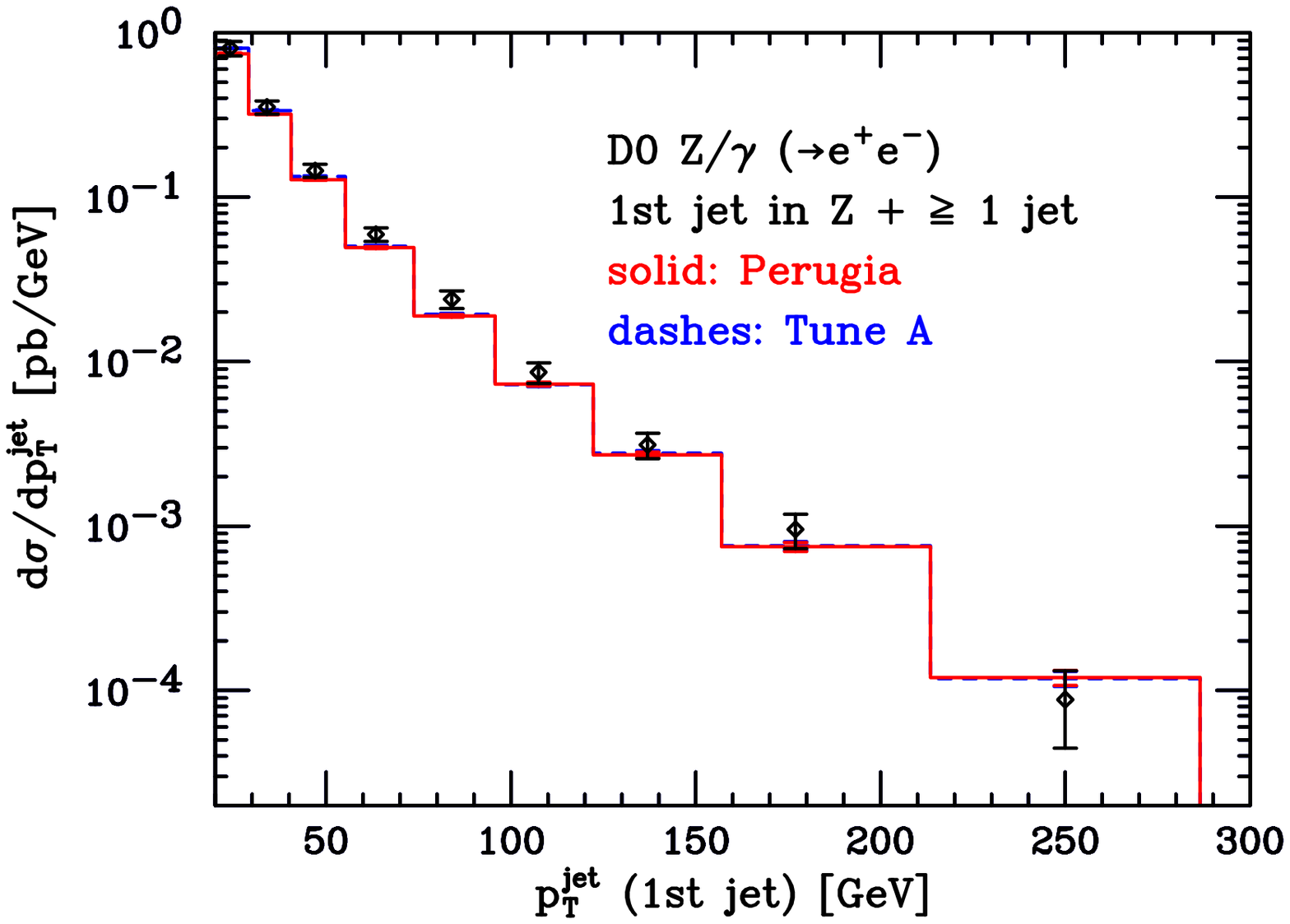,width=0.48\textwidth}
\epsfig{file=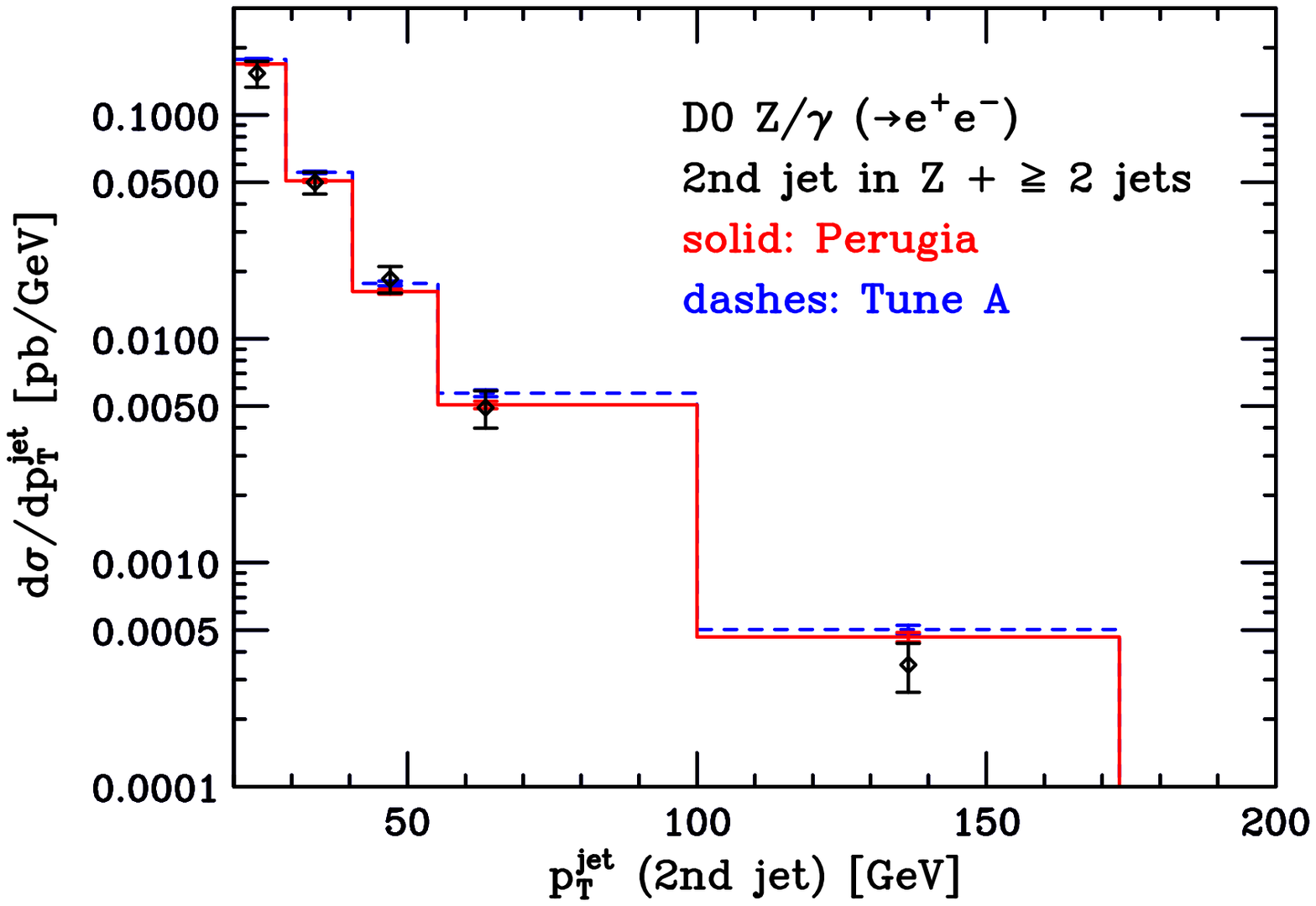,width=0.48\textwidth}\\
\epsfig{file=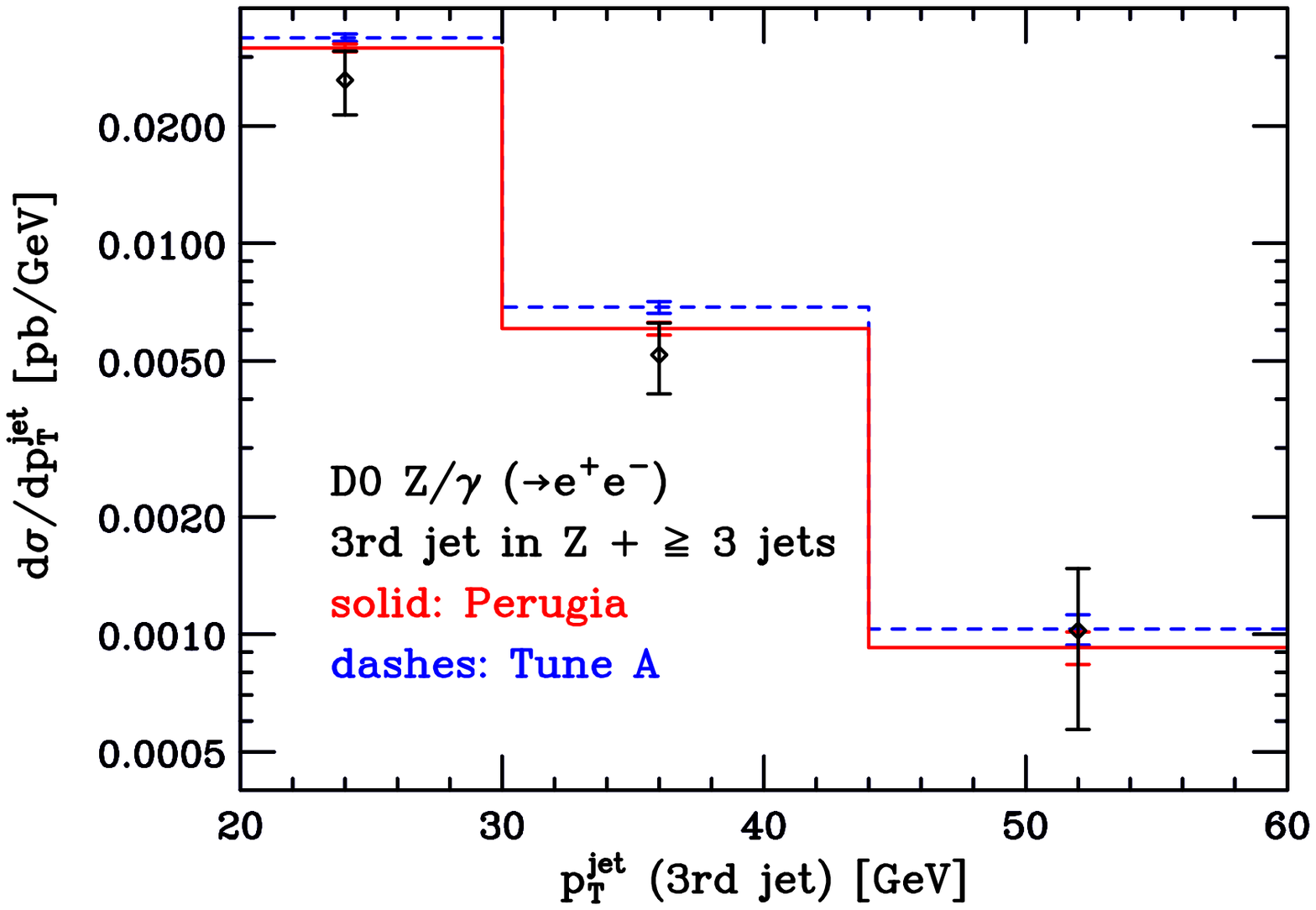,width=0.48\textwidth}\hfill
\end{center}
\caption{\label{fig:d0epem}
$\pt$ distributions of the hardest, next-to-hardest and
next-to-next-to-hardest jet.}
\end{figure}
In fig.~\ref{fig:d0epem} we have plotted the $\pt$ distributions of the
hardest, next-to-hardest and next-to-next-to-hardest jet. No information is
given in the paper on the value of the total inclusive $Z$ cross section used
to normalized the distributions. It is reasonable to use the same total cross
section of eq.~(\ref{eq:sigZ}).

The cuts that we have applied for this analysis were the following
\beqn
&&65~{\rm GeV} < M_{ee} < 115~{\rm GeV},\quad p_{\sss T}^e > 25~{\rm GeV},
\quad |\eta^{e}|< 1.1 \ {\rm or}\  1.5 < |\eta^{e}|< 2.5\,,
\nonumber\\
&&|y^{\rm jet}|<2.5\,, \quad p_{\sss T}^{\rm jet} > 20~{\rm GeV}\,.
\eeqn
We can see quite a good agreement among the data and the two \POWHEG{}
distributions for the hardest and next-to-hardest jet. Unexpectedly, there is
quite a good agreement for the next-to-next-to-hardest jet too: in fact, this
jet is a shower jet, so we do not expect it to be correct away from the
collinear limit.

\subsection{LHC results}
In this section, we show a few results for the LHC at 14~TeV. We have
produced 1.400.000 events, generated with the \wneg{} flag set to 1, with
$\kgen=5$~GeV and with a 5-10-5 folding. With this folding, the fraction of
negative weights is around 1.6\%.  We have then showered the events generated
by \POWHEG{} with \PYTHIA, using the tune~A and the Perugia~0 tuning.
\begin{figure}[htb]
\begin{center}
\epsfig{file=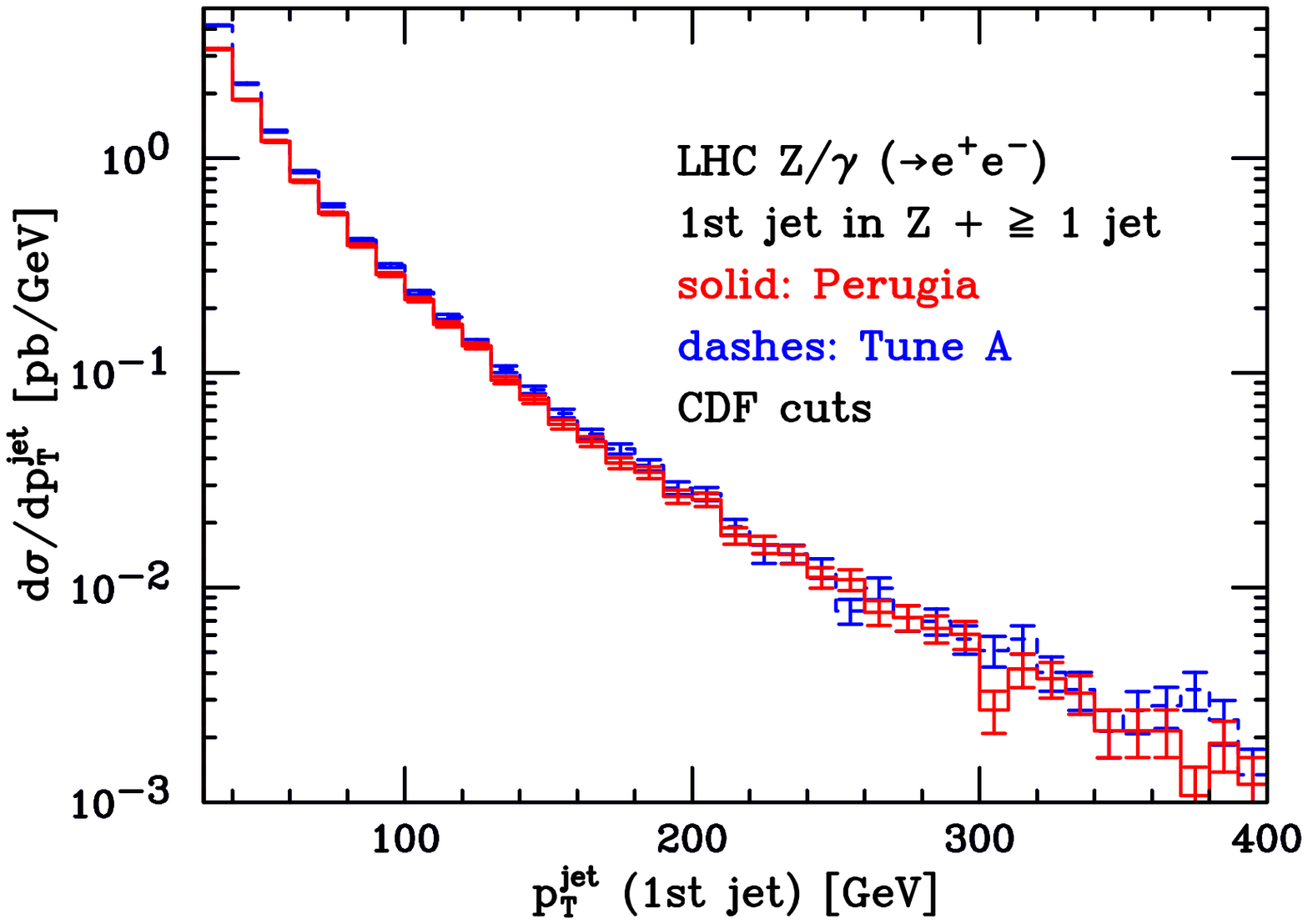,width=0.48\textwidth}
\epsfig{file=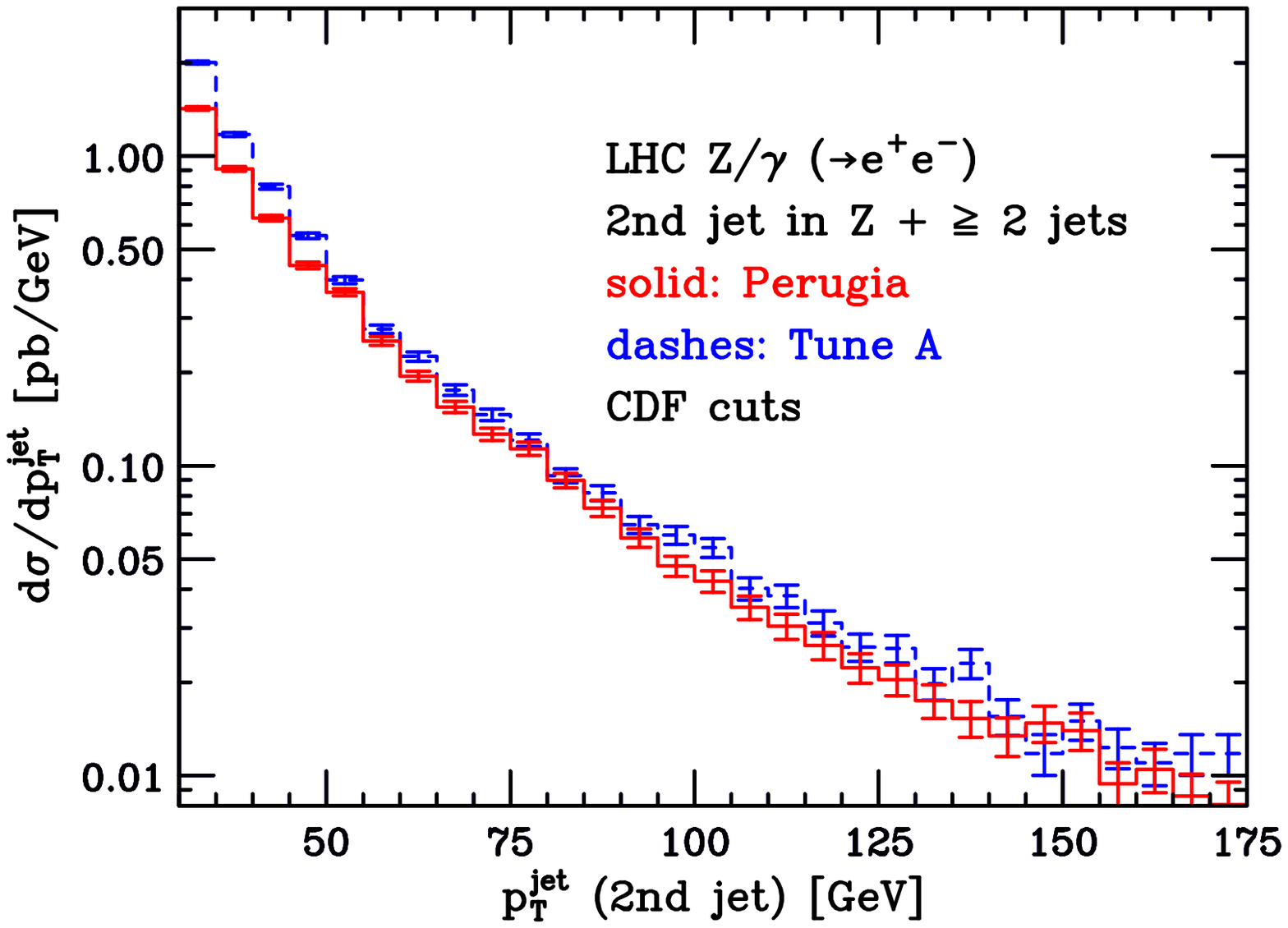,width=0.48\textwidth}\\
\epsfig{file=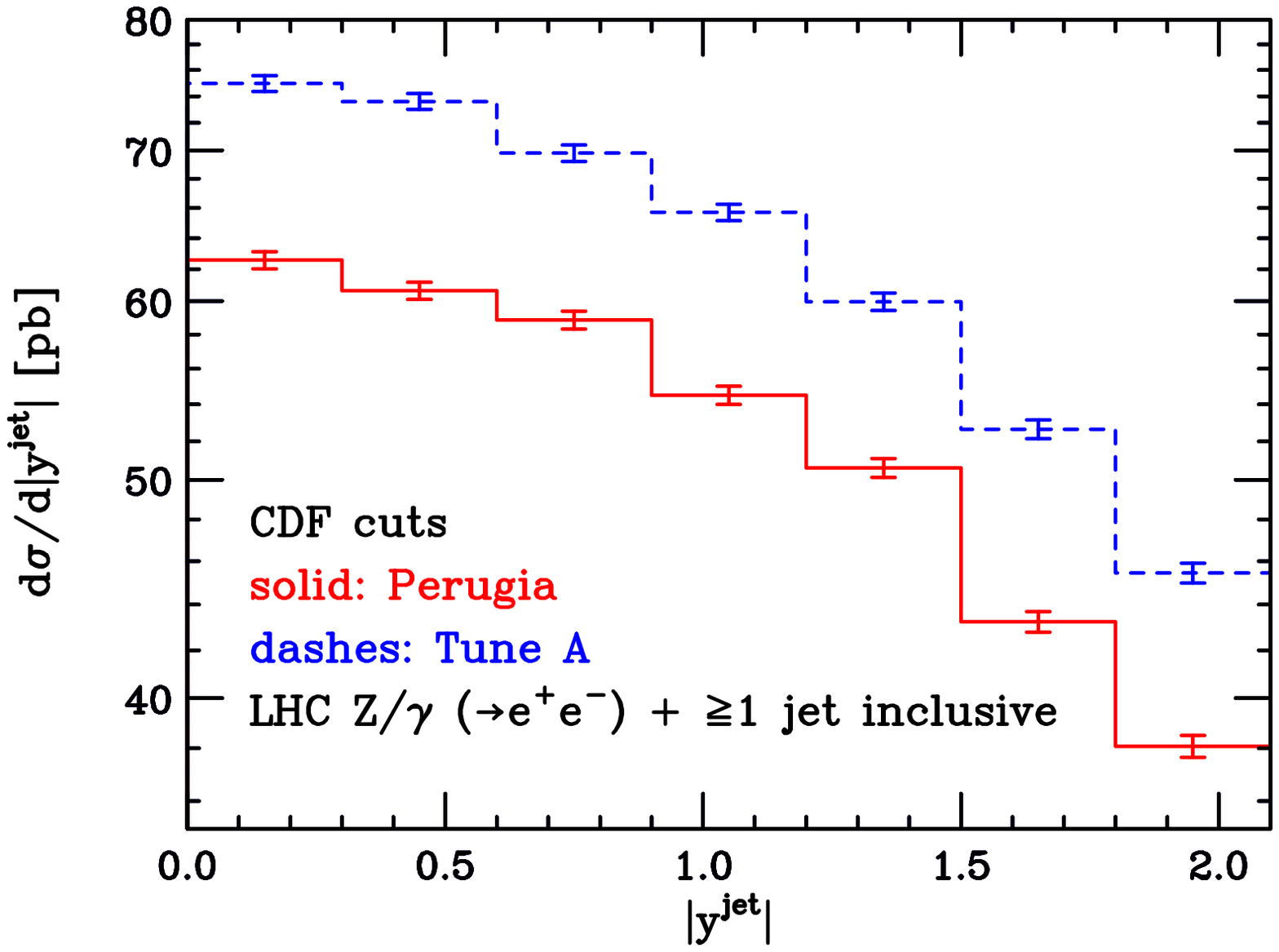,width=0.48\textwidth}
\epsfig{file=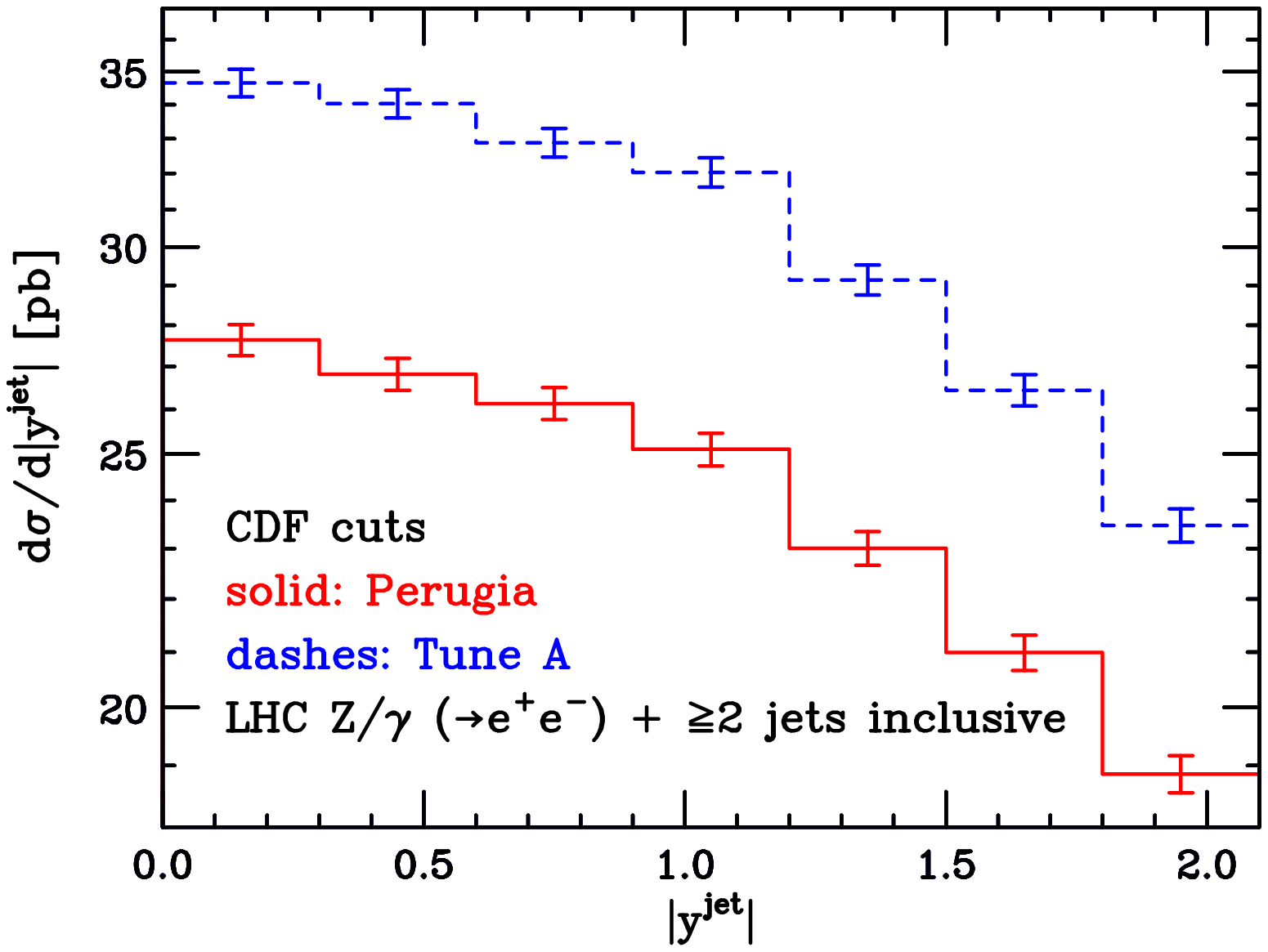,width=0.48\textwidth}
\end{center}
\caption{\label{fig:LHC_CDF_epem} $\pt$ distributions of the hardest and
  next-to-hardest jet and the inclusive rapidity distributions for events
  with at least one and two jets at the LHC at 14~TeV, with the cuts of
  eq.~(\ref{eq:CDF_epem_cuts}).}
\end{figure}
\begin{figure}[htb]
\begin{center}
\epsfig{file=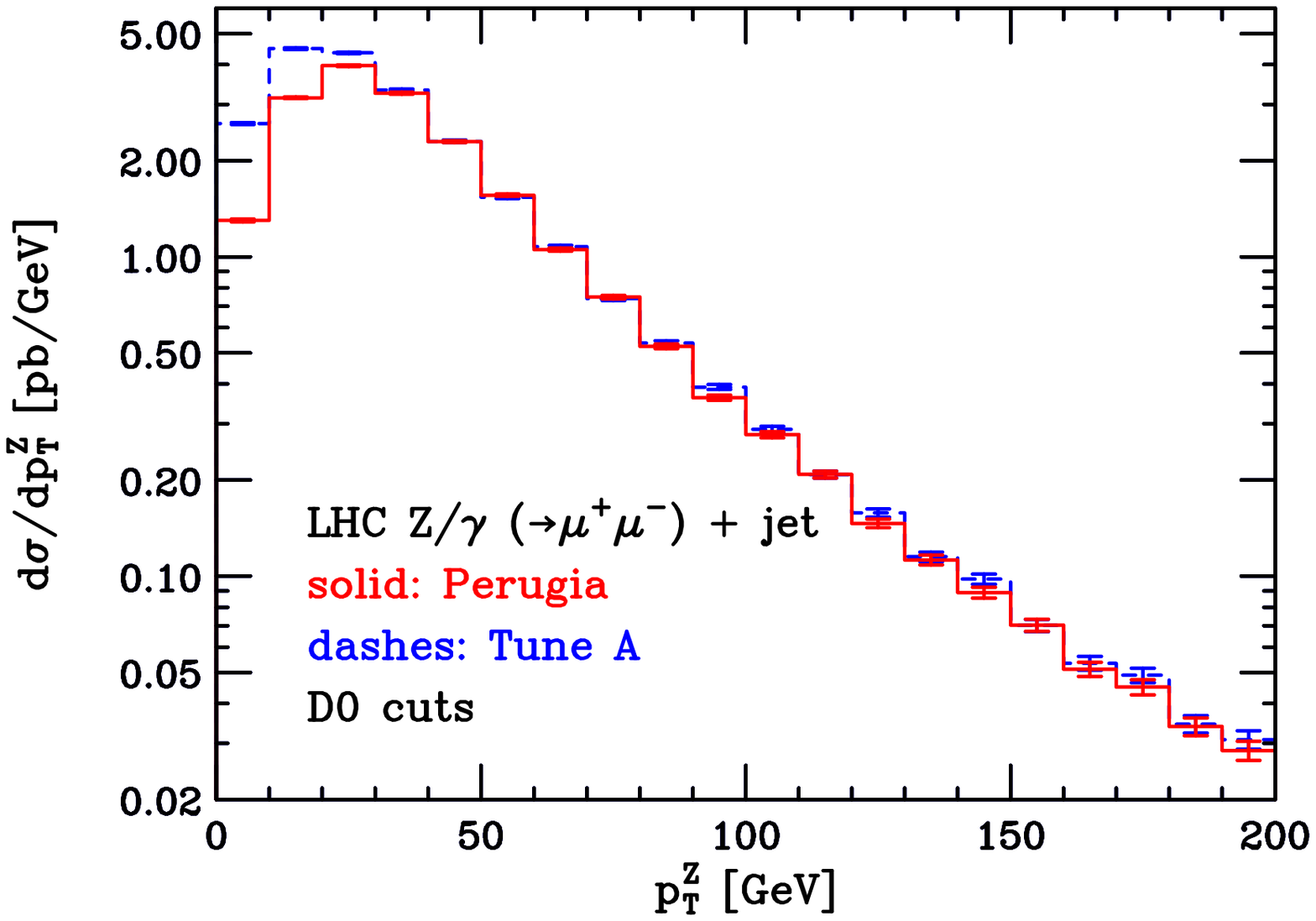,width=0.48\textwidth}
\epsfig{file=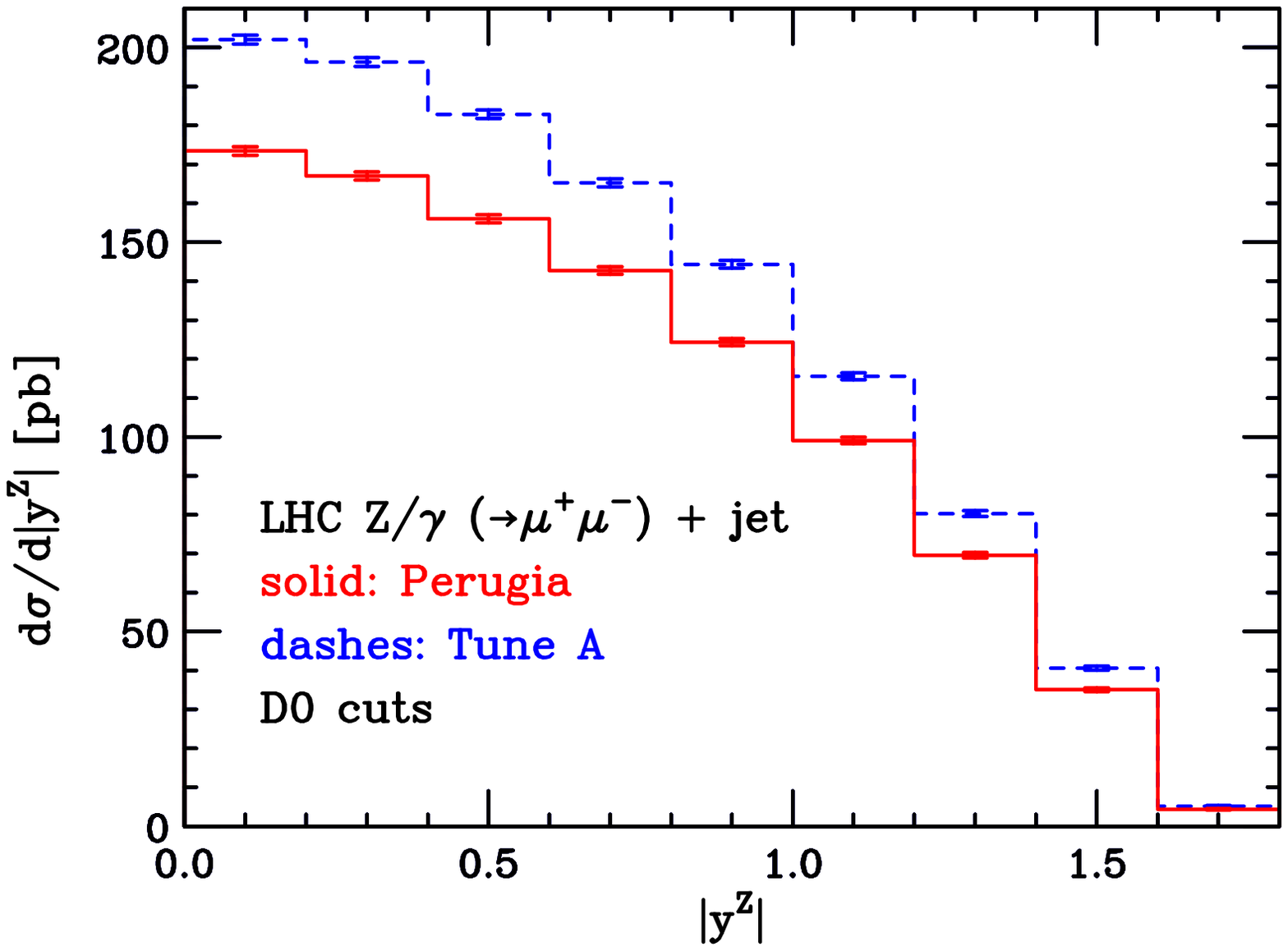,width=0.48\textwidth}\\
\epsfig{file=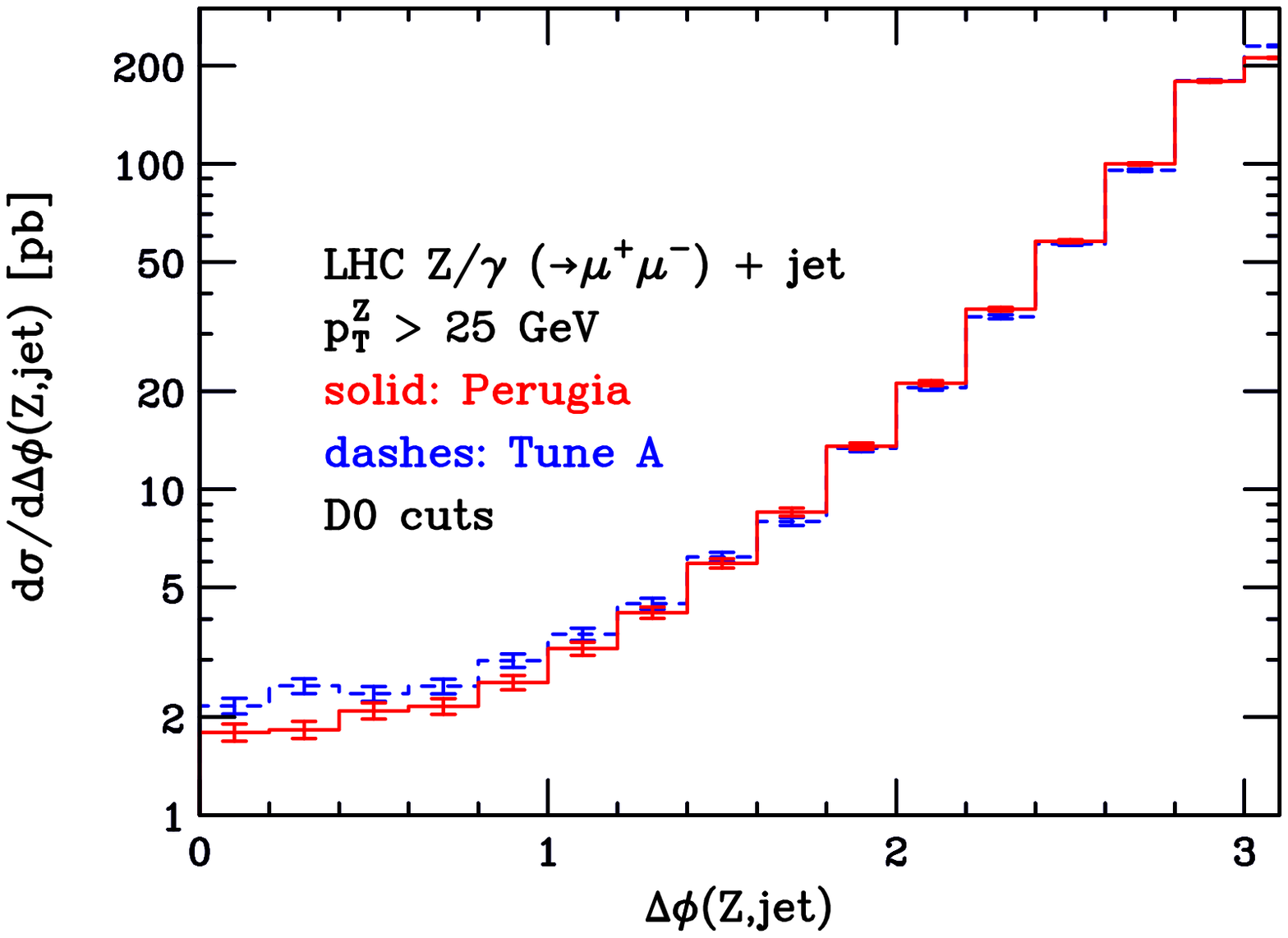,width=0.48\textwidth}
\epsfig{file=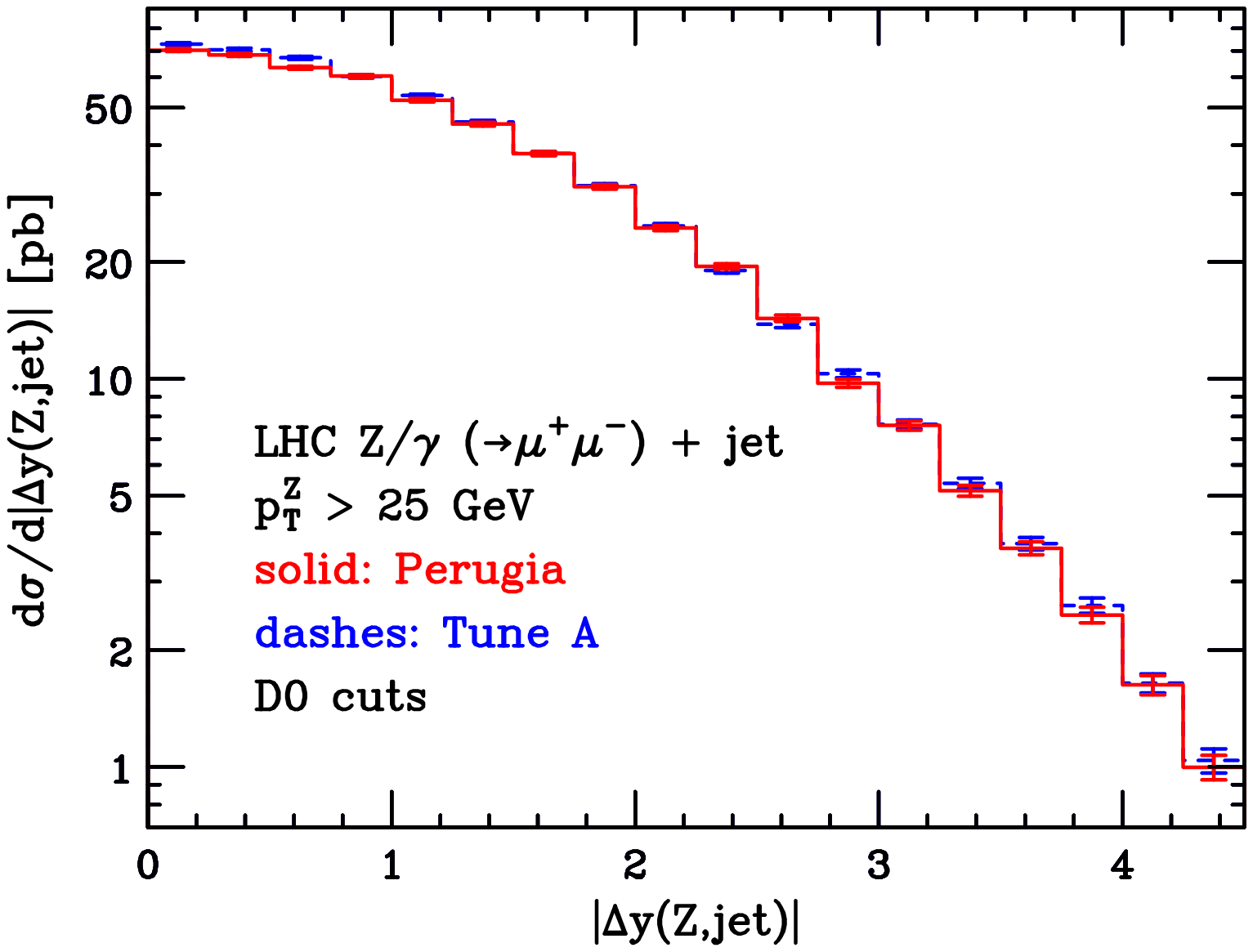,width=0.48\textwidth}
\end{center}
\caption{\label{fig:LHC_D0_mupmum} Distributions of the transverse momentum
  and rapidity of the $Z$ boson and azimuthal separation and rapidity
  separation between the $Z$ boson and the leading jet, for events with
  $p_{\sss T}^Z > 25$~GeV, at the LHC at 14~TeV, with the cuts of
  eq.~(\ref{eq:D0_mupmum_cuts}). }
\end{figure}
In figs.~\ref{fig:LHC_CDF_epem} and~\ref{fig:LHC_D0_mupmum} we have plotted a
few distributions obtained for the LHC, using the same cuts and the same jet
algorithms used by the two Collaborations at the Tevatron: in
fig.~\ref{fig:LHC_CDF_epem}, we have applied the cuts of
eq.~(\ref{eq:CDF_epem_cuts}) that CDF used for the $Z/\gamma \ (\to e^+ e^-)
+1j$ analysis, while in fig.~\ref{fig:LHC_D0_mupmum}, we have applied the
cuts of eq.~(\ref{eq:D0_mupmum_cuts}), that D0 used for the $Z/\gamma \ (\to
\mu^+\mu^-)+1j$ study.

In fig.~\ref{fig:LHC_CDF_epem} we can see a pattern of the two tunes similar
to the one depicted in the corresponding fig.~\ref{fig:cdfepem3}: small
differences for the distribution of the hardest jet, some differences both at
low and high $\pt$ for the next-to-hardest jet, and increasing differences in
their inclusive rapidity distributions, with peaks of $\sim\! 20$\%{}.
Similarly, fig.~\ref{fig:LHC_D0_mupmum} should be compared with the
corresponding panels in figs.~\ref{fig:D0mupmum1} and~\ref{fig:D0mupmum2}.
Tune~A gives a higher cross section at low transverse momentum and central
rapidity of the vector boson with respect to the Perugia~0 tuning, while they
agree very well in the azimuthal separation and rapidity separation between
the $Z$ boson and the leading jet, for events with $p_{\sss T}^Z > 25$~GeV.

\section{Conclusions}
\label{sec:conclusions}
In this paper, we have presented, for the first time, a calculation of
vector boson plus one jet NLO cross section interfaced to a Shower Monte
Carlo program, within the \POWHEG{} framework.  More specifically, we have
implemented this process using the \POWHEGBOX{}, a general computer code
framework for embedding NLO calculations in shower Monte Carlo programs
according to the \POWHEG{} method.

We have compared the \POWHEG{} results for $Z/\gamma+1j$ with the data
available from the CDF and the D0 Collaborations, using two different tunes
for \PYTHIA{}. We notice that differences between the \POWHEG{} results and
the data is of the same order of the differences between the two tunes, thus
suggesting that, by directly tuning the \POWHEG{} results to data, one may
get an even better agreement.

Finally, we have presented similar results for the LHC, running at 14~TeV,
showing that, in some distributions, the difference between the two tunes is
of the order of 20\%.

The code of our generator can be accessed in the \POWHEGBOX{} svn repository:\\
\url{svn://powhegbox.mib.infn.it/trunk/POWHEG-BOX}, with username {\tt
  anonymous} and password {\tt anonymous}.

\acknowledgments We thank John Campbell for helping us to access the virtual
corrections in the \MCFM{} code. We also thank Gavin Hesketh, Sabine Lammers
and Henrik Nilsen for useful discussions on the experimental
side. E.R.~thanks Daniel Ma\^{\i}tre for his help comparing \BlackHat{} and
\MCFM{} results.  The work of S.A.~has been supported in part by the Deutsche
Forschungsgemeinschaft in SFB/TR 9.  
S.A.~and E.R.~would like to thank DESY Zeuthen for the use of the computing
facilities and the hospitality, and CERN, where part of this work was
performed.

\bibliography{paper}

\providecommand{\href}[2]{#2}\begingroup\raggedright\begin{thebibliography}{10}

\bibitem{Aaltonen:2007cp}
{\bf CDF - Run II} Collaboration, T.~Aaltonen {\em et~al.}, {\it {Measurement
  of inclusive jet cross-sections in $Z/\gamma^* (\to e^{+} e^{-})$ + jets
  production in $p \bar{p}$ collisions at $\sqrt{s}$ = 1.96~TeV}},  {\em Phys.
  Rev. Lett.} {\bf 100} (2008) 102001,
  [\href{http://xxx.lanl.gov/abs/0711.3717}{{\tt 0711.3717}}].

\bibitem{Abazov:2006gs}
{\bf D0} Collaboration, V.~M. Abazov {\em et~al.}, {\it {Measurement of the
  ratios of the $Z/\gamma^* + \ge n$ jet production cross sections to the total
  inclusive $Z/\gamma^*$ cross section in $p\bar{p}$ collisions at $\sqrt{s} =
  1.96$~TeV}},  {\em Phys. Lett.} {\bf B658} (2008) 112--119,
  [\href{http://xxx.lanl.gov/abs/hep-ex/0608052}{{\tt hep-ex/0608052}}].

\bibitem{Abazov:2008ez}
{\bf D0} Collaboration, V.~M. Abazov {\em et~al.}, {\it {Measurement of
  differential $Z / \gamma^{*}$ + jet + $X$ cross sections in $p \bar{p}$
  collisions at $\sqrt{s} = 1.96$~TeV}},  {\em Phys. Lett.} {\bf B669} (2008)
  278--286, [\href{http://xxx.lanl.gov/abs/0808.1296}{{\tt 0808.1296}}].

\bibitem{Abazov:2009av}
{\bf D0} Collaboration, V.~M. Abazov {\em et~al.}, {\it {Measurements of
  differential cross sections of $Z /\gamma^\ast$+jets+X events in proton
  anti-proton collisions at $\sqrt{s}$=1.96 TeV}},  {\em Phys. Lett.} {\bf
  B678} (2009) 45--54, [\href{http://xxx.lanl.gov/abs/0903.1748}{{\tt
  0903.1748}}].

\bibitem{Abazov:2009pp}
{\bf D0} Collaboration, V.~M. Abazov {\em et~al.}, {\it {Measurement of $Z /
  \gamma^\ast +jet+X$ angular distributions in $p \bar{p}$ collisions at
  $\sqrt{s}=1.96$ TeV}},  {\em Phys. Lett.} {\bf B682} (2010) 370--380,
  [\href{http://xxx.lanl.gov/abs/0907.4286}{{\tt 0907.4286}}].

\bibitem{Buttar:2008jx}
C.~Buttar {\em et~al.}, {\it {Standard Model Handles and Candles Working Group:
  Tools and Jets Summary Report}},
  \href{http://xxx.lanl.gov/abs/0803.0678}{{\tt 0803.0678}}.

\bibitem{Nason:2004rx}
P.~Nason, {\it {A new method for combining NLO QCD with shower Monte Carlo
  algorithms}},  {\em JHEP} {\bf 11} (2004) 040,
  [\href{http://xxx.lanl.gov/abs/hep-ph/0409146}{{\tt hep-ph/0409146}}].

\bibitem{Frixione:2007vw}
S.~Frixione, P.~Nason, and C.~Oleari, {\it {Matching NLO QCD computations with
  Parton Shower simulations: the POWHEG method}},  {\em JHEP} {\bf 11} (2007)
  070, [\href{http://xxx.lanl.gov/abs/0709.2092}{{\tt 0709.2092}}].

\bibitem{Alioli:2010xd}
S.~Alioli, P.~Nason, C.~Oleari, and E.~Re, {\it {A general framework for
  implementing NLO calculations in shower Monte Carlo programs: the POWHEG
  BOX}},  {\em JHEP} {\bf 06} (2010) 043,
  [\href{http://xxx.lanl.gov/abs/1002.2581}{{\tt 1002.2581}}].

\bibitem{Hagiwara:1985yu}
K.~Hagiwara and D.~Zeppenfeld, {\it {Helicity Amplitudes for Heavy Lepton
  Production in $e^+ e^-$ Annihilation}},  {\em Nucl. Phys.} {\bf B274} (1986)
  1.

\bibitem{Hagiwara:1988pp}
K.~Hagiwara and D.~Zeppenfeld, {\it {Amplitudes for Multiparton Processes
  Involving a Current at $e^+ e^-$, $e^\pm p$, and Hadron Colliders}},  {\em
  Nucl. Phys.} {\bf B313} (1989) 560.

\bibitem{Campbell:2002tg}
J.~M. Campbell and R.~K. Ellis, {\it {Next-to-leading order corrections to $W
  +2$ jet and $Z+ 2$ jet production at hadron colliders}},  {\em Phys. Rev.}
  {\bf D65} (2002) 113007, [\href{http://xxx.lanl.gov/abs/hep-ph/0202176}{{\tt
  hep-ph/0202176}}].

\bibitem{Bern:1997sc}
Z.~Bern, L.~J. Dixon, and D.~A. Kosower, {\it {One-loop amplitudes for $e^+
  e^-$ to four partons}},  {\em Nucl. Phys.} {\bf B513} (1998) 3--86,
  [\href{http://xxx.lanl.gov/abs/hep-ph/9708239}{{\tt hep-ph/9708239}}].

\bibitem{Giele:1991vf}
W.~T. Giele and E.~W.~N. Glover, {\it {Higher order corrections to jet
  cross-sections in $e^+ e^-$ annihilation}},  {\em Phys. Rev.} {\bf D46}
  (1992) 1980--2010.

\bibitem{Alioli:2008gx}
S.~Alioli, P.~Nason, C.~Oleari, and E.~Re, {\it {NLO vector-boson production
  matched with shower in POWHEG}},  {\em JHEP} {\bf 07} (2008) 060,
  [\href{http://xxx.lanl.gov/abs/0805.4802}{{\tt 0805.4802}}].

\bibitem{Boos:2001cv}
E.~Boos {\em et~al.}, {\it Generic user process interface for event
  generators},  \href{http://xxx.lanl.gov/abs/hep-ph/0109068}{{\tt
  hep-ph/0109068}}.

\bibitem{Alwall:2006yp}
J.~Alwall {\em et~al.}, {\it A standard format for les houches event files},
  {\em Comput. Phys. Commun.} {\bf 176} (2007) 300--304,
  [\href{http://xxx.lanl.gov/abs/hep-ph/0609017}{{\tt hep-ph/0609017}}].

\bibitem{Pumplin:2002vw}
J.~Pumplin {\em et~al.}, {\it {New generation of parton distributions with
  uncertainties from global QCD analysis}},  {\em JHEP} {\bf 07} (2002) 012,
  [\href{http://xxx.lanl.gov/abs/hep-ph/0201195}{{\tt hep-ph/0201195}}].

\bibitem{Frixione:2007nw}
S.~Frixione, P.~Nason, and G.~Ridolfi, {\it {A Positive-Weight
  Next-to-Leading-Order Monte Carlo for Heavy Flavour Hadroproduction}},  {\em
  JHEP} {\bf 09} (2007) 126, [\href{http://xxx.lanl.gov/abs/0707.3088}{{\tt
  0707.3088}}].

\bibitem{Frixione:2007nu}
S.~Frixione, P.~Nason, and G.~Ridolfi, {\it {The POWHEG-hvq manual version
  1.0}},  \href{http://xxx.lanl.gov/abs/0707.3081}{{\tt 0707.3081}}.

\bibitem{Nason:2007vt}
P.~Nason, {\it {MINT: a Computer Program for Adaptive Monte Carlo Integration
  and Generation of Unweighted Distributions}},
  \href{http://xxx.lanl.gov/abs/0709.2085}{{\tt 0709.2085}}.

\bibitem{Frixione:2002ik}
S.~Frixione and B.~R. Webber, {\it {Matching NLO QCD computations and parton
  shower simulations}},  {\em JHEP} {\bf 06} (2002) 029,
  [\href{http://xxx.lanl.gov/abs/hep-ph/0204244}{{\tt hep-ph/0204244}}].

\bibitem{mcfm}
\texttt{http://mcfm.fnal.gov}.

\bibitem{Berger:2008sj}
C.~F. Berger {\em et~al.}, {\it {An Automated Implementation of On-Shell
  Methods for One-Loop Amplitudes}},  {\em Phys. Rev.} {\bf D78} (2008) 036003,
  [\href{http://xxx.lanl.gov/abs/0803.4180}{{\tt 0803.4180}}].

\bibitem{Sjostrand:2006za}
T.~Sjostrand, S.~Mrenna, and P.~Skands, {\it Pythia 6.4 physics and manual},
  {\em JHEP} {\bf 05} (2006) 026,
  [\href{http://xxx.lanl.gov/abs/hep-ph/0603175}{{\tt hep-ph/0603175}}].

\bibitem{Cacciari:2005hq}
M.~Cacciari and G.~P. Salam, {\it {Dispelling the $N^3$ myth for the $k_T$
  jet-finder}},  {\em Phys. Lett.} {\bf B641} (2006) 57--61,
  [\href{http://xxx.lanl.gov/abs/hep-ph/0512210}{{\tt hep-ph/0512210}}].

\bibitem{Abulencia:2005yg}
{\bf CDF} Collaboration, A.~Abulencia {\em et~al.}, {\it {Measurement of the
  inclusive jet cross section in $p\bar{p}$ interactions at $\sqrt{s} =
  1.96$~TeV using a cone-based jet algorithm}},  {\em Phys. Rev.} {\bf D74}
  (2006) 071103, [\href{http://xxx.lanl.gov/abs/hep-ex/0512020}{{\tt
  hep-ex/0512020}}].

\bibitem{cdf:webpage}
\texttt{http://www-cdf.fnal.gov/physics/new/qcd/QCD.html}.

\bibitem{D0:runIIjetalgo}
G.~C. Blazey and et~al., {\it {Proceedings of the Workshop: QCD and Weak Boson
  Physics in Run~II}}. Edited by U.~Baur, R.~K.~Ellis and D.~Zeppenfeld,
  Fermilab-Pub-00/297 (2000).

\end{thebibliography}\endgroup

\end{document}